 \documentclass[preprint2,twoside]{myaastex}

\usepackage{amssymb}                       
\usepackage{amsmath}                       
\usepackage{wasysym}
\newcommand{\doverdt}[1]{\frac{\partial #1}{\partial t}}

\newcommand{\subscr}[1]{_\mathrm{#1}}
\newcommand{\superscr}[1]{^\mathrm{#1}}
\newcommand{\gv}[1]{\boldsymbol{#1}}      
\newcommand{\ud}{\mathrm{d}}               
\newcommand{\refEqt}[1]{Equation~(\ref{#1})}   
\newcommand{\refeqt}[1]{equation~(\ref{#1})}   
\newcommand{\refeqs}[2]{equations~(\ref{#1}) 
                        and (\ref{#2})}
\newcommand{\refeqp}[1]{eq.~[\ref{#1}]}        
\newcommand{\refFgt}[1]{Figure~\ref{#1}}       
\newcommand{\refFgp}[1]{Fig.~\ref{#1}}         
\newcommand{\refTab}[1]{Table~\ref{#1}}        
\newcommand{\refSec}[1]{Section~\ref{#1}}      
\newcommand{\refsec}[1]{\S~\ref{#1}}           
\newcommand{\gAA}{\textsc{Paper~I}}            
\newcommand{\gApJ}{\textsc{Paper~II}}           
\newcommand{\MSun}{\mbox{$M_{\odot}$}}
\newcommand{\MStar}{\mbox{$M_*$}}
\newcommand{\MEarth}{\mbox{$M_{\oplus}$}}
\newcommand{\MJup}{\mbox{$M_{\mathrm{J}}$}}
\newcommand{\Mp}{\mbox{$M_{\mathrm{p}}$}}
\newcommand{\Md}{\mbox{$M_{\mathrm{D}}$}}
\newcommand{\dMp}{\dot{M}_{\mathrm{p}}}          
\newcommand{\RH}{R_{\mathrm{H}}}                 
\newcommand{\AU}{\mbox{\textrm{AU}}}

\newcommand{\dunits}{\mbox{$\mathrm{g}\,\mathrm{cm}\superscr{-3}$}}
\newcommand{\sdunits}{\mbox{$\mathrm{g}\,\mathrm{cm}\superscr{-2}$}}
\newcommand{\kelvin}{\mbox{$\mathrm{K}$}}      
\newcommand{\nunits}{\mbox{$\mathrm{cm}\superscr{2}\,%
                            \mathrm{s}\superscr{-1}$}}
\newcommand{\dpl}{s}                           

\received{year Month day}
\revised{year Month day}
\accepted{year Month day}
\ccc{code}
\cpright{none}{73 AC}

\slugcomment{\small{September 2003}}

\shorttitle{Thermo-Hydrodynamics of Disks with Planets}
\shortauthors{D'Angelo, Henning, \& Kley}
\begin{document}


\title{\textbf{%
       Thermo-Hydrodynamics of Circumstellar Disks\\ 
       with High-mass Planets\footnote{%
       To appear in \textsc{The Astrophysical Journal} 
       (v599 n1 December 10, 2003 issue).
       Also available as ApJ preprint doi:
       \texttt{10.1086/379224}.}}}


\author{\textsc{Gennaro D'Angelo\footnote{UKAFF Fellow.}}}
\affil{School of Physics,
       University of Exeter, 
       Stocker Road,
       Exeter EX4 4QL, 
       United Kingdom}
\email{gennaro@astro.ex.ac.uk\\[2mm]}

\and
\author{\vspace*{-5mm}\textsc{Thomas Henning}}
\affil{Max-Planck-Institut f\"ur Astronomie,
    K\"onigstuhl 17,
    D-69121 Heidelberg,
    Germany}
\email{henning@mpia-hd.mpg.de\\[2mm]}

\and
\author{\vspace*{-5mm}\textsc{Willy Kley}}
\affil{Computational Physics,
       Auf der Morgenstelle 10,
       D-72076 T\"ubingen, Germany}
\email{wilhelm.kley@uni-tuebingen.de\\[-20mm]}



\small

\begin{abstract}
With a series of numerical simulations, we analyze the 
thermo-hydrodynamical evolution of circumstellar disks 
containing Jupiter-size protoplanets.  
In the framework of the two-dimensional approximation,
we consider an energy equation that includes viscous heating 
and radiative effects in a simplified, yet consistent form.
Multiple nested grids are used in order to study both global 
and local features around the planet.
By means of different viscosity prescriptions, we investigate
various temperature regimes.
A planetary mass range from $0.1$ to $1\;\MJup$ is examined.
Computations show that gap formation is a general property  
which affects density, pressure, temperature, optical thickness, 
and radiated flux distributions.
However, it remains a prominent feature only when the kinematic 
viscosity is on the order of $10^{15}\;\nunits$ or lower, though 
it becomes rather shallow for $0.1\,\MJup$ perturbers. 
Around accreting planets, a circumplanetary disk forms that has
a surface density profile decaying exponentially with the distance
and whose mass is 
$5$--$6$ orders of magnitudes smaller than Jupiter's mass.
Circumplanetary disk temperature profiles decline roughly as the 
inverse of the distance from the planet, matching the values measured 
in the gap toward the border of the Roche lobe. 
Temperatures range from some $10$ to $\sim 1000\;\kelvin$. 
Moreover, circumplanetary disks are generally opaque, with optical thickness
larger than $1$ and aspect ratios around a few tenths.
Non-accreting protoplanets provide quite different scenarios, with
a clockwise, i.e., reversed flow rotation around low-mass bodies. 
Planetary accretion and migration rates depend on the viscosity
regime, with discrepancies within an order of magnitude.
Coorbital torques increase as viscosity increases.
For high viscosities, Type~I migration may extend to larger
planetary masses.
Estimates of growth and migration time scales inferred by these models
are on the same orders of magnitude as those previously obtained with
locally isothermal simulations, both in two and three dimensions.
\end{abstract}



\section{Introduction}
\label{introduction}

Since the very first attempt of modeling
a protoplanet interacting with its primitive nebula by \citet{miki1982},
two decades have passed. Seventeen years had to go by for tackling
this problem with the strength of new and more powerful computational tools
\citep{bryden1999,kley1999,lubow1999,miyoshi1999}.
As a consequence of the mounting interest in extrasolar
planets, boosted by eight years of uninterrupted new detections\footnote{%
Data archives on extra-solar planets, with the latest information, 
can be found at the \textit{Extrasolar Planet Encyclopedia} 
(\url{http://www.obspm.fr/planets})
and at the \textit{California \& Carnegie Planet Search} 
(\url{http://exoplanets.org/}).},
the scientific community has made an extraordinary effort,
over the past four years, to analyze
disk-planet interactions by means of numerical calculations.

Yet,
regardless of the processes accounted for in the implemented numerical
models, limitations and restrictions remain an issue to deal with.
As a matter of fact, numerical calculations still lag behind the
present knowledge of the possible physical effects which occur in a
circumstellar environment containing massive bodies. 
Thereupon, simplifying assumptions are always demanded.
Nonetheless, steps forward have been made in many directions 
to reduce their number.

Three-dimensional (3D) disks containing Jupiter-size objects have been
investigated by \citet{kley2001} and, including also magnetic fields, by
\citet{papa2003}, \citet{rnelson2003}, and \citet{winters2003}. 
\citet[][hereafter \gAA]{g2002} carried out global two-dimensional (2D)
high-resolution
simulations of Jupiter-mass as well as Earth-mass planets,
resolving also the flow structure within the planetary's Roche lobe
by applying a resolution enhancement strategy. 
This study has been lately extended to the 3D space by 
\citet[][hereafter \gApJ]{g2003}
and by \citet{bate2003}, who used instead a single grid with variable
step size. \citet{anelson2003a,anelson2003b} performed 2D
computations, considering the effects due to disk self-gravity.
\citet{tanaka2003} have investigated the
interactions between a planet and an optically thin disk with local 
simulations in the shearing sheet approximation, in order to determine
the one-sided gravitational torque acting on the planet.

With the present work we intend to
relax the widely adopted local-isothermal approximation, i.e., that of
treating a disk as a system having a fixed temperature distribution
which depends neither on time nor on any other system variable.
Yet, we will preserve the global viewpoint which we believe to be
crucial because disk-planet interactions can be highly non-local.

Many of the sophisticated accretion disk studies adopt 2D or 
1+1D schemes due to the difficulty of dealing with all the necessary 
ingredients in a full 3D space. 
Until very recently, in fact,
radiation \citep[e.g.,][]{dullemond2002}, 
convection  \citep[e.g.,][]{agol2001}, 
magnetic fields \citep[e.g.,][]{matt2002} , 
mixtures of gas and dust  \citep[e.g.,][]{suttner2001}, and 
chemical evolution \citep[e.g.,][]{martin2002} 
have been treated via two-dimensional models.
Furthermore, numerical simulations generally involve parameter studies, 
because many physical quantities are poorly known. Thereby,
with a restricted number of dimensions, one is at least able to acquire
a spectrum of possible approximate solutions.

In order to overcome the local-isothermal hypothesis and to maintain
a global approach at the same time, we introduce
an energy transport in 2D ($r$--$\varphi$) models. 
We include all the supposedly major causes responsible for generation,
transfer, and loss of energy in low-temperature circumstellar disk
environments.
We take advantage
of the small aspect ratio of the disk, and assume that all the 
radiation transport is effective only in the vertical direction.
Such approximation works reasonably well in accretion disks, away from 
the boundary and surface layers 
\cite[see, e.g.,][]{pringle1981}. Though such an approach
relies on the ``slimness'' assumption, it is as well 
appropriate in the local environment around a protoplanet. 
In fact, around there, the disk becomes even thinner when the
planet's gravitational action is accounted for. Hence, the amount of
energy transported by radiation in the vertical direction still 
overwhelms that transported horizontally.
Although this represents a simplistic kind of thermal description,
it allows to investigate processes which have not been considered
thus far, i.e., the joint thermo-hydrodynamical evolution of 
disk-planet interactions. 
Furthermore, since these simulations benefit of a
nested-grid numerical technique (\gAA, \gApJ), though relying on
the global approach we are capable of achieving resolutions
large enough to accurately describe the system within the planetary's
Roche lobe.

\pagestyle{myheadings}
\markboth{\hfill Thermo-Hydrodynamics of Disks with Planets \hfill}%
       {\hfill \textsc{G. D'Angelo, Th. Henning, \& W. Kley} \hfill}
The outline of the paper is the following. In the next section
we introduce the physical formulation of the problem, focusing on the 
equation for the thermal energy density and related issues. 
\refSec{sec:esolver} concerns the numerical method
utilized to solve the energy equation along with a series
of tests.
In \refsec{sec:parameters} we specify the adopted parameters
and other various details. 
Sections~\ref{sec:global} and \ref{sec:pstructure} 
describe, respectively, global and local properties of models with different
masses and viscosity regimes.
In \refsec{sec:accmig} we report the
results for mass accretion and orbital migration.
Our conclusions are presented in \refsec{sec:conclusions}.
\section{Physical Formalism}
\label{sec:model}

As mentioned before, we are
going to tackle the problem of energy transfer in an accretion disk
with an embedded planet.
Since solving the full set of equations in three dimensions and
simultaneously achieving sufficient resolution around the protoplanet
is computationally difficult at the moment, 
we restrict to two-dimensional disk models.

The reason for this choice is two-folded. On the one hand, 
with affordable computational times and sufficiently high
resolutions, 
it allows to investigate the coupling between hydrodynamics and 
thermodynamics both in the disk and nearby the embedded body. 
On the other, it permits
to adopt a strategic assumption in writing the energy equation,
i.e., that the horizontal energy transport can be neglected
compared to the vertical one. In other words, in a 2D disk 
one can assume that radiation transport is only effective along
the direction perpendicular to the equatorial plane of the system.

Hence, let us describe the disk material through the Navier-Stokes
equations for the surface density $\Sigma$, the linear momentum
(density) $\Sigma\,u_r$, and the angular momentum (density) 
$r\,\Sigma\,u_\varphi$, where we choose as working 
coordinate frame a cylindrical one $\{O; r, \varphi, z\}$.
The origin is located at the center of mass of the star and the 
planet while the disk lies in the plane $z=0$.
The hydrodynamics equations have been already stated, by using the
same notations, in \gAA.

As in most of the computations performed so far, 
we avoid dealing with the Poisson equation and rather assume that the
gravitational field is determined only by the star and the planet,
neglecting the self-gravity of the disk material. 
This is allowed by the circumstance that the Toomre $Q$-parameter is
well beyond $1$ in all of our models.
\citet{anelson2003a,anelson2003b} found that some effects indeed arise
when disk self-gravity is taken into account. However, we note that they
consider disk masses ten times as large as those considered here.

Thereby, we rely on a smoothed point-mass gravitational potential,
which well reproduces more realistic protoplanetary gravitational
potentials (\gApJ),
\begin{equation}
    \Phi =
         - \frac{G\,\MStar}{| \gv{r} - \gv{r}_{*} |}
         -  \frac{G\,\Mp}{\sqrt{|\gv{r}-\gv{r}_\mathrm{p}|^2 + \delta^2}},
         \label{eq:potential}
\end{equation}
where $\gv{r}_{*}$ and $\gv{r}_\mathrm{p}$
are the radius vectors indicating the positions of the star and 
the planet, respectively. As done in \gApJ,
we set the parameter $\delta$ as a constant length 
(see \refsec{sec:parameters} for details).

The thermal structure of the fluid has been usually fixed through
a power law of the distance $r$. Thereupon,
the set of equations has been closed by adding an
equation of state which connects the local gas
(two-dimensional) pressure to the surface density by means of a
local isothermal sound speed $c_s^{\mathrm{iso}}$
\begin{equation}
    c_s^{\mathrm{iso}} = h\,\sqrt{\frac{G\,\MStar}{r}}.
         \label{eq:csiso}
\end{equation}
Thus, the Mach number of the flow is considered constant throughout
the system's evolution, constrained solely by the disk aspect ratio $h=H/r$
\citep[e.g.,][]{lubow1999,miyoshi1999,rnelson2000,tanaka2002,masset2002,rnelson2003,bate2003,anelson2003b}.
Alternatively, \citet{kley1999} and \citet{tanigawa2002} did some
simulations using a polytropic equation 
of state of the type $P=\bar{K}\,\Sigma^{\gamma}$. However, the
evolution of the thermal properties of the system have not been
generally considered.

In the present calculations we use an ideal equation of state
which directly ties the gas pressure $P$ to the thermal energy density
(energy per unit area) $E$
\begin{equation}
    P = (\gamma -1)\,E,
         \label{eq:es}
\end{equation}
where the adiabatic index $\gamma$ is a constant.
Furthermore, if we suppose that disk material behaves as a perfect gas, 
then the temperature is
\begin{equation}
    T = \left(\frac{\mu\,m_\mathrm{H}}{k}\right)\frac{P}{\Sigma},
         \label{eq:T}
\end{equation}
where $\mu$ is the mean molecular weight, $m_\mathrm{H}$ is the
hydrogen mass, and $k$ the Boltzmann constant. 
The adiabatic sound speed is given by
\begin{equation}
    c_s = \sqrt{\gamma\,\left(\frac{k\,T}{\mu\,m_\mathrm{H}}\right)}.
         \label{eq:cs}
\end{equation}

\refEqt{eq:es} implicitly assumes that radiation
pressure $P_{\mathrm{rad}}$ is negligibly small compared to gas pressure.
Such hypothesis is connected to the relatively low temperature regimes 
we deal with
(in fact $P_{\mathrm{rad}}\propto T^4$) and its validity was checked 
afterward. In all of the models presented here, the ratio 
$P_{\mathrm{rad}}/P_{\mathrm{gas}}$ never exceeds
$10^{-4}$.

\subsection{Energy Equation}
\label{ssec:energyeq}

Equations for energy transport differ according to the processes that 
have to be included for a correct description of the energy budget
of fluid elements. In our case, we suppose that a parcel of disk
fluid can gain or lose thermal energy only because of flow advection, 
compressional work, viscous dissipation, and dissipative effects due to
radiation transport. In this sense, rather than strictly treating 
the transfer of radiation in the disk, we will account only for the 
cooling effects that radiation causes.
Then, the energy equation takes the following form
\begin{equation}
 \doverdt{E} + \nabla \cdot (E\,\gv{u})
                =  -P \, \nabla\cdot\gv{u} + \Upsilon -\Lambda.
\label{eq:eneq}
\end{equation}
In \refeqt{eq:eneq} we indicated with $\Upsilon$ the dissipation
function and with $\Lambda$ the radiated energy. In two dimensions,
either of these quantities is an energy flux. 
Both functions are always positive, thereby the former acts
as a heating source, whereas the latter as a cooling term.
In this study we do not consider convection because we do not expect
it to be a major energy source \citep{paola1998}. 
Furthermore, we neglect irradiation from the central star which
is presumably an effective heating mechanism only in the upper layers
of circumstellar disks \citep[see, e.g.,][]{paola1998}, whose effects
however are reduced by small disk scale heights.
More importantly, as a consequence of the gap formation and the 
protoplanet's gravity, circumplanetary material is 
shielded from stellar radiation, as it will be clarified later.

The function $\Upsilon$ can be directly
computed from the components of the viscous stress tensor  
$S_{rr}$, $S_{\varphi\varphi}$, and $S_{r\varphi}$
\citep{m&m,collins1998}
\begin{equation}
 \Upsilon = \frac{1}{2\,\nu\,\Sigma}\left(%
 S^2_{rr} + S^2_{\varphi\varphi} + S^2_{r\varphi}
\right) + \frac{2\,\nu\,\Sigma}{9}\,(\nabla\cdot\gv{u})^2.
  \label{eq:upsilon}
\end{equation}
The divergence term $\nabla\cdot\gv{u}$ in the equation arises because
a three-dimensional definition of stress tensor 
\textbf{\textsf{S}} is adopted.
In fact, the reduction from three to two dimensions is merely achieved
by setting the vertical component of the flow field $u_z$ and the
derivative operator $\partial/\partial z$ to zero.

In a three-dimensional disk, the last term on the right-hand
side of \refeqt{eq:eneq} is equal to $\nabla\cdot\gv{F}$, where
$\gv{F}$ is the frequency-integrated radiation flux
\begin{equation}
\gv{F}= -\frac{16\,\sigma_{\mathrm{R}}}{3\,\kappa\,\rho}\, T^3\, \nabla T.
  \label{eq:radflux}
\end{equation}
In the above equation $\sigma_{\mathrm{R}}$ is the Stefan-Boltzmann
constant, $\kappa$ is a frequency-integrated opacity coefficient and 
$\rho$ is the mass density.
As mentioned before, we suppose that the amount of energy
transported by radiation in the vertical direction is much larger
than that transported horizontally, i.e., $|F_r|$ and $|F_\varphi| \ll
|F_z|$. The validity of this statement
holds as long as the vertical extent of the disk remains very small
compared to the disk characteristic size in the other directions. 
Thus, in a two-dimensional cylindrical disk we have
\begin{equation}
\Lambda = \int_{-\infty}^{+\infty} \nabla\cdot\gv{F}\, \ud z \simeq
          \int_{-\infty}^{+\infty} \frac{\partial F_z}{\partial z}\,
          \ud z.
  \label{eq:lambda1}
\end{equation}
Since the vertical disk structure is not meant to be accounted for,
it is possible to equate the pressure scale height to the photospheric scale
height and assume that all the radiation is liberated at $z=\pm H$.
Thereupon, \refeqt{eq:lambda1} becomes
\begin{equation}
\Lambda = \int_{-H}^{+H} \frac{\partial F_z}{\partial z}\, \ud z
        = F(H) - F(-H) = 2\,F(H).
  \label{eq:lambda2}
\end{equation}
In 2D disks it is useful to measure the emitted flux by means of
an \textit{effective temperature}:
$F(H)=\sigma_{\mathrm{R}}\,T_\mathrm{eff}^4$, therefore
\begin{equation}
\Lambda = 2\,\sigma_{\mathrm{R}}\,T_\mathrm{eff}^4.
  \label{eq:lambda3}
\end{equation}
The factor $2$ indicates that in a disk radiation escapes from both
of its sides.

A simple relation between the midplane temperature and the
emergent radiation flux can be found by writing
\begin{eqnarray}
\Lambda&=&2\,\int_{0}^{H} \frac{\partial F_z}{\partial z}\, \ud z \nonumber\\
       &=&2\,\left[F(H) - F(0)\right] \nonumber\\
       &\approx&
                            \frac{8}{3}
                            \frac{\sigma_{\mathrm{R}}}{\kappa\,\rho\,H}
                            \left[T^4(0)-T^4(H)\right].
  \label{eq:teff_a}
\end{eqnarray}
In the inner parts of a circumstellar disk
the inequality  $T^4(0) \gg T^4(H)$ generally holds
\citep[e.g.,][]{bell1997,paola1998}, 
hence \refeqt{eq:teff_a} yields
\begin{equation}
\Lambda \approx \frac{2\,\sigma_{\mathrm{R}}\,T^4}{(3/4)\,\tau},
  \label{eq:teff_c}
\end{equation}
in which we introduced the optical thickness 
$\tau=\kappa\,\rho\,H=\frac{1}{2}\,\kappa\,\Sigma$ 
(we generally adopt a Rosseland mean opacity).
We notice that
with the previous definition the total disk optical thickness
is $2\,\tau$. From now on, the quantity $T$ refers to the disk
midplane temperature.

\refEqt{eq:teff_c} represents a fairly good approximation when
the medium is very optically thick, i.e., $\tau\gg 1$.
This is indeed the case in those regions of unperturbed accretion disks
which we simulate, because $\Sigma$ is large enough.
Yet, because of the action of massive bodies, in our case 
deep density gaps form where material is very diluted. In addition there
are zones, where the disk spirals interact with the circumplanetary
disk spirals, which can have very low densities 
(see \gAA, Fig.~12 and 13). 
Such conditions
make \refeqt{eq:teff_c} not always applicable.
\citet{hubeny1990} found a more rigorous relation between the
effective and midplane accretion disk temperature, which represents a
generalization of the \textit{gray model} of classical stellar
atmospheres in local thermodynamic equilibrium. Following Hubeny's
theory, in a circumstellar disk we can write
\begin{equation}
\Lambda = 2\,\sigma_{\mathrm{R}}\,T^4%
               \left[%
                \frac{3\,\tau}{8} +
                \frac{\sqrt{3}}{4}+
                \frac{\varepsilon_\mathrm{H}}{4\,\tau}
                \right]^{-1}.
  \label{eq:Lambda}
\end{equation}
According to \refeqt{eq:Lambda}, the emitted flux is inversely
proportional to $\tau$ in optically thick disk portions, whereas it
becomes proportional to $\tau$ in the opposite limit.
The quantity $\varepsilon_\mathrm{H}$ is equal to the ratio between 
the Rosseland and the Planck mean opacities. This quantity can be also roughly
interpreted as the ratio of \textit{total extinction} (i.e.,
absorption plus scattering coefficient) to the pure absorption
coefficient: $\kappa_\mathrm{ext}=\kappa_\mathrm{abs}+\kappa_\mathrm{sc}$. 
In these computations, we set
$\varepsilon_\mathrm{H}= 1$ because we checked that we can neglect 
radiation scattering (see \refsec{ssec:tests}). Thus, no distinction
is made between the extinction and the absorption coefficients.

\refEqt{eq:Lambda} mimics the radiative losses of both optically
thick and optically thin media, and therefore is very suitable to
our purposes. In fact, it has been successfully applied to many studies
on accretion disks \citep{popham1993,godon1996,collins1998,hure2001}.

\subsection{Opacity Table}
\label{ssec:opacity}

\begin{figure}[!t]
\epsscale{1.0}
\plotone{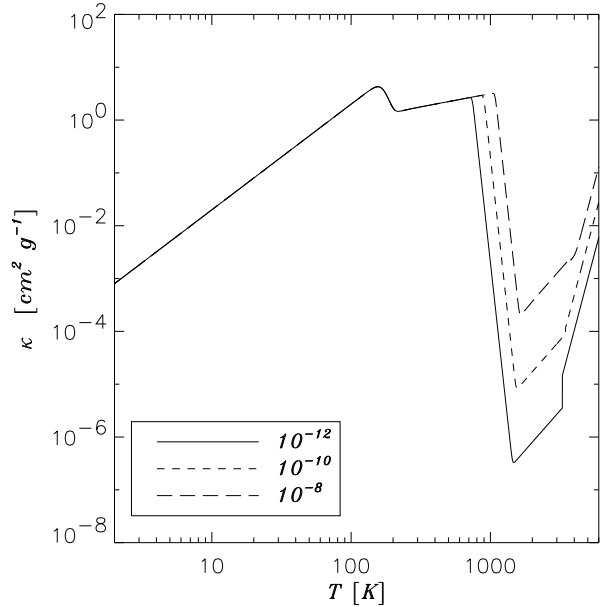}
\caption{\small{Rosseland opacity coefficient $\kappa$ as function of the
temperature for three different values of the mass
density $\rho$ \citep[from][]{bell1994}.
The legend quotes the density magnitude in \dunits.}
\label{f:kappavst}}
\end{figure}
In order to calculate the disk optical semi-thickness $\tau$, we adopt
the opacity formulas derived by \citet{bell1994} (see
\refFgp{f:kappavst}), as an improvement of those by \citet{lin1985}.
Eight temperature regimes are identified according to the dominant
processes active in each of them. 
Contributions from dust grains, molecules, atoms, and ions are 
accounted for. 
Since we simulate a distance range where disk material can be 
relatively cold, dust opacity is 
crucial for an accurate estimation of radiative losses \citep{lin1981}. 
Because of this,
Bell's opacity includes grain absorption as tabulated by 
\citet{alexander1989}. 

However, for the sake of comparison we ran some test cases (see
\refsec{ssec:tests}) with a new opacity coefficient developed by
Semenov and collaborators (2002, private communication) 
and based on the improved grain opacity tables provided by \citet{henning1996}.
They aimed at coupling gas and dust opacities, focusing on the
temperature range proper to protostellar disk environments.

In either case, the midplane temperature $T$ and mass density 
$\rho$ have to
be provided in cgs units. In turn $\kappa=\kappa(T,\rho)$, has
units of square centimeters per gram.
Within the
infinitesimally thin disk approximation, dynamics variables are
vertically integrated. Thus, the mass density $\rho$ is not
directly available. It must be retrieved from the surface
density and the disk semi-thickness $H$. 
For this purpose, we assume that
\begin{equation}
\rho=\frac{\Sigma}{2\,H}.
\end{equation}
In the next section we explain how the disk scale height $H$ is
calculated in a fashion such to account also for the gravitational
influence of the planet. This guarantees a more refined and consistent
modeling of the system.
 
\subsection{Disk Scale Height}
\label{ssec:height}
Inside a circumstellar disk it is natural to assume that material is
in hydrostatic equilibrium along the vertical direction 
\citep{pringle1981,frank1992}.
Let $g_z$ be the gravitational acceleration in the $z$-direction
within the disk, then the hydrostatic equilibrium reads
\begin{equation}
\frac{P}{\Sigma} = 2 \int_{0}^{H} g_z\, \ud z.
\label{eq:H1}
\end{equation}
The factor $2$ is necessary because the midplane pressure
sustains both sides of the disk.

When the temperature distribution is not coupled to the rest of the
dynamical variables, \refeqt{eq:H1} is used to obtain 
the disk pressure from the density and the pressure scale height,
assuming that $g_z=G\,\MStar/(r^2 + z^2)$ and omitting the planet
gravitational attraction.

In the present study, we compute the pressure directly from the
thermodynamical processes occurring in the gas (\refeqp{eq:es}), 
thus we can use
\refeqt{eq:H1} to estimate the disk scale height $H$ in a consistent 
fashion. This quantity, in fact, is
needed to obtain the mass density from the surface density, 
as explained above. 
Since the left-hand side
of \refeqt{eq:H1} is equal to $c_s^2/\gamma$, 
we can integrate the right-hand side in order to get an implicit
function of $H$. Furthermore, the effects due to the gravitational
field of the protoplanet can be included in this evaluation.
If we set $q=\MStar/\Mp$ and $\dpl =| \gv{r} - \gv{r}_{\mathrm{p}} |$,
\refeqt{eq:H1} generates the implicit function
\begin{equation}
 \frac{H^2}{2\,r^2} - \frac{q}{\sqrt{(\dpl/r)^2 + (H/r)^2}} = 
 \frac{r}{2\,\gamma}\left(\frac{c_s^2}{G\,\MStar}\right) -
 q\,\left(\frac{r}{\dpl}\right).
 \label{eq:H2}
\end{equation}

Although there are no constraints on the ratio $\dpl/r$, because with nested
grid computations that ratio can be very small, the aspect disk ratio
$H/r$ is smaller than one, by working hypothesis, otherwise the 2D
geometry would not be consistent. Therefore, the second term on the
left-hand side can be expanded in a binomial series in $H/r$ (up to
the its second power), obtaining $H$ as an explicit function
\begin{equation}
  H^2 = \frac{r}{\gamma}\left(\frac{c_s^2}{G\,\MStar}\right)
        \left[\frac{1}{r^2} +  q \left(\frac{r}{\dpl^3}\right)\right]^{-1}.
 \label{eq:H3}
\end{equation}
We note that, in the limit $q \rightarrow 0$ or $\dpl \gg r$, 
\refeqt{eq:H3} yields $\sqrt{\gamma}\,H=c_s/\Omega_\mathrm{K}$,
as in regular accretion disks.
Close to the planet, i.e., in the circumplanetary disk where
$\dpl \ll r$, the above relation reduces to 
$\sqrt{\gamma}\,H=c_s/\sqrt{G\,\Mp/\dpl^3}$, which resembles the
previous limit because the circumstellar Keplerian angular velocity is
replaced by the circumplanetary one.
Rearranging the terms in \refeqt{eq:H3}, a more compact expression
can be written
\begin{equation}
  H^2 = \frac{1}{\gamma}\left(\frac{c_s}{\Omega_\mathrm{K}}\right)^2
        \left[1 +  q\,\left(\frac{r}{\dpl}\right)^3\right]^{-1}.
 \label{eq:H}
\end{equation}
\refEqt{eq:H} has a singularity at $\dpl = 0$. This arises from
the corresponding singularity in the gravitational potential $\Phi$,
which is overcome by introducing the smoothing length 
(\refeqp{eq:potential}). In our computations, for consistency
reasons, the distance $\sqrt{s^2 + \delta^2}$ substitutes
$\dpl$ in \refeqt{eq:H}. 

\subsection{Artificial Viscosity}
\label{ssec:aviscosity}

Non-linear effects in wave propagation inevitably lead to shock
formation. This is indeed the case in disk-planet simulations. 
In ideal fluids shocks are mathematical
discontinuities. Therefore, in finite-differencing schemes, they must
extend only over one or two grid points. In order to provide the 
correct jump conditions, ahead and behind a shock front, dissipative 
terms have to be present in the equations. This is usually done
by introducing a non-linear viscous pressure otherwise known as
\textit{artificial viscosity}.
Since the pressure gradient is no longer proportional to the density 
gradient, shocks stronger than those observed in local-isothermal 
computations (see, e.g., \gAA) may develop.
Therefore, the physical 
viscosity might not be sufficient to provide the correct jump conditions
across shock fronts.

The most rigorous treatment of shocks in a multidimensional space, 
with a generic metric tensor, requires the definition of an isotropic 
viscous stress tensor \textbf{\textsf{Q}}
\citep{winkler1986}, which can be written as \citep{m&m,stone1992a}
\begin{equation}
\mbox{\textbf{\textsf{Q}}}= \mu_\mathrm{Q}\,\left[\nabla \gv{u} - \frac{1}{3}
  (\nabla\cdot\gv{u})\,\mbox{\textbf{\textsf{I}}}\right],
\label{eq:Q}
\end{equation}
being \textbf{\textsf{I}} the unit tensor.
Here we choose to discard the off-diagonal, i.e., shear tensor components 
because they may lead to artificial angular momentum transport
\citep{stone1992a}. 
The coefficient of artificial viscosity is defined as
\begin{equation}
\mu_\mathrm{Q} = - \mathcal{L}^2\,\Sigma\,\min{(\nabla\cdot\gv{u}, 0)},
\label{eq:muQ}
\end{equation}
where $\mathcal{L}$ represents the length over which the
shock is smeared out. This is usually fixed to the maximum grid spacing.
The coefficient
$\mu_\mathrm{Q}$ is positive only for compression and zero for
expansion, so the artificial viscosity is large in shocks and
negligibly small elsewhere. 

Since the artificial viscous tensor acts as a pseudo-pressure, it
affects both momentum and energy equations through the terms
$\nabla\cdot\mbox{\textbf{\textsf{Q}}}$ and 
$\mbox{\textbf{\textsf{Q}}}\cdot\nabla \gv{u}$ (meant as a tensor
product), respectively. 
The aforementioned equations are updated, by using
the correct components $Q_{ij}$, as explained in \citet{stone1992a}.
One severe side-effect
is that \textbf{\textsf{Q}} can reduce the viscous
time step by a factor which is proportional to 
$\left[\mathcal{L}^2\,\nabla\cdot\gv{u}\right]^{-1}$.
\section{Energy Equation Solver}
\label{sec:esolver}

The general numerical method employed to solve the
hydrodynamical equations on a hierarchy of nested grids,
applied to simulations of disk-planet interactions, has been explained in
\gAA. 
In contrast to previous calculations, thermal 
energy density $E$ now appears as an independent variable. 
Here we outline how the source 
terms (right-hand side of \refeqp{eq:eneq}) are dealt with.
In the framework of the nested-grid scheme, whenever required, $E$ is 
interpolated from a finer to a coarser grid and \textit{vice versa},
according to the procedures outlined in \gAA\
(see also \gApJ).

The energy equation (\refeqp{eq:eneq}) is solved by means of
a multi-step operator splitting method. The first step takes
care of the energy advection and this is done in the same
fashion as the advection of the other dynamical quantities, i.e.,
through the van Leer's algorithm.
Viscous dissipation and radiative cooling are treated
by means of a predictor-corrector scheme, which is second-order
accurate in time,
where heating and cooling terms are considered simultaneously
for better accuracy and stability.

During the course of the first experiments, it was discovered that,
due to very high pressures and large values of the flow divergence
(in the wake of the circumplanetary spirals) the compressional work 
could lower the thermal energy by a large amount. Consequently, the
predictor-corrector procedure applied to this equation term
occasionally occurred to be unstable, producing negative energy values.
Therefore we decided to take advantage of \refeqt{eq:es} and use an
analytic solution for updating the energy.

In the framework of the operator-splitting approach, in order to correct
the energy because of the gas compression/dilation, we have to
solve numerically the equation
\begin{equation}
 \doverdt{E} =  -P \, \nabla\cdot\gv{u}.
\end{equation}
With the aid of the state equation, this becomes
\begin{equation}
 \doverdt{E} =  -(\gamma-1)\,E \, \nabla\cdot\gv{u}.
 \label{eq:step1}
\end{equation}
\refEqt{eq:step1} has an analytic solution, given by
$E=E_0\,\exp{[-(\gamma-1)\,\Delta t\,\nabla\cdot\gv{u}]}$, where
$\Delta t$ is a time interval over which the divergence 
$\nabla\cdot\gv{u}$ has not appreciably changed. 
Hence, thermal energy density can be updated as follows:
\begin{equation}
 E_{n+1}=E_n\,\exp{[-(\gamma-1)\,\Delta t_n\,\nabla\cdot\gv{u}_n]}.
 \label{eq:compup}
\end{equation}
 
We notice that,
expanding the exponential of \refeqt{eq:compup} in a Taylor
series and keeping all the terms up to the second order in
$\Delta t_n$, one finds
\begin{eqnarray}
 \frac{E_{n+1}-E_n}{\Delta t_n}& = & -P_n\,\nabla\cdot\gv{u}_n \nonumber \\ 
                               & + & 
 \frac{1}{2}\,(\gamma-1)\,P_n\,\Delta t_n\,(\nabla\cdot\gv{u}_n)^2,
 \label{eq:pceqiv}
\end{eqnarray}
where $P_{n} = (\gamma-1)\,E_{n}$.
It is easy to prove that \refeqt{eq:pceqiv} is equivalent to a 
predictor-corrector scheme. The above equation also shows that the
second-order correction always increase the thermal energy.

Although the procedure in \refeqt{eq:compup} is not so general as the
predictor-cor\-rec\-tor
algorithm, it is more accurate and unconditionally stable. 
Furthermore, contrary to what may seem at first glace, 
from the computational
viewpoint it is faster than the other
because it requires less mathematical operations.

\subsection{Solver Test}
\label{ssec:tests}

\begin{figure*}[!t]
\epsscale{1.8}
\plottwo{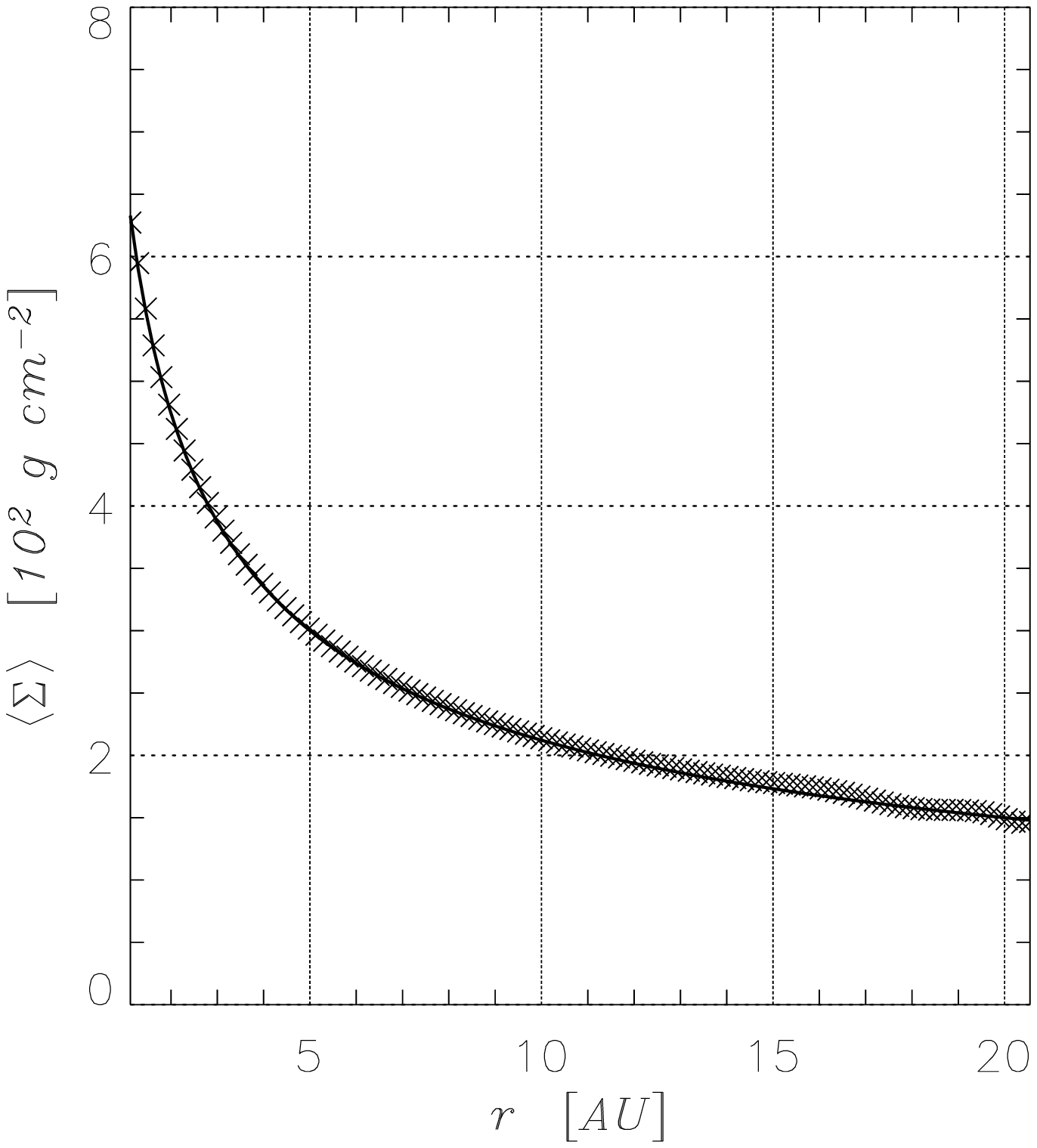}{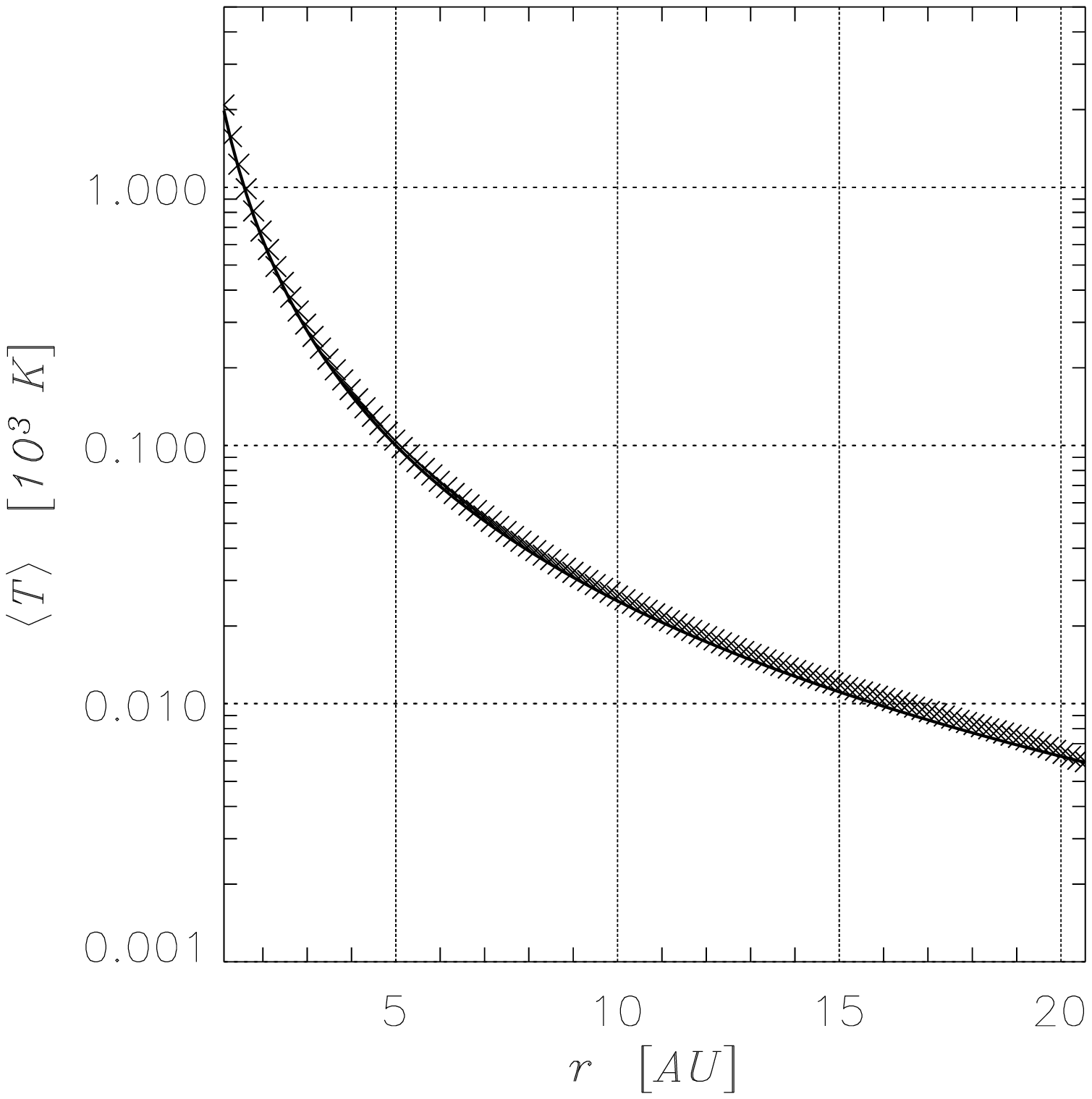}
\caption{\small{Test simulation of an unperturbed disk in which
       equations~(\ref{eq:simpleUpsilon}),
       (\ref{eq:simpleLambda}), and (\ref{eq:morfill})
       are implemented.
       \textit{Left}:
         Crosses represent the computed $\Sigma$, averaged
         along the azimuth. The solid line indicates
         the power-law $\Sigma=\Sigma_0\,\sqrt{r_{\mathrm{p}}/r}$ 
         (see text).
       \textit{Right}:
         The analytical formula for the midplane temperature
         (\refeqp{eq:Tan}) is indicated in the plot by a
         solid line.
         Crosses represent the computed average temperature.}
\label{f:Mtest}}
\end{figure*}
We have tested the equation energy solver from both the numerical
and the physical point of view. Here we present some of these tests. 

The first test is intended to check whether the equation
of energy furnishes physically consistent results by comparing 
the computed temperature to that derived from some analytical
solutions of \refeqt{eq:eneq}.
If we deal with a stationary Keplerian
disk then the energy budget simplifies enormously. In fact,
in a pure Keplerian flow both energy advection and compressional
work are negligibly small. Therefore \refeqt{eq:eneq} reduces to
\citep[e.g., see][]{pringle1981}
\begin{equation}
   \Upsilon -\Lambda = 0.
   \label{eq:simplebudget}
\end{equation}
From the same hypotheses, it follows that the dissipation
function can be written in a simple form
\begin{equation}
   \Upsilon = \Sigma\,\nu\,\left(%
              r\,\frac{\ud \Omega_\mathrm{K}}{\ud r}\right)^2
            = \frac{9}{4}\,\Sigma\,\nu\,\Omega_\mathrm{K}^2.
   \label{eq:simpleUpsilon}
\end{equation}
In fact, in a Keplerian flow, the terms $(\partial u_r/\partial r)^2$
and $(u_r/r)^2$ are both much less than 
$(r\,\partial \Omega_\mathrm{K}/\partial r)^2$.
If we additionally assume that the disk is optically thick
then the emitted flux can be approximated to
\begin{equation}
   \Lambda = 2\,\sigma_{\mathrm{R}}\,T_\mathrm{eff}^4
           \simeq \frac{2\,\sigma_{\mathrm{R}}\,T^4}{(3/8)\,\tau},
   \label{eq:simpleLambda}
\end{equation}
as implied by \refeqt{eq:Lambda}, when $\tau\gg 1$.
Recalling that $\tau=\frac{1}{2}\,\kappa\,\Sigma$, then, equating
\refeqt{eq:simpleUpsilon} and \refeqt{eq:simpleLambda} yields
\begin{equation}
   T^4 = \frac{27}{128}\,\frac{\kappa\,\nu}{\sigma_{\mathrm{R}}}%
         \,\Sigma^2\,\Omega_\mathrm{K}^2.
   \label{eq:Tan1}
\end{equation}
Now we assume that disk material has an opacity which we can be 
cast in the form
\begin{equation}
 \kappa=\kappa_0\,T^2\;\mathrm{cm}^2\,\mathrm{g}^{-1}.
 \label{eq:morfill}
\end{equation} 
with $\kappa_0=2\times10^{-6}$ \citep{morfill1985}.
Placing \refeqt{eq:morfill} in \refeqt{eq:Tan1}, one finds
\begin{equation}
   T = \sqrt{\frac{27}{128}\,\frac{\kappa_0\,\nu}{\sigma_{\mathrm{R}}}}%
         \,\Sigma\,\Omega_\mathrm{K}.
   \label{eq:Tan}
\end{equation}
Such expression allows a direct comparison with the temperature 
distribution obtained from simulations, whose setup involves 
\refeqt{eq:simpleUpsilon}, \refeqt{eq:simpleLambda}, and the 
opacity law~(\ref{eq:morfill}). 

In order to carry out a comparison of this kind, we simulated an
unperturbed disk (i.e., without any embedded body), with
borders at $1$ and $20\,\AU$, surrounding a Solar-mass star. 
The disk mass is $\Md=0.03\;\MStar$ and the kinematic viscosity
is $\nu=5\times10^{16}\;\nunits$. 
Both the initial surface density and temperature are constant:
$\Sigma(t=0)=197\;\sdunits$ and $T(t=0)=352\;\kelvin$.
Once the system has reached a stationary state,
the surface density is expected to decay as $1/\sqrt{r}$,
because $\nu$ is constant \citep{lyndenbell}. 
Indeed, this is what one can observe in
\refFgt{f:Mtest} (left panel), where the azimuthally averaged,
computed surface 
density $\langle\Sigma\rangle$ (crosses) is fitted by the power-law
\begin{equation}
 \Sigma= 300\,\sqrt{\frac{5\;\AU}{r}}\;\sdunits.
 \label{eq:Sfit}
\end{equation}
On the other hand, \refeqt{eq:Tan} states that 
$T\sim \Sigma\,\Omega_\mathrm{K}\sim 1/r^2$, or
more precisely, $T=104\,(5\,\AU/r)^2\;\kelvin$.
\refFgt{f:Mtest} (right panel)
shows how the calculated averaged temperature $\langle T \rangle$
(crosses) fits to this analytic estimation.

For the sake of completeness, we repeated the same type of test
in which, instead of the Morfill et al.'s relation (\refeqp{eq:morfill}),
we chose the Kramer's law
$\kappa=6.6\times
10^{22}\,\rho\,T^{-3.5}\;\mathrm{cm}^2\,\mathrm{g}^{-1}$, 
as in \citet{frank1992}, obtaining the same good agreement between
analytical and numerical results.

Although not reported here, we also tested the complete algorithm against
convergence stability, i.e., we checked that it always converges to the
same solution whatever the initial conditions are. Two simulations were set up
in which the full form of \refeqt{eq:eneq}
is solved along with \refeqt{eq:upsilon}, \refeqt{eq:Lambda},
and the Bell's opacity coefficient (\refsec{ssec:opacity}).
The disk radial domain is the same as before, but now
the disk mass is smaller ($\Md=0.01\;\MStar$) and so is the viscosity
($\nu=10^{16}\;\nunits$). The initial $\Sigma$
is imposed to be constant and equal to $66\;\sdunits$ for both models.
As initial temperature, in one model $T(t=0)=14.5\;\kelvin$, while in
the other $T(t=0)=580\;\kelvin$. The ``cold'' as well as the ``hot''
model soon evolve toward a stationary state.
In each of the cases, the solutions exactly match, which
indicates that initial values play no role in the 
equilibrium solution attained after the system has relaxed.
The relaxation time somehow depends on the initial $\Sigma$, whereas
scarcely any dependency on $T$ at $t=0$ is observed.

\begin{figure}[!t]
\epsscale{1.0}
\plotone{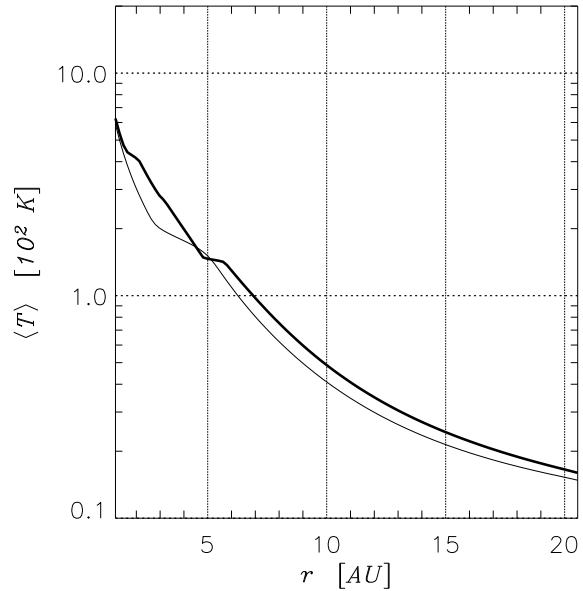}
\caption{\small{Comparison between two protostellar disk models which
       differ only on the choice of the opacity tables. The solid
       thin line is produced by the model run with Bell's opacity
       formulas \citep{bell1994},
       whereas the thick line comes from the model
       executed with the new opacity tables by Semenov and
       collaborators (2002, private communication).
       The temperature is slightly larger
       in this second case with respect to the first one.
       Discrepancies generally stay within the 20\%.}
\label{f:DBtest}}
\end{figure}
Finally, we examine some implications related to the opacity tables.
A simulation was performed with the same parameterization as the
``hot'' model mentioned above. 
But instead
of choosing Bell's opacity, this model is based on the new tables
by Semenov and collaborators (2002, private communication). 
The resulting surface density
distribution, at equilibrium, is hardly distinguishable from that obtained
with the ``hot'' model. 
Furthermore, not much difference is 
measured between the stationary averaged temperature profiles, as
demonstrated in \refFgt{f:DBtest}.
The result is not surprising, since already \citet{liu1997} tested
the influence of different opacity tables on the vertical structure
of accretion disks and found that the change in disk structure, due 
to an improved opacity coefficient, is hardly perceivable when 
compared to the uncertainties connected with the general disk 
parameterization.

Hence, throughout these simulations we use the opacity coefficient by
\citet{bell1994}.
It is worthy to note that this opacity table is well tested and 
has been extensively adopted in accretion disk studies 
\citep[see, e.g.,][]%
{godon1996,bell1997,klahr1999,armitage1999,papa1999,nomura2002}.

\refEqt{eq:Lambda} contains the quantity $\varepsilon_\mathrm{H}$ which
can be interpreted as the ratio of extinction to the absorption coefficient. 
However, the physical conditions of the protostellar environment
are such that it is allowed to neglect radiation scattering and
write $\kappa_\mathrm{ext}\simeq\kappa_\mathrm{abs}$, and hence
$\varepsilon_\mathrm{H}\simeq 1$. 
In this phase, we checked such approximation by repeating the calculation 
addressed in \refFgt{f:DBtest} and setting 
$\varepsilon_\mathrm{H}=\kappa_\mathrm{ext}/\kappa_\mathrm{abs}$,
in \refeqt{eq:Lambda}.
The resulting temperature profile is hardly distinguishable from the
thick line in \refFgt{f:DBtest}.
Therefore, the hypothesis $\varepsilon_\mathrm{H}=1$
is quite correct in our applications.

\section{Model Parameters}
\label{sec:parameters}
\begin{figure*}[!th]
\epsscale{2.0}
\begin{center}
\mbox{%
\plottwo{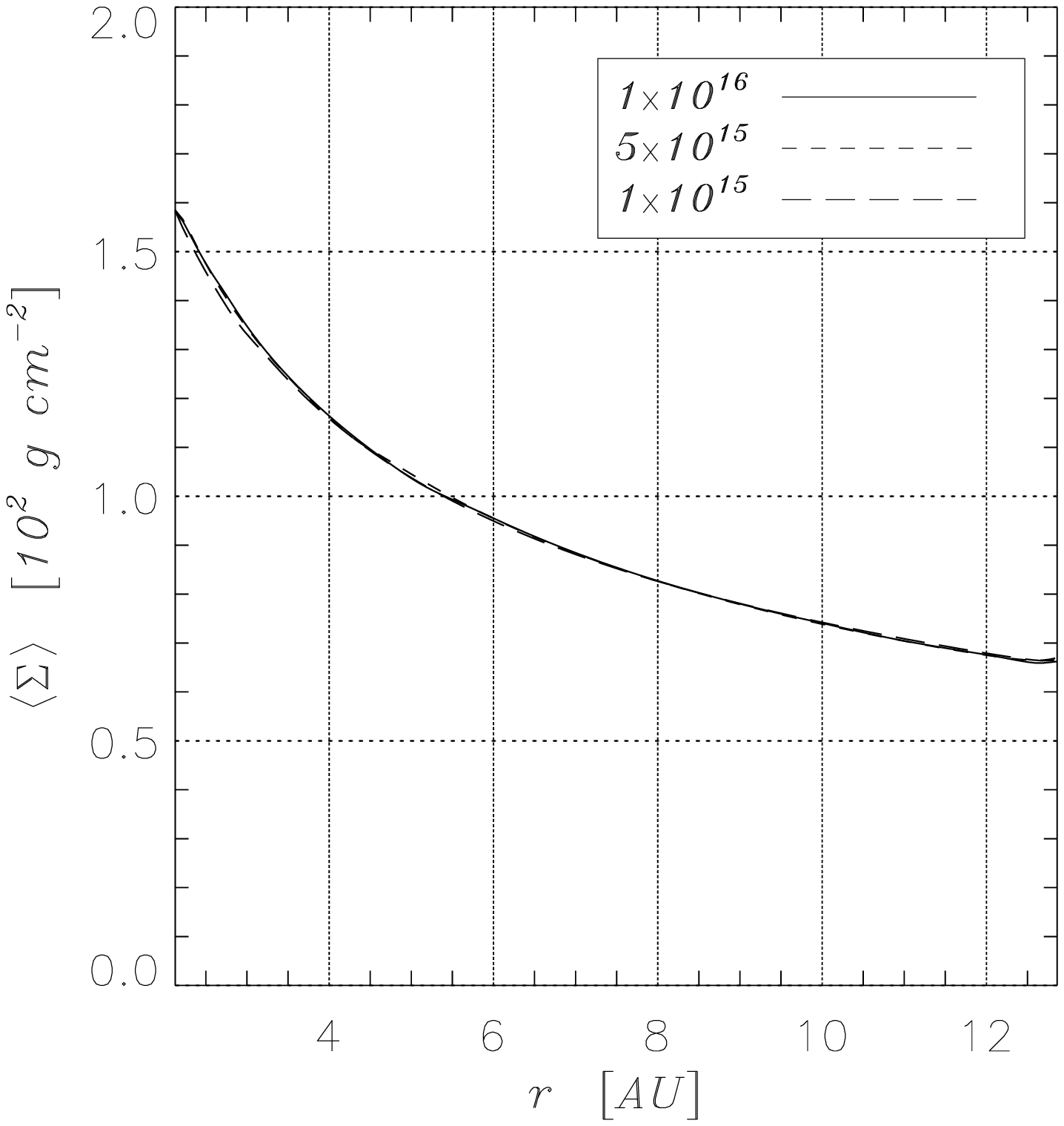}{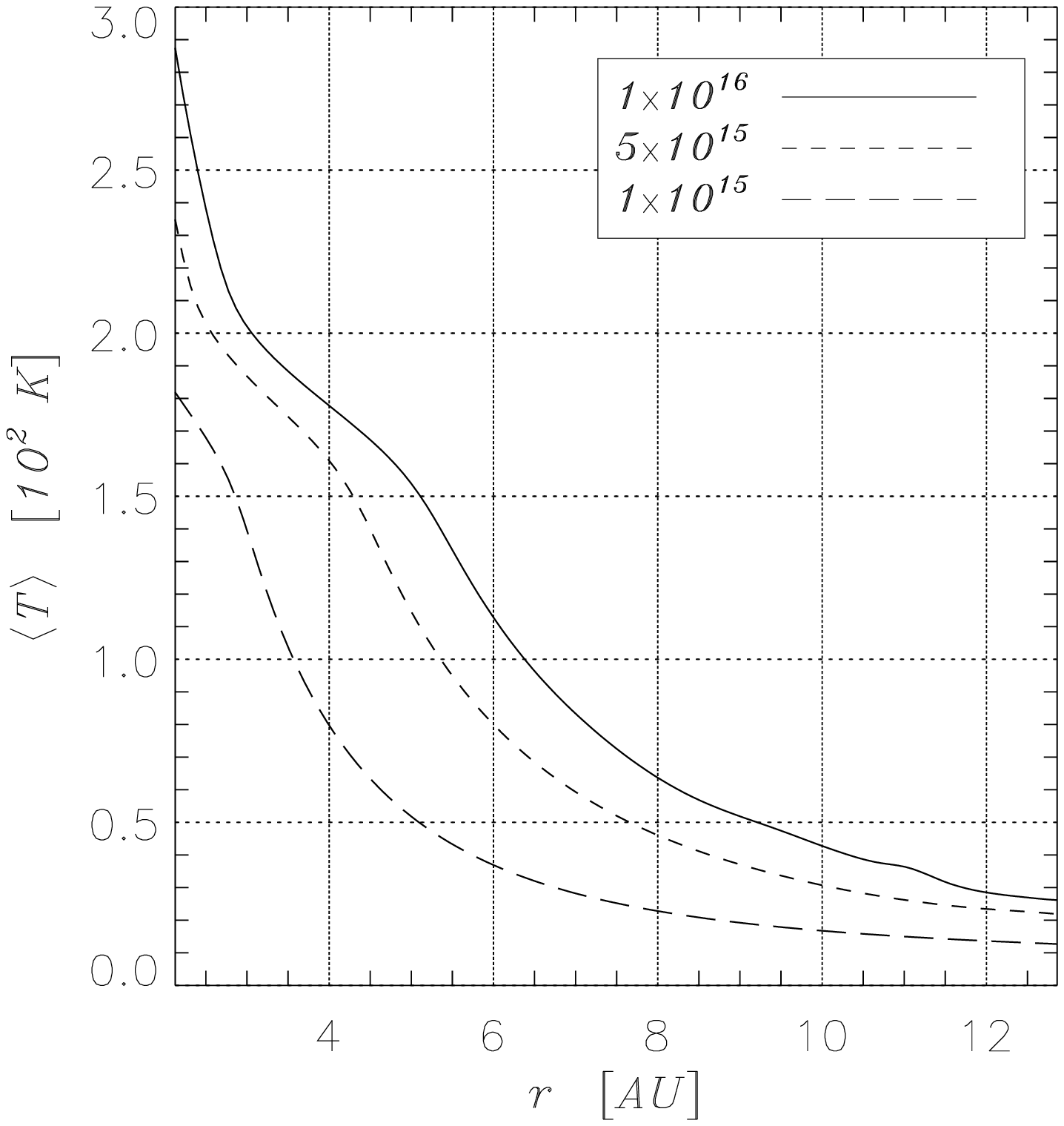}}
\mbox{%
\plottwo{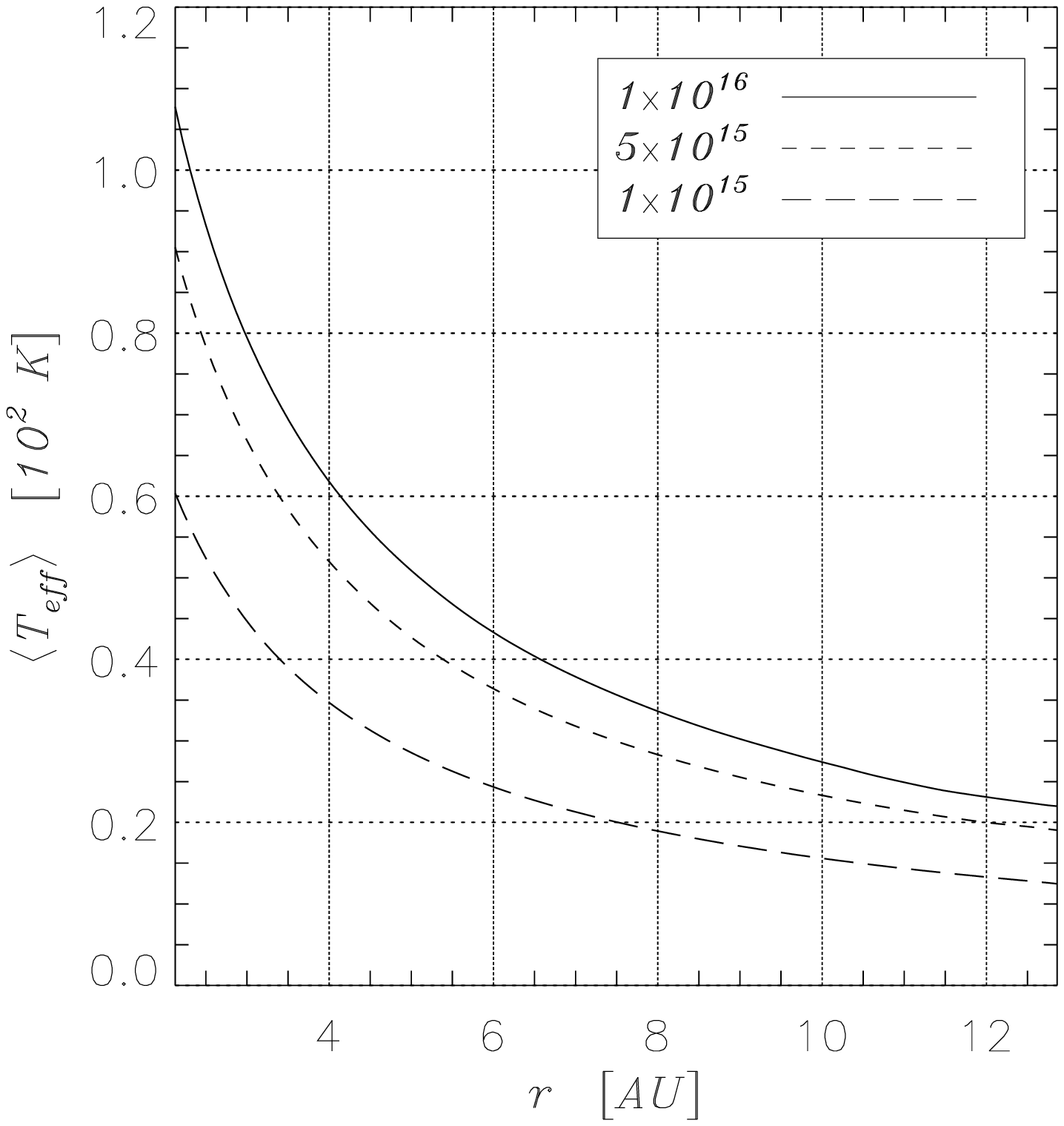}{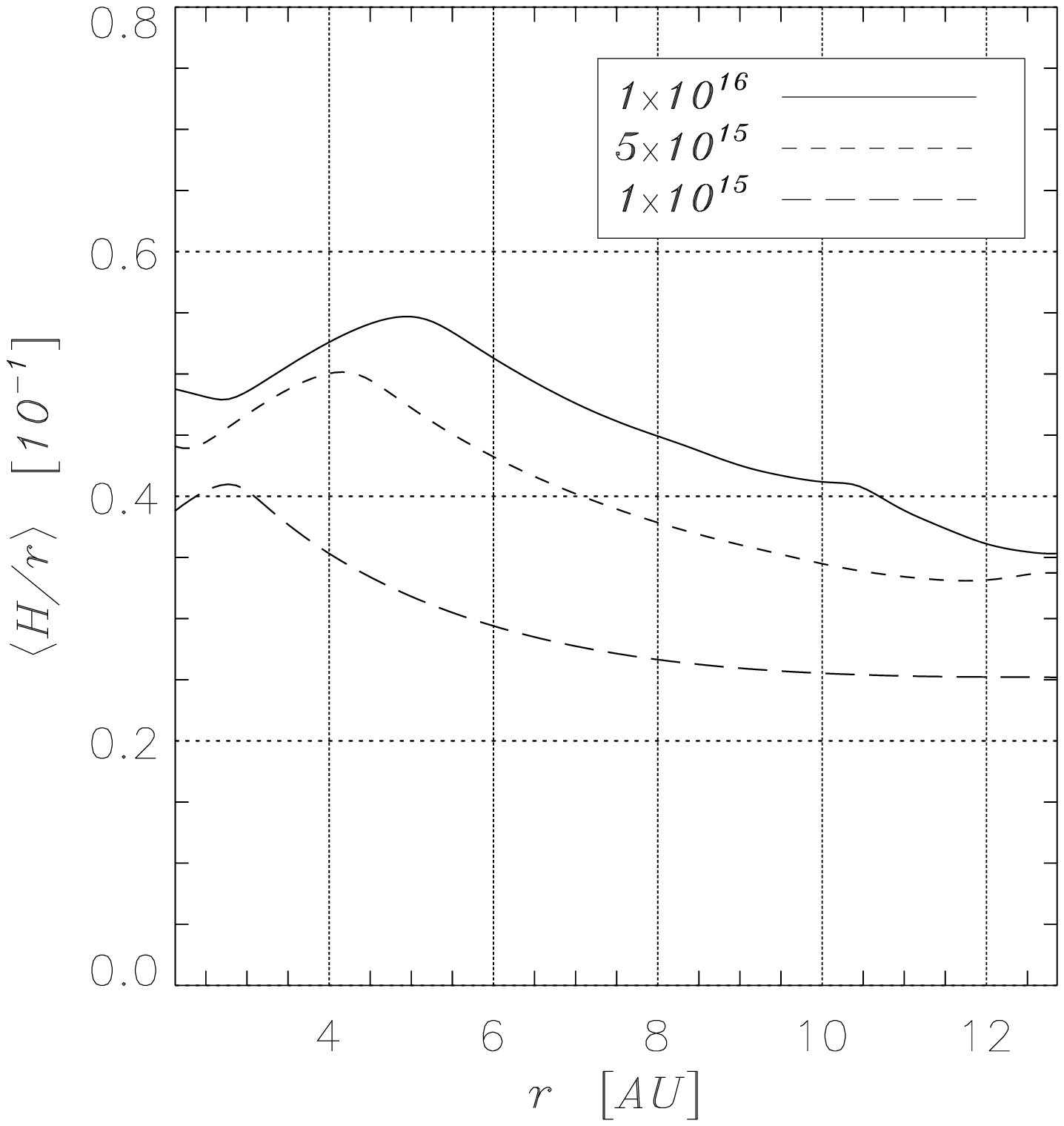}}
\end{center}
\caption{\small{Some characteristics of radiative,
       unperturbed and stationary disks,
       as function of the magnitude of kinematic viscosity (shown in the
       legend in \nunits).
       Radial boundaries are closed in order to achieve a strict
       stationarity by preventing mass losses.
       \textit{Top left}: Surface density.
       \textit{Top right}: Midplane temperature.
       \textit{Bottom left}: Disk effective temperature (\refeqp{eq:lambda3}).
       \textit{Bottom right}: Disk aspect ratio.}
\label{f:Tini}}
\end{figure*}
Model parameterization deserves more attention in these computations
than it did in the simulations where energy transfer by radiation is
not taken into account.
The reason for this resides in the nature of the opacity coefficient
$\kappa$, that is always in cgs units.
Hence, lengths and masses are to be fixed in physical units and
consequently outcomes will not be scale-free, as they were in other
circumstances (see, e.g., \gAA; \gApJ).

As in previous simulations, we model a circumstellar disk, orbiting
a one-solar-mass star ($\MStar=1\;\MSun$), whose radial borders are
$r_\mathrm{min}=2\;\AU$ and $r_\mathrm{max}=13\;\AU$.
The initial mass enclosed in this domain is 
$\Md=4.8\times10^{-3}\;\MStar$, i.e., $0.01\;\MStar$ within $20\;\AU$
\citep{paola1998}. 
Since in stationary Keplerian disks, the energy budget is regulated by the
balance between viscous dissipation and radiative losses 
(\refeqp{eq:simplebudget}), 
the magnitude of viscosity may play an important role. Hence, we will
cover various viscosity regimes. The bulk of the simulations were run with
a constant kinematic viscosity coefficient, by setting
$\nu=1.0\times10^{15}$,
$5.0\times10^{15}$, and $1.0\times10^{16}\;\nunits$ \citep{bell1997}.
Besides physical viscosity, we consider an artificial viscosity, which
is active only in compression regions, according to the tensor
formulation explained in \refsec{ssec:aviscosity}. The length
$\mathcal{L}$ is chosen to be equal to the maximum grid spacing, on
each grid level \citep{stone1992a,ziegler1998}.
The mean molecular weight is $\mu=2.39$ \citep{morfill1985}
and the adiabatic index $\gamma=1.4$.

\refFgt{f:Tini} shows characteristic averaged quantities of non-perturbed,
stationary disks with the aforementioned characteristics. To
distinguish among the different viscosity (and temperature) disk
regimes, models will be referred to as ``Hot'' (H), 
``Warm'' (W), and ``Cold'' (C).

To further explore the parameter space, two models were executed applying the 
Shakura-Sunyaev viscosity prescription $\nu=\alpha\,c_s\,H$, with 
$\alpha=10^{-2}$ and $10^{-3}$ (models A2 and A3). Hence, in such
models $\nu$ is a function of space and time.

An embedded planet is placed on a fixed circular orbit at 
$r_\mathrm{p}=5.2\;\AU$.
The azimuthal position of the planet is kept fixed ($\varphi_\mathrm{p}=\pi$)
by the employment of a rotating grid.
In order to obtain reliable outcomes in the two-dimensional geometry,
we consider a planetary mass range extending from $\Mp=33\;\MEarth$ to
$\Mp=1\;\MJup$ or, in terms of mass ratios $q=\Mp/\MStar$, we have
models with $q=1\times10^{-4},2\times10^{-4},5\times10^{-4}$, and
$q=1\times10^{-3}$. 
It has been recently demonstrated (\gApJ) that, concerning
gravitational torques and mass accretion, 3D models agree quite
well with 2D ones in that mass range. 

\begin{figure*}[!t]
\epsscale{0.7}
\begin{center}
\mbox{\plotone{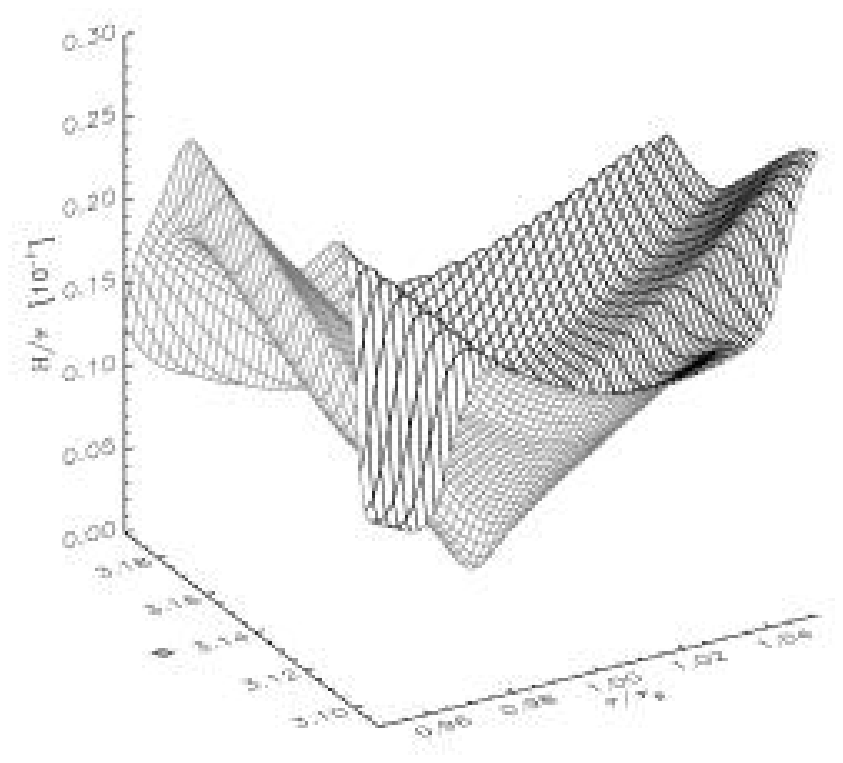}\plotone{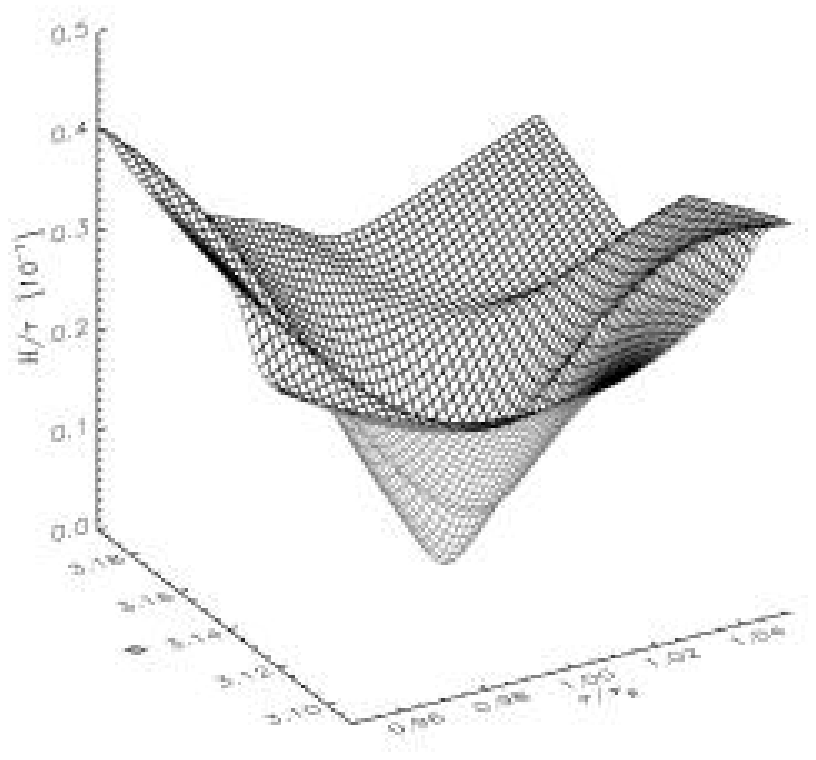}\plotone{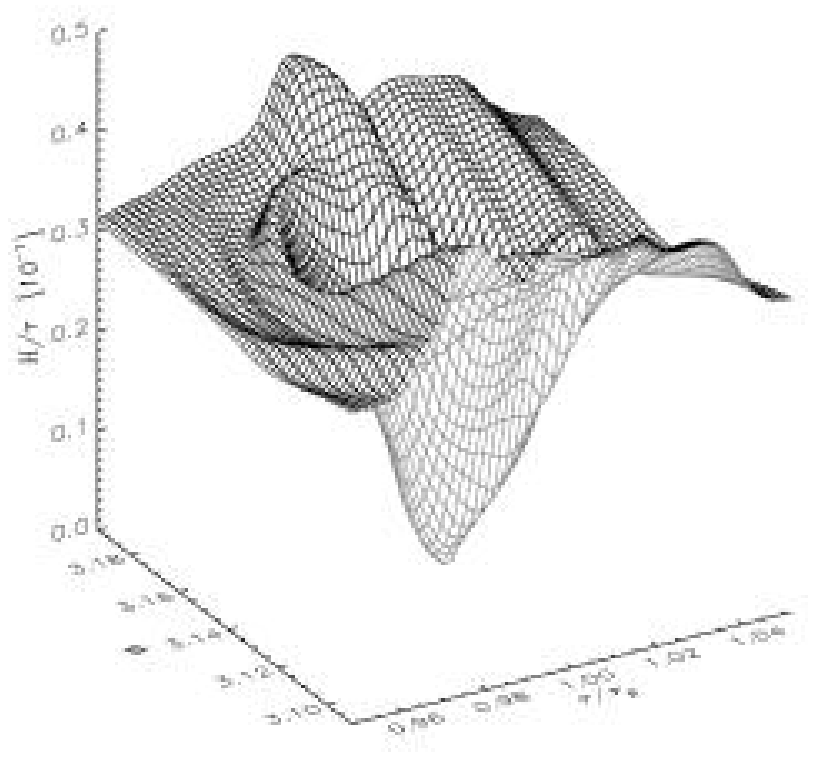}}
\end{center}
\caption{\small{Because of the gravitational field of the embedded planet,
       the disk scale height may reduce considerably in its proximity,
       as prescribed by \refeqt{eq:H}.
       This effect is shown here for models with different planetary
       masses:
       $\Mp=1\,\MJup$ with $\nu=1.0\times10^{16}$\;\nunits (\textit{left}); 
       $\Mp=0.1\,\MJup$ with $\nu=1.0\times10^{16}$\;\nunits (\textit{center});
       $\Mp=20\;\MEarth$  with $\nu=1.0\times10^{15}$\;\nunits (\textit{right}).
       For comparison, in the lowest mass case, $\RH/r_\mathrm{p}=0.027$.}
\label{f:H20E}}
\end{figure*}

However, since the disk thickness is sufficiently small, 
even for low protoplanetary masses (see \refFgp{f:Tini}), 
we go as far as to simulate an additional C-model with $\Mp=20\;\MEarth$.
\refFgt{f:H20E} shows the disk semi-thickness in the planet's
neighborhood for three selected masses. 
Indicating with $\RH=r_\mathrm{p}\,\left(q/3\right)^{1/3}$ the
planet's Hill radius,
one can clearly see that the condition $H\lesssim \RH$ locally
holds even in the smallest mass case. The figure also shows that
protoplanets dwell in cavities and thus radiation from the
central star cannot directly reach the surrounding matter. 

The smoothing length $\delta$ is set to $2\times10^{-2}\,\RH$
(see \refeqp{eq:potential}).
All of the models are executed in an accreting and a non-accreting mode.
When accretion onto the planet is allowed, the procedure outlined in
\gAA\ is utilized. The accretion region has a radius
$\kappa_\mathrm{ac}=0.1\,\RH$, which should be short enough to let the whole
accretion procedure be almost independent of the removal characteristic
time, as proved by \citet{tanigawa2002}.

\begin{figure*}[!t]
\epsscale{1.9}
\plottwo{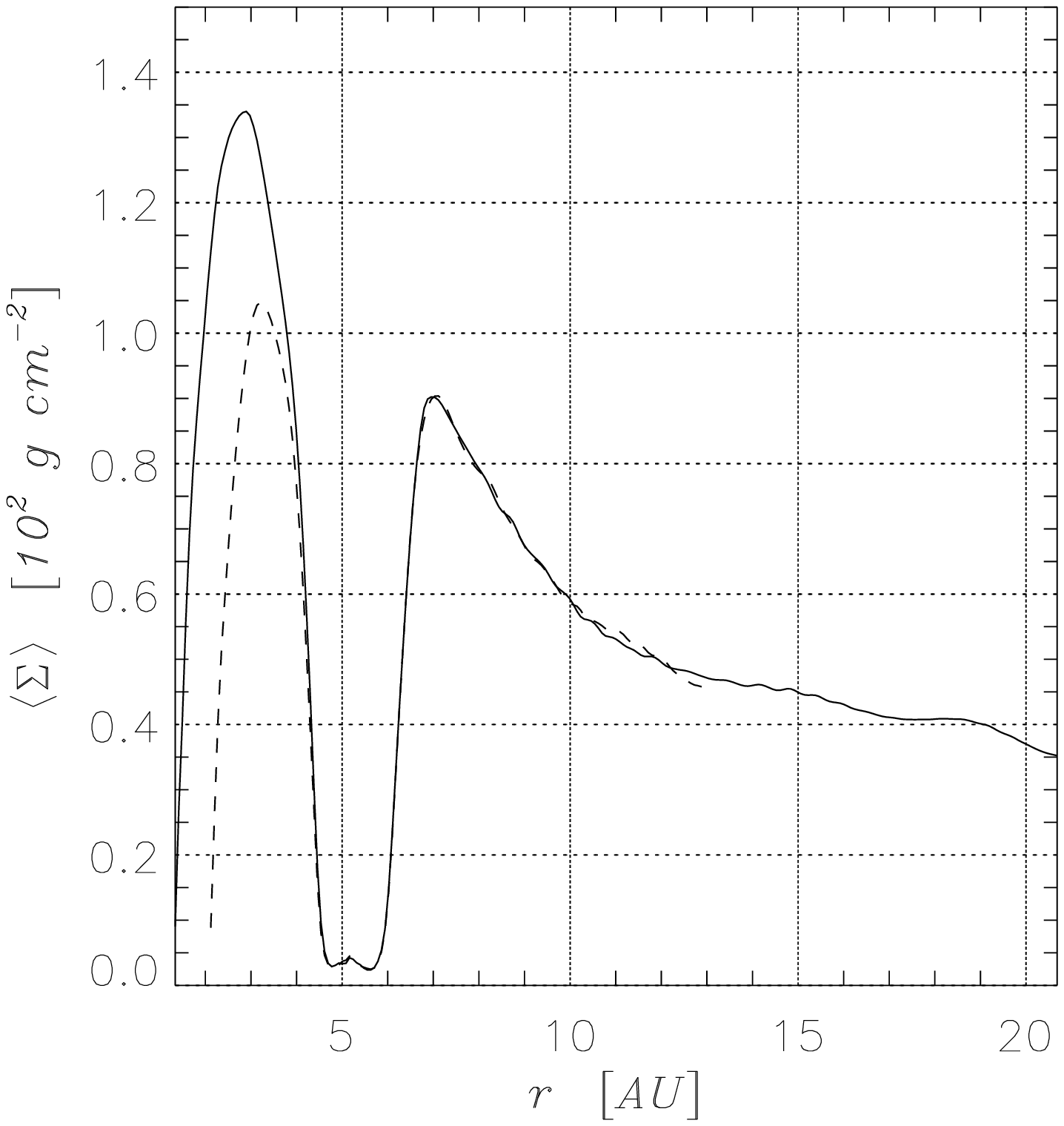}{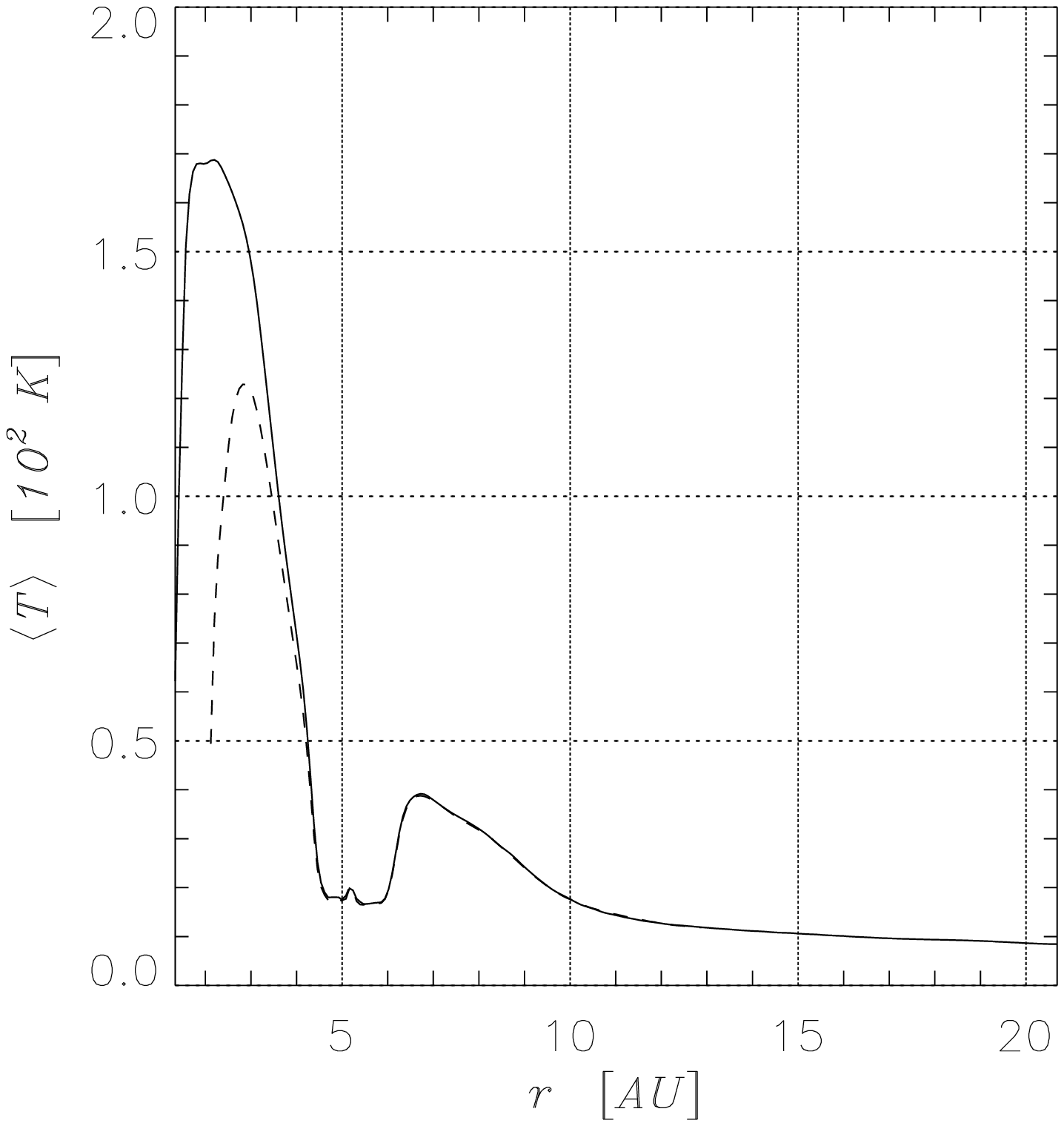}
\caption{\small{Average surface density (\textit{left})
       and temperature (\textit{right}) profiles
       belonging to two Jupiter-mass test models which differ only
       by the extension of the radial domain. The solid line
       refers to a model for which $r_\mathrm{min}=1.3\;\AU$ and
       $r_\mathrm{max}=20.8\;\AU$ ($\Md=7.5\times10^{-3}\;\MStar$),
       whereas the dashed line represents
       a standard radial domain model $[2, 13]\;\AU$
       ($\Md=3.5\times10^{-3}\;\MStar$).
       The kinematic viscosity corresponds to that employed in C-models.}
\label{f:Jradial}}
\end{figure*}

\refFgt{f:Jradial} illustrates the influence of the extent of the 
domain's radial extent for Jupiter C-models. 
Simulations provide quite consistent outcomes over the matching domain.
Differences exist only close to the
inner border, due to inflow boundary conditions.
\subsection{Initial and Boundary Conditions}
\label{ssec:bci}

Initial conditions require the assignment of the three functions
$\Sigma(t=0)$, $T(t=0)$, and  $\gv{u}(t=0)$ that
we suppose to be axi-symmetric because we start from a stationary disk.

The theoretical surface density profile of a stationary 
accretion disk, with constant $\nu$, is a power-law of 
the radial distance $r$ \citep{lyndenbell}, as also found
numerically in \refsec{ssec:tests}. 
Thus, as starting density distribution for our simulations, 
we set
\begin{equation}
 \Sigma(t=0)= \Sigma_0\,\sqrt{\frac{r_\mathrm{p}}{r}}.
 \label{eq:Sini}
\end{equation}
where $\Sigma_0$ is determined by the disk mass value $\Md$.
 
\begin{deluxetable}{cccc}
\tablewidth{0pt}
\tablecaption{Initial Temperatures versus Viscosity and Disk Mass.
           \label{tb:T0nu}}
\tablehead{%
\colhead{$\nu\;[\nunits]$} &
\colhead{$T_0\;[\kelvin]$} &
\colhead{$T_\mathrm{max}\;[\kelvin]$} &
\colhead{$T_\mathrm{min}\;[\kelvin]$}
}
\startdata
 $1.0\times10^{15}$ & $ 50$ &  $180$  &  $13$ \\
 $5.0\times10^{15}$ & $110$ &  $235$  &  $22$ \\
 $1.0\times10^{16}$ & $150$ &  $295$  &  $26$ \\
\enddata
 
\tablecomments{Values of the initial temperature as function the three
       values of the kinematic viscosity. Temperatures are
       sampled at $r=r_\mathrm{p}=5.2\;\AU$ ($T_0$),
                  $r=r_\mathrm{max}=2.1\;\AU$ ($T_\mathrm{max}$),
              and $r=r_\mathrm{min}=13\;\AU$ ($T_\mathrm{min}$).
       Models with $\alpha$-viscosity A2 and A3 are initiated as
       C-model.}
\end{deluxetable}
Calculations carried out for different viscosity magnitudes
imply different temperature regimes (see \refFgp{f:Tini}).
According to them, the initial temperature distribution is 
prescribed. 
The behavior of the profiles in the top-right panel of \refFgt{f:Tini} 
can be roughly reproduced by the power-law
\begin{equation}
 T(t=0)= T_0(\nu)\,\left(\frac{r_\mathrm{p}}{r}\right)^{1.8}.
 \label{eq:Tini}
\end{equation}
The initial temperature at $r=r_\mathrm{p}$ is given in
\refTab{tb:T0nu} for three values of $\nu$.
Finally, the initial circumstellar flow $\gv{u}(t=0)$ is a 
Keplerian one, corrected for the grid rotation.

As for the boundary conditions, all models are run with a
partially open inner radial border and a reflective outer one. 
Thus, matter is free to flow out of the computational domain,
at $r=r_\mathrm{min}$,
but the opposite is not allowed. This is the same expedient
invoked in \gAA\ and \gApJ\
to artificially mimic the mass accretion toward the central star
and avoid spurious wave reflection at the inner domain edge, which
is the closer to the planet.
The flow field is Keplerian both at $r_\mathrm{min}$ and 
$r_\mathrm{max}$: $\gv{u}=[0,r\,(\Omega_\mathrm{K}-\Omega_\mathrm{p})]$.

\subsection{Numerical Specifications}
\label{ssec:numerics}
The basic computational mesh is made of $N_r\times N_\varphi=143\times423$
cells, while sub-grid patches are $64\times64$. All of the computations are
based on a five-grid hierarchy. Thus, the finest resolution
ranges from $1.3\times 10^{-2}\,\RH$ (if $\Mp=1\;\MJup$) to 
$2.9\times 10^{-2}\,\RH$ (if $\Mp=33\;\MEarth$).
Further details on numerical issues have been presented in
\gAA\ and \gApJ.

Despite we have not implemented a sophisticated energy transport treatment, 
these simulations take between 35 and 40\%
longer than those discussed in \gAA, for equal size grid
hierarchies. 
 
\section{Global Model Properties}
\label{sec:global}

\begin{figure*}[!t]
\epsscale{2.0}
\plottwo{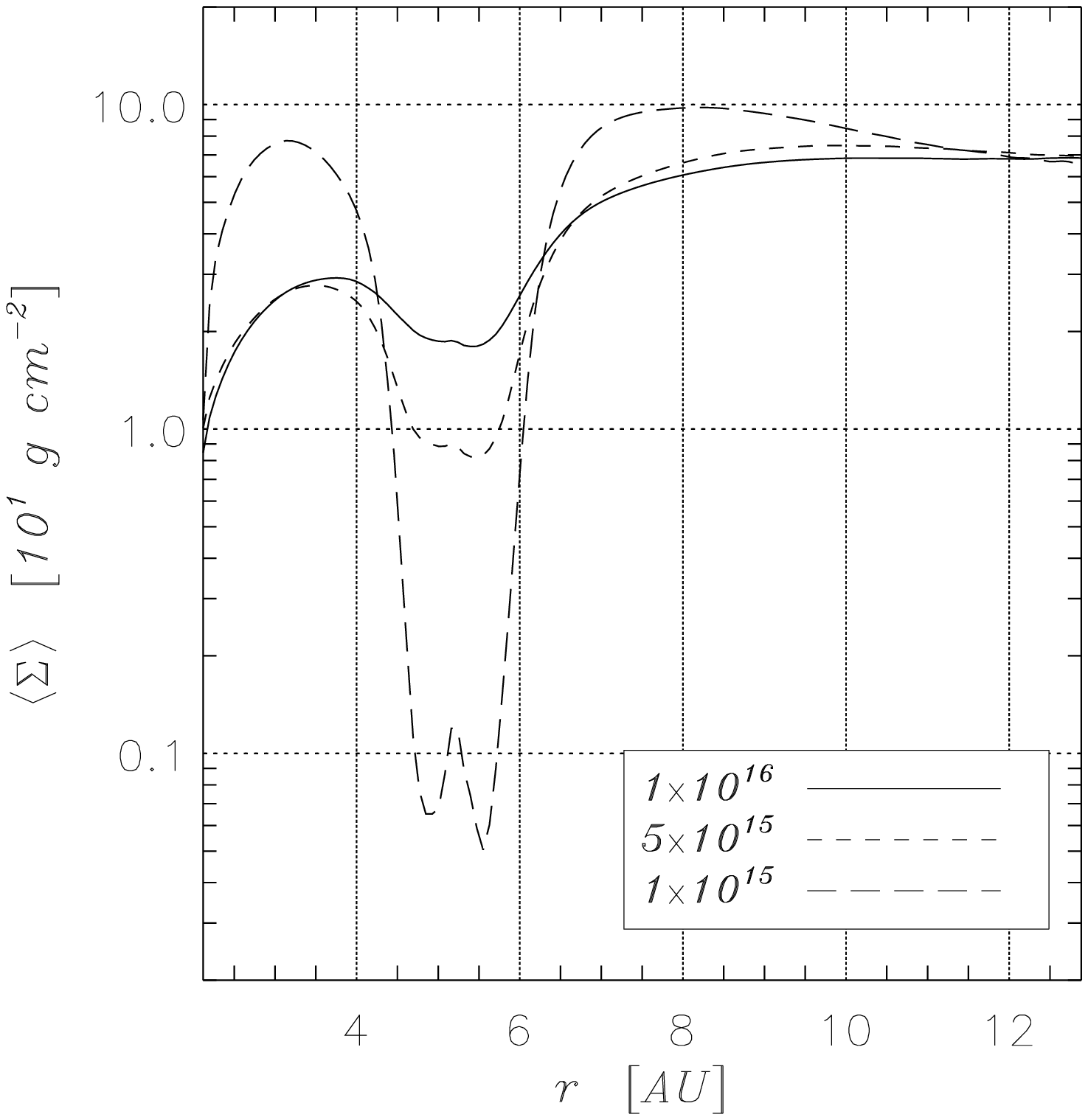}{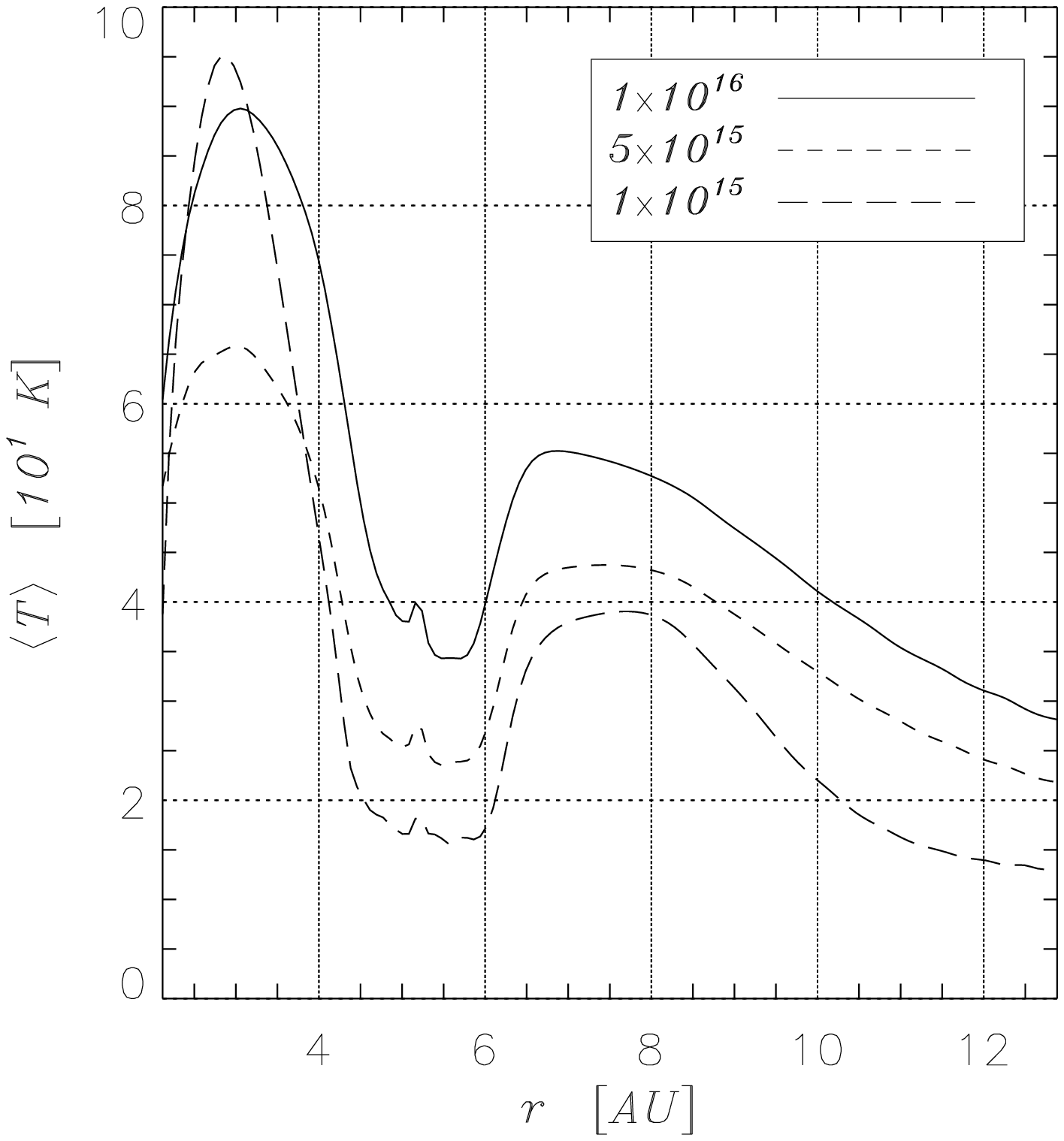}
\caption{\small{Average surface density (\textit{left}) and temperature
       (\textit{right}) in Jupiter-mass models after $240$
       orbital periods. By this evolutionary time, models have settled
       on a quasi-equilibrium state.
       The different
       profiles refer to different values of $\nu$, as indicated in
       the legends. The density
       gap, in the C-model ($\nu=10^{15}$\;\nunits), is more than
       an order of magnitude deeper (long-dash line) than it is
       in the other models.
       The temperature gap follows the same trend, though differences
       are smaller.}
\label{f:gst}}
\end{figure*}
\begin{figure}[!t]
\epsscale{1.0}
\plotone{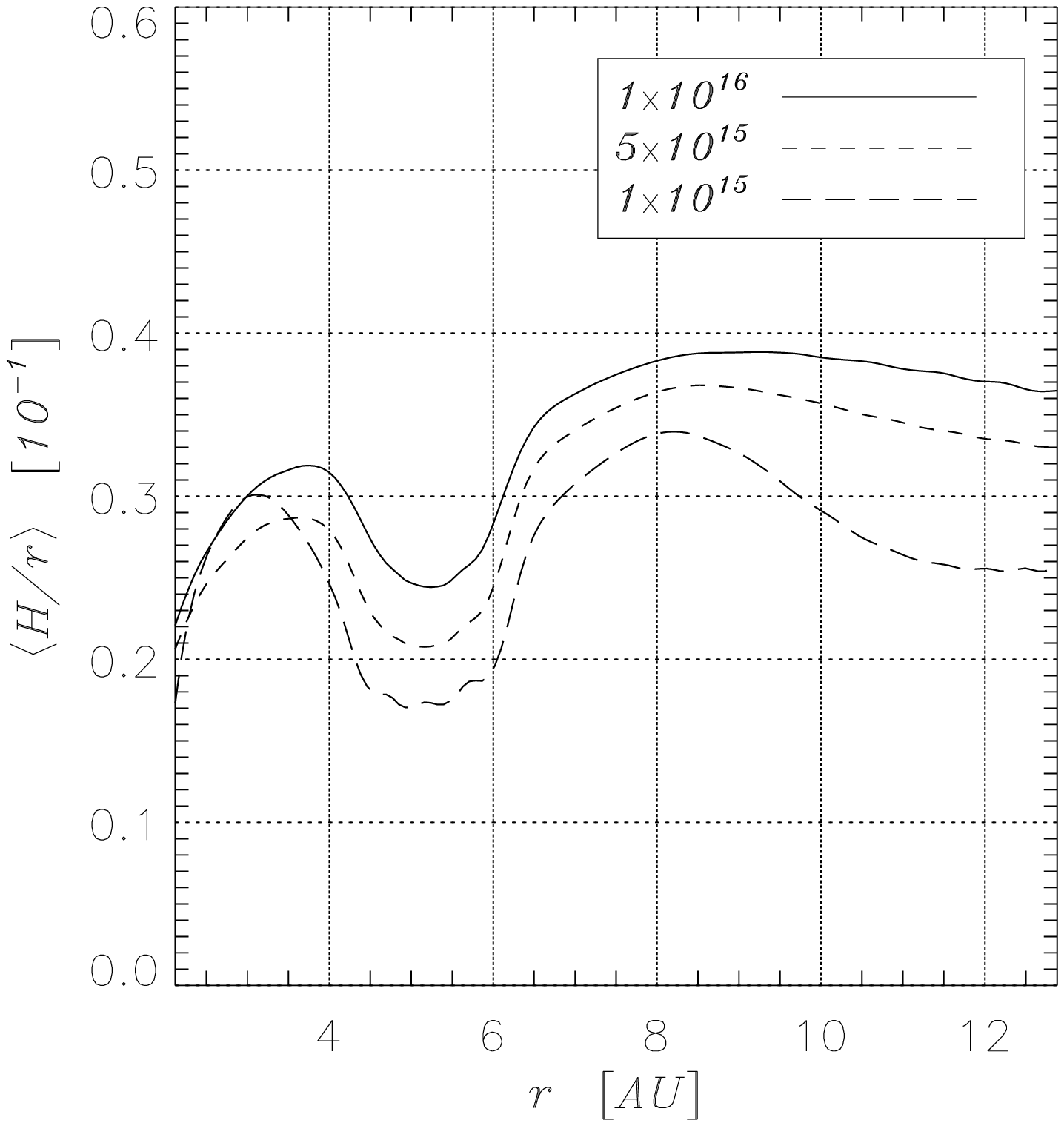}%
\caption{\small{Average pressure scale-height (\refeqp{eq:H}), normalized to 
       the distance from the star, in Jupiter-mass models
       according to different viscosity values, as indicated in the
       legend (in \nunits)}.
\label{f:gsh}}
\end{figure}
We now examine the behavior of global quantities for some
selected models because the large scale look of disks with planets
is, observationally, of extreme importance. 
\refFgt{f:gst} shows both the averaged surface
density and temperature in Jupiter-mass computations with different
kinematic viscosities.
We note that matter (left panel) is more uniformly distributed in the H- 
(solid line) and W-model (short-dash line), with respect to the
C-model (long-dash line).
In these cases, the density gap is not very deep and actually looks
like a depression,
because of the larger viscosity coefficient and the depletion of the
inner disk. 
On the one hand, in fact,
a high viscosity prevents the formation of wide and deep gaps.
On the other hand, a large viscosity facilitates the inner-disk depletion,
since $\dot{M}_\mathrm{D}\propto\nu\,\Sigma$ \citep{lyndenbell}.
The disk depletion rate that we measure in these simulations ranges from
few times $10^{-8}$ to $10^{-7}\,\MSun\,\mathrm{yr}^{-1}$. 

In contrast, in the low viscosity regime (C-model) a well-defined and
deep gap is carved in, where the density
is more than an order of magnitude lower than it is in the other two
models. 
The density peak visible in the middle of the gap is the signature
of the circumplanetary disk.
Since material removed from gap regions is disposed at short
distances, due to the small damping length of the waves launched 
by the planet, gap shoulders have
high densities. It has been argued that these density
enhancements might trigger planet formation.
If so, low viscosity disks should be more efficient in doing that.

Following the reduction of thermal energy in the low density regions, 
a temperature gap accompanies the density gap (right panel, \refFgp{f:gst}). 
In the C-model, temperatures drop below $20\;\kelvin$ and it generally
stays below $40\;\kelvin$. As expectable, the temperature gap has a
counterpart in the pressure scale height, as seen in \refFgt{f:gsh}.
Such a deep trough represents an additional evidence that, 
as long as an inner disk exists, protoplanets are shaded against
stellar radiation.

\begin{figure*}[!hb]
\epsscale{2.0}
\begin{center}
\mbox{%
\plottwo{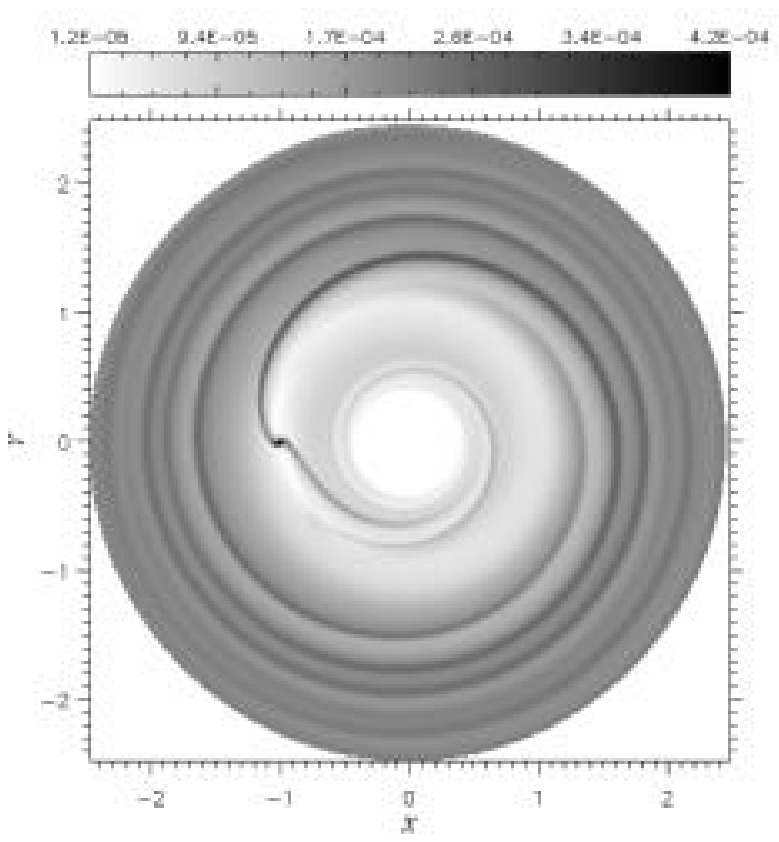}{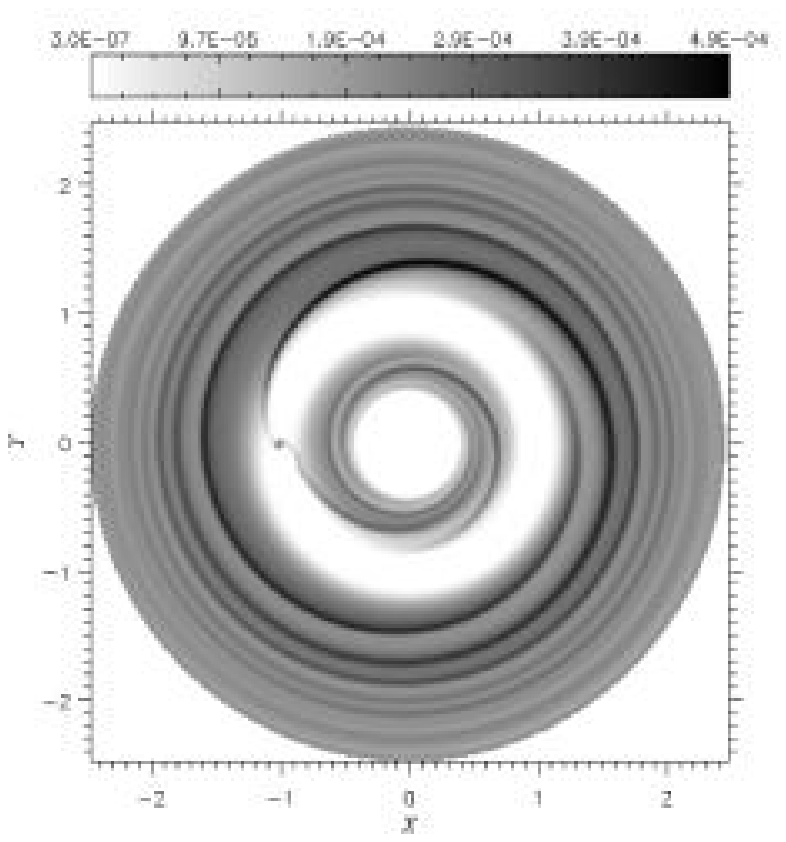}}
\mbox{%
\plottwo{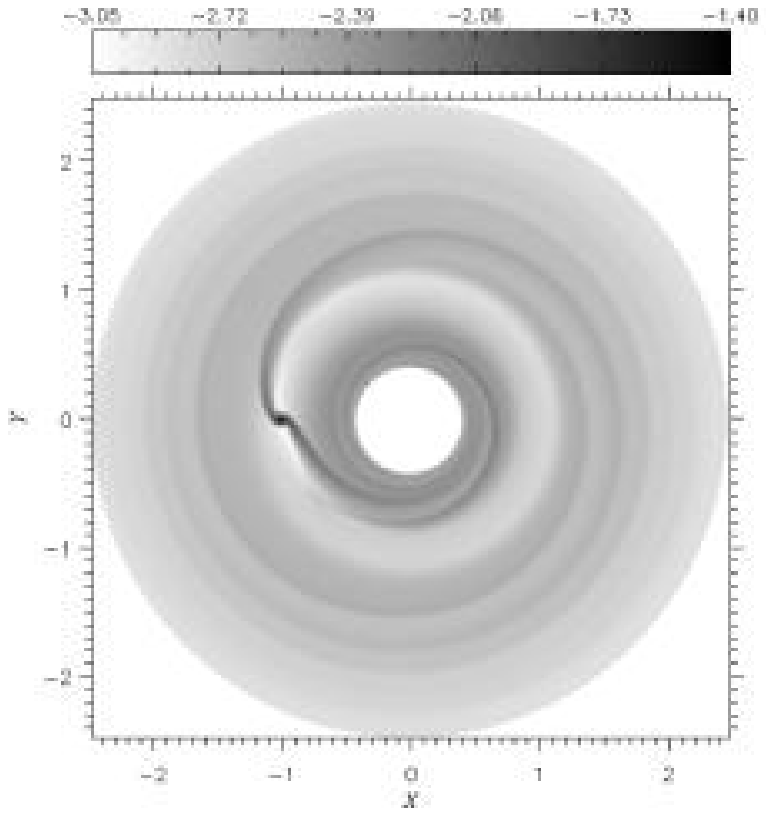}{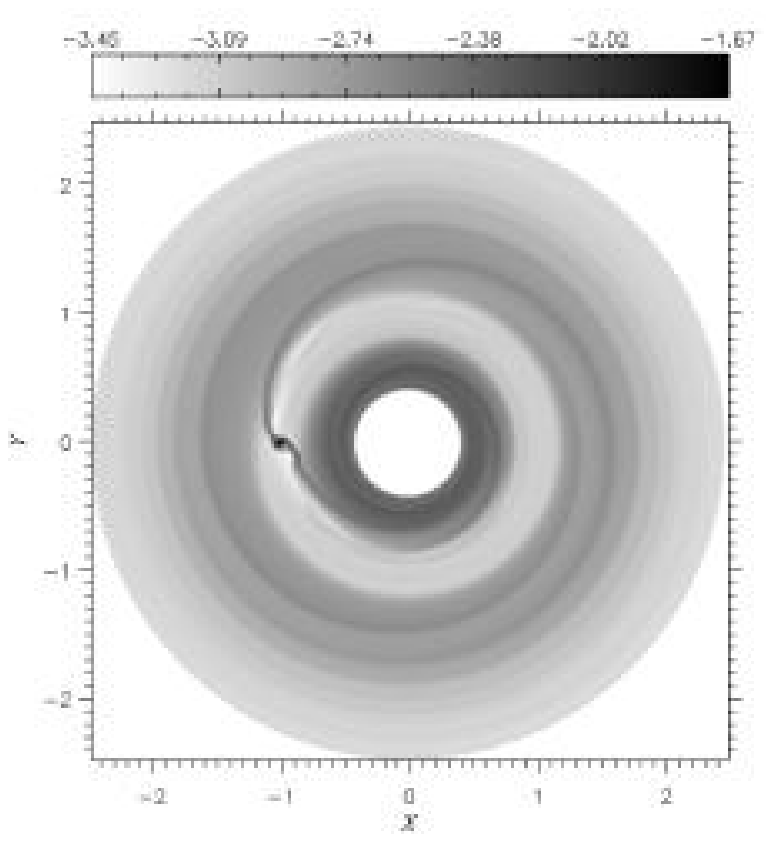}}
\end{center}
\caption{\small{Two-dimensional surface density (\textit{top}) and
       temperature (\textit{bottom})
       distributions for Jupiter-mass models. In the top panels,
       the color scale is linear; in the bottom ones the colors
       scale logarithmically.
       In this plot, $\Sigma=10^{-4}$ corresponds to
       $32.9\;\sdunits$, whereas $T=10^{-2}$ equals $198\;\kelvin$.
       Left panels correspond to H-models whereas right panels refer
       to C models. Density distributions
       show clear evidences that the a well-defined gap needs a cold
       (i.e., low viscosity)
       environment to be established. The same consideration applies 
       to the temperature distribution.}
\label{f:gpic1}}
\end{figure*}

Global views of both density and temperature 2D-distributions are
displayed in \refFgt{f:gpic1}. Top panels illustrate the surface
density, while the bottom ones show the temperature structure.
From the density maps one can see that disk spirals get closer to each
other toward the outer border of the domain.
In fact, circumstellar spirals propagate at a velocity equal to
the sound speed, which
is proportional to $\sqrt{T}$ (see \refeqp{eq:cs}). 
Thus, the propagation velocity reduces
as $r$ increases and therefore faster waves overtake slower
ones. Past the gap region, temperature decays more rapidly in the
C-model (\refFgp{f:gst}, right panel), hence this phenomenon is
more evident. From the bottom panels one can also deduce that 
spiral perturbations in the thermal energy and mass densities are 
quite phase coherent and thus disturbances in the thermal structure 
are rather weak.

\begin{figure*}[!t]
\epsscale{2.0}
\begin{center}
\mbox{%
\plottwo{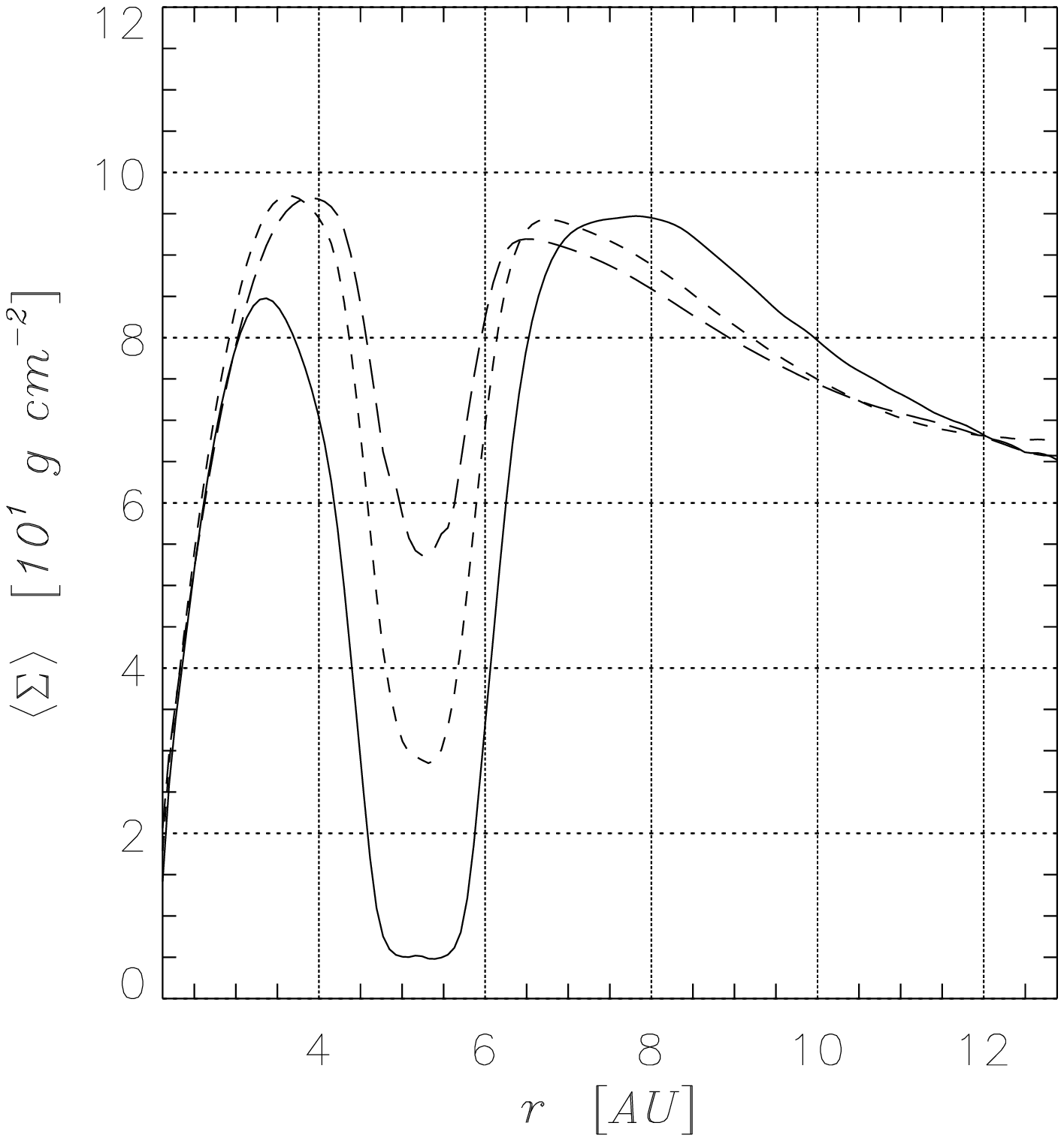}{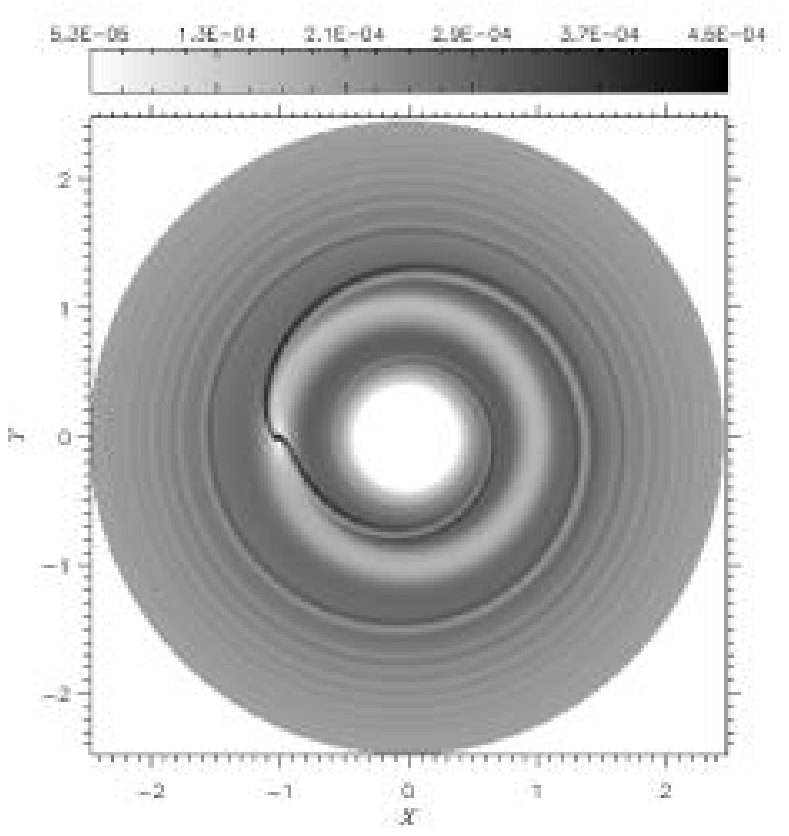}}
\mbox{%
\plottwo{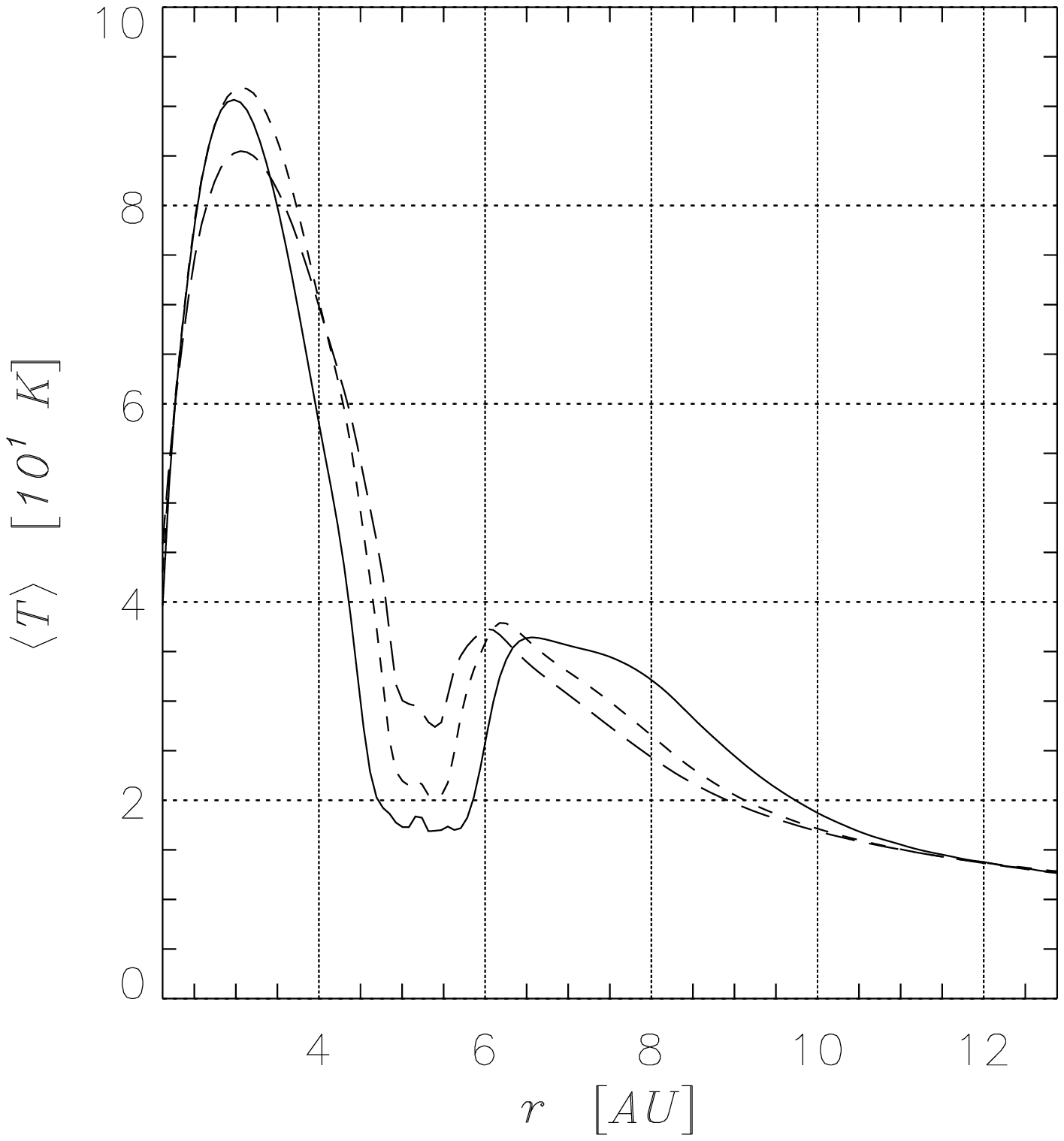}{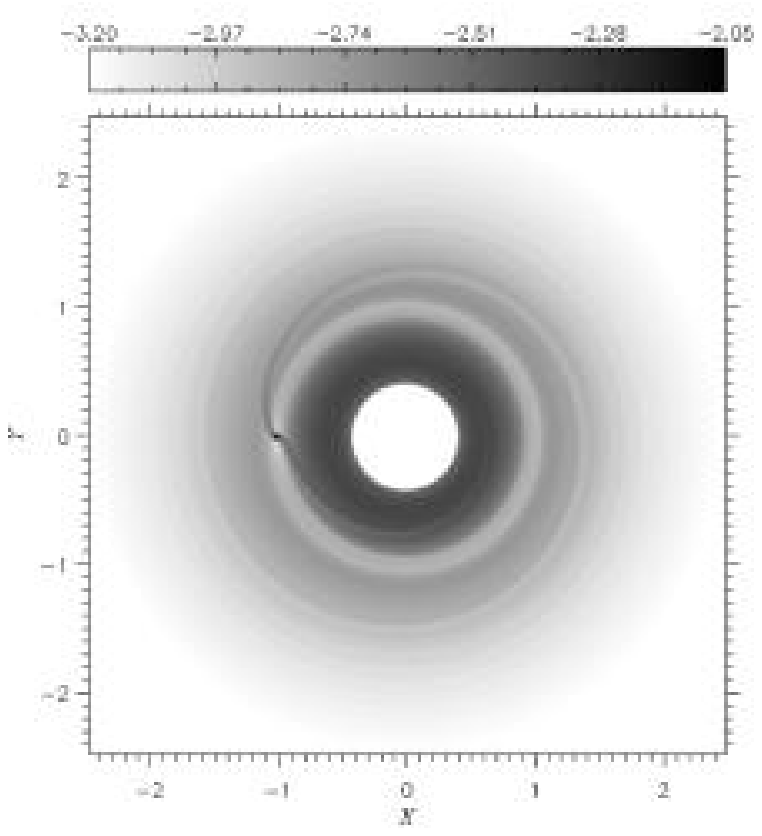}}
\end{center}
\caption{\small{\textit{Left}. Averaged surface density and temperature
       in C-models ($\nu=10^{15}$\;\nunits) with different masses:
       $\Mp=0.5\;\MJup$ (solid line), $\Mp=0.2\;\MJup$ (short-dash line),
       and $\Mp=0.1\;\MJup$ (long-dash line).
       \textit{Right}. 2D-distribution of the surface density (top) and
       temperature (bottom) of a C-model with $\Mp=0.1\;\MJup$. The density
       is scaled linearly, while temperature is scaled
       logarithmically. Conversion factors are: 
       $\Sigma=10^{-4}\rightarrow 32.9\;\sdunits$, 
       $T=10^{-2}\rightarrow 198\;\kelvin$.}
\label{f:gvsmass}}
\end{figure*}
Shifting toward smaller masses, the gap is gradually
refilled. This is visible
in \refFgt{f:gvsmass}, where C-models of different mass are compared. 
At $\Mp=0.1\;\MJup$ (i.e., $\Mp=33\;\MEarth$), the gap is one order 
of magnitude as shallow as
it is when $\Mp=0.5\;\MJup$. We will further address the issue of the
gap structure in \refsec{sec:GPS}, since gaps are considered promising
features for detecting embedded planets. 
As regards the circumstellar disk spirals, only feeble traces are
left, when $\Mp=0.1\;\MJup$, both in the
temperature and density distributions.
The absence of global disk features produced
by an embedded body invalidates direct imaging observations as
a detection tool in the intermediate-mass range, i.e., around a few
times of Uranus' mass.

\begin{figure*}[!t]
\epsscale{1.52}
\plottwo{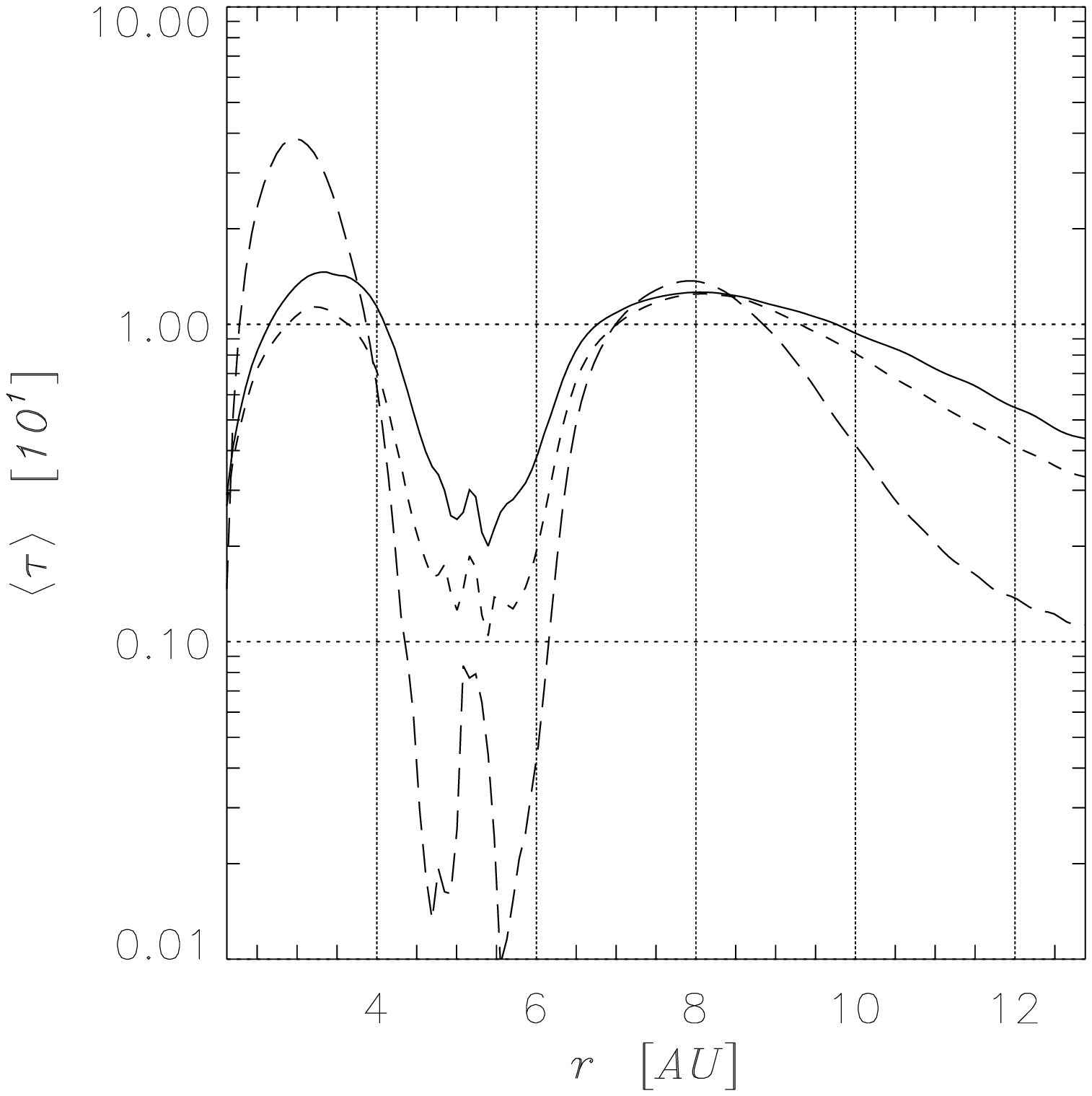}{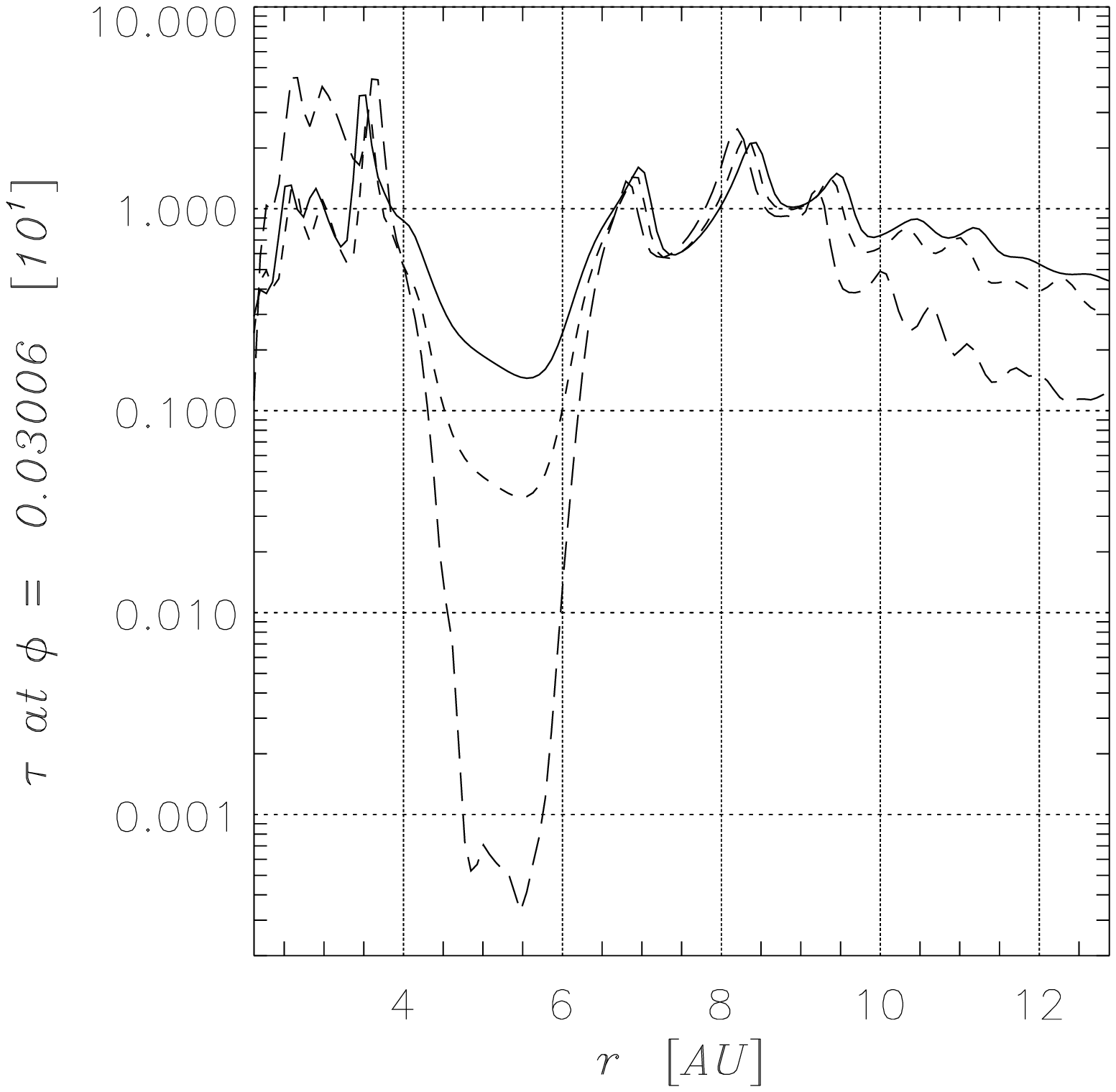}
\caption{\small{Circumstellar disks are usually optically thick. This
       plot shows that this is generally the case even when a disk
       contains a Jupiter-mass body. The line sequence is the same
       as in \refFgt{f:gst} (data refer to the same models).
       \textit{Left}. Optical thickness averaged around the star.
       Only in the C-model, $\langle\tau\rangle$ drops below
       $1$.
       \textit{Right}. Profiles at $\varphi\simeq 0$, i.e.,
       $180^\circ$ away from the planet location. The medium becomes
       very optically thin ($\tau\ll 1$) only in the gap region of the
       C-model.}
\label{f:gtau}}
\end{figure*}
Concerning the emitted energy, we can notice that the optical 
thickness of the models is generally larger than one, throughout
the computational domain (\refFgp{f:gtau}, left panel). The averaged $\tau$
stays above $1$ also at the outer disk edge.
This is in agreement with what was found for accretion disks around T 
\textit{Tauri} stars \citep{paola1998}.
The density and temperature gap reflects in a similar feature in
the distribution of $\tau$, as shown in
\refFgt{f:gtau} (left panel). Nonetheless, even within the gap region, 
material is, on the average, optically thick ($\tau\gtrsim 1$). 
The exception is represented by the C-model, where gap optical 
thickness reaches values around $0.01$ at $\varphi=\varphi_{\mathrm{p}}+\pi$
(see \refFgp{f:gtau}, right panel). The most diluted portions of the
gap, situated where the circumplanetary disk merges into the gap, 
have values of $\tau$ on the order of $10^{-2}$--$10^{-3}$.

\begin{figure*}[!b]
\epsscale{1.56}
\begin{center}
\mbox{%
\plottwo{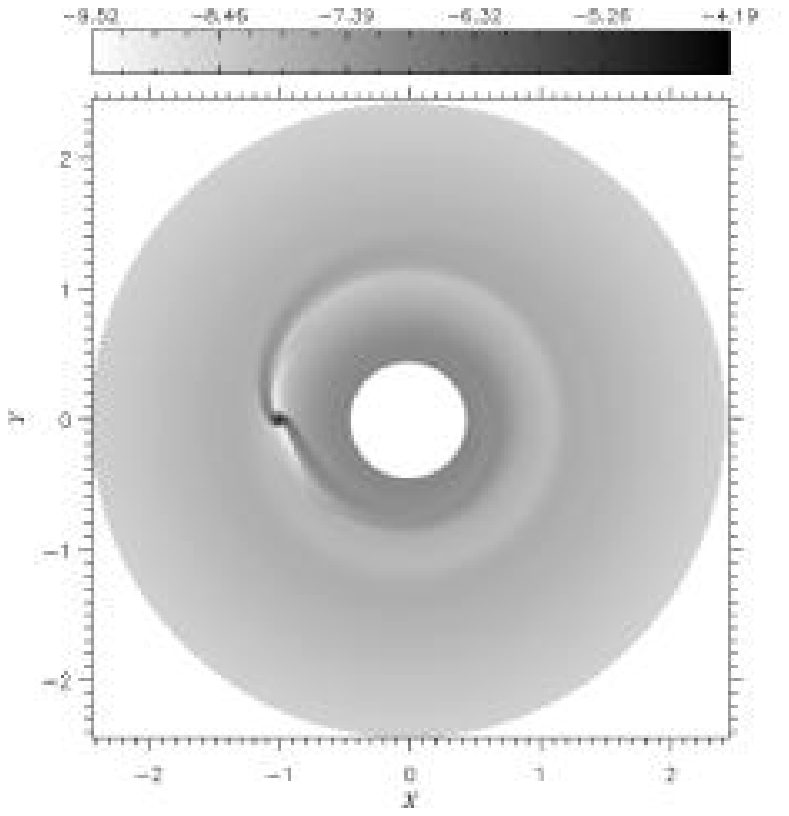}{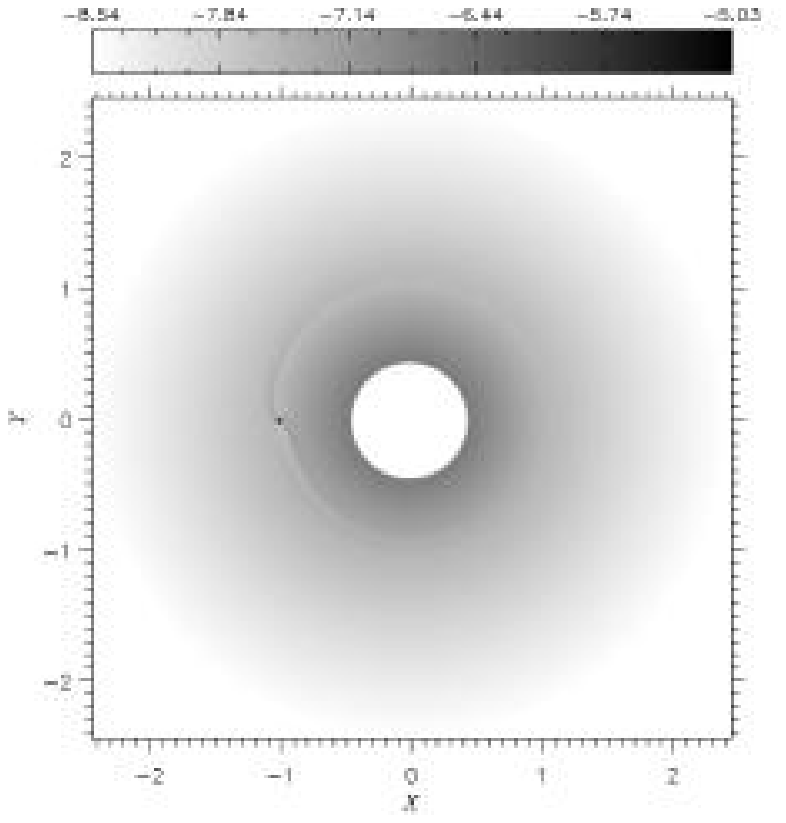}}
\mbox{%
\plottwo{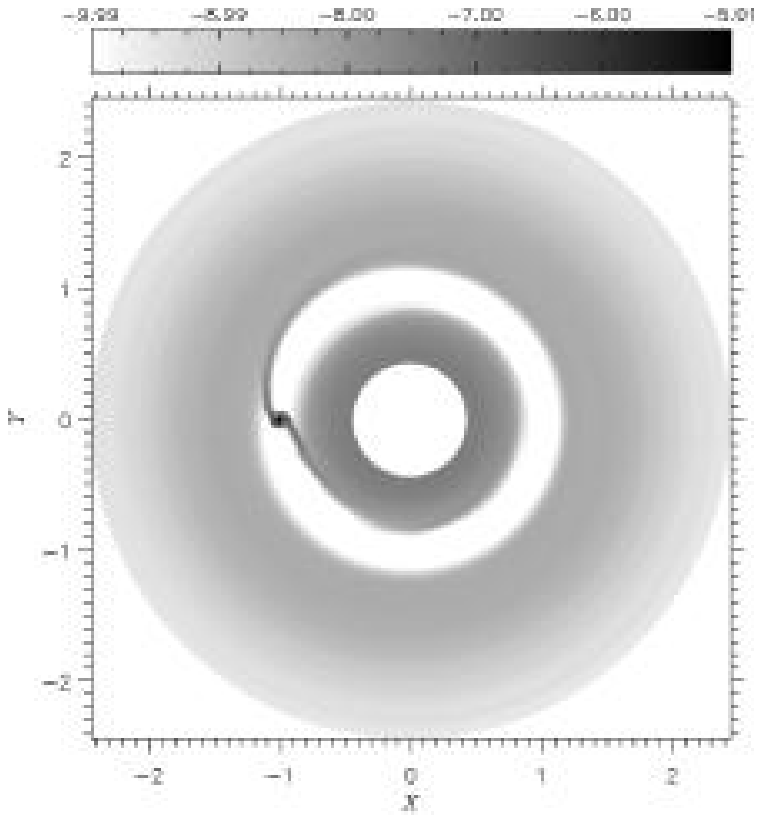}{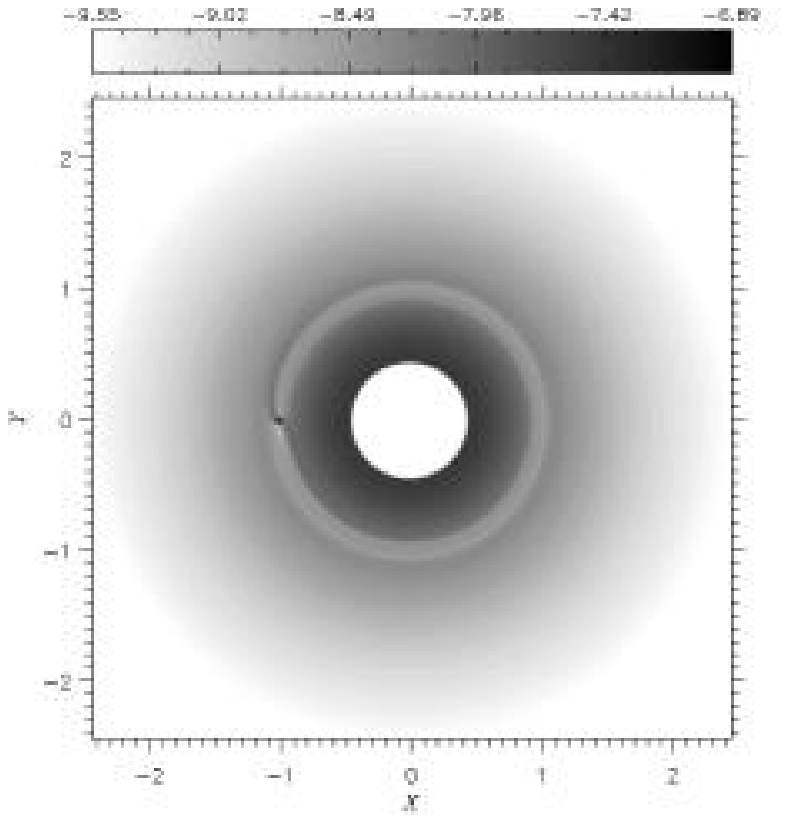}}
\end{center}
\caption{\small{Flux radiated (i.e., $\Lambda$, see \refsec{ssec:energyeq}
       for details) by H-models
       (\textit{top}) and C-models (\textit{bottom}).
       Left panels regard simulations of Jupiter-mass objects,
       whereas right panels describe models with $\Mp=0.1\;\MJup$
       ($\Mp=33\;\MEarth$).
       The color scale in both panels is logarithmic.
       In plot units, $10^{-5}$ means an energy flux of
       $4.7\times10^{4}\;\mathrm{erg}\,\mathrm{cm}^{-2}\,\mathrm{s}^{-1}$.
       For comparison, the surface flux of the Sun is
       $6.3\times10^{10}\;\mathrm{erg}\,\mathrm{cm}^{-2}\,\mathrm{s}^{-1}$.} 
\label{f:gLambda}}
\end{figure*}
Finally, we examine the emitted flux $\Lambda$, computed
according to
\refeqt{eq:Lambda} (see \refFgp{f:gLambda}).
From the observational point of view, disk gaps could represent
a probe for protoplanet detection. In fact, this is by far the
most extensive imprint that a planet leaves on a circumstellar disks.
Prospective studies on observability of gaps due to Jupiter-mass 
protoplanets, have already been presented by \citet{wolf2002,stein2003}, and 
speculations on a developing gap around the \textit{T Tauri} 
star
\textit{TW Hya} have been reported by \citet{calvet2002}.
Here we do not intend to address the issue of whether gaps are really
observable or to what extent they are. 
However, we have to notice that a necessary but not sufficient
condition is that they must be wide, deep, and with sharp edges 
in order for the flux emitted from this region to have the
strongest contrast with respect to the surrounding environment 
\citep{rice2003}. 
From the bottom panels of
\refFgt{f:gLambda}, it appears clear that low kinematic viscosities 
would favor this kind of investigation. Disk spirals are probably too 
elusive to be captured by present-day ground-based instruments,
as shown by the flux maps. 
We also argue that it would seem rather
unlikely to detect planetary masses smaller than Jupiter's, by
means of these measurements. Right panels of \refFgt{f:gLambda}
display the emitted flux in case of a $\Mp=0.1\;\MJup$ model. 
The quantity $\Lambda$ furnished by high-viscosity model (top) looks 
quite smooth and only some inhomogeneities appear in C-model (bottom).
\subsection{Models with $\alpha$-viscosity}
\label{sec:amodels}
\begin{figure*}[!t]
\epsscale{2.0}
\begin{center}
\mbox{%
\plottwo{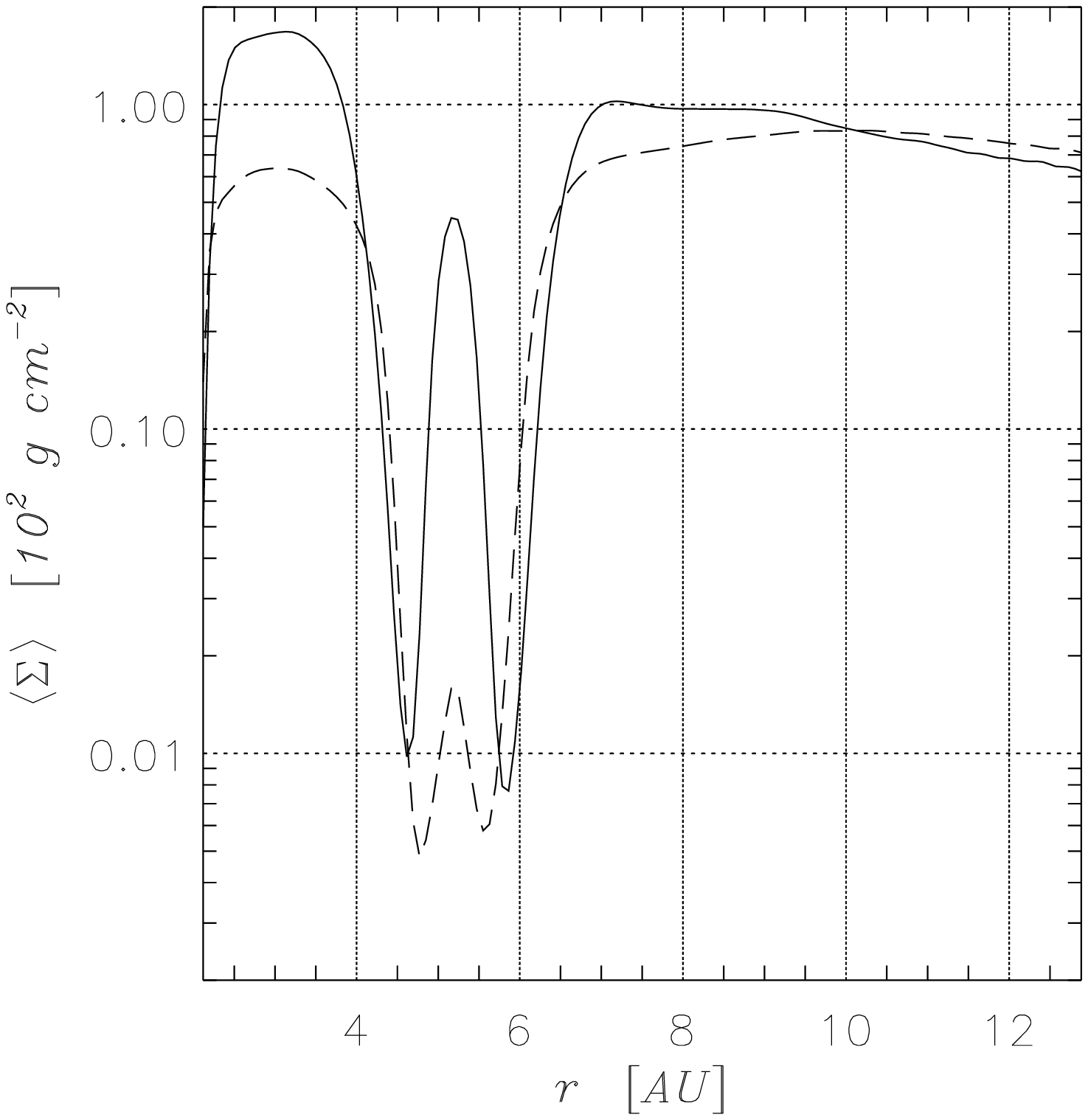}{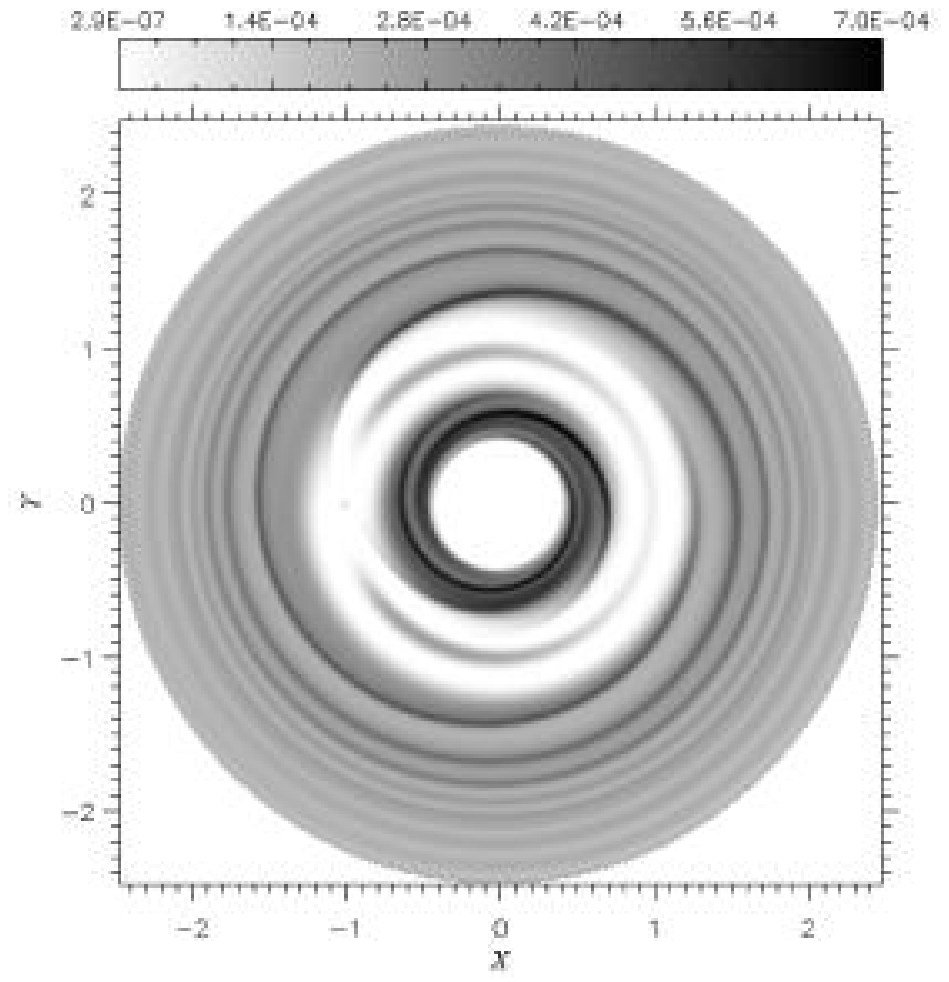}}
\mbox{%
\plottwo{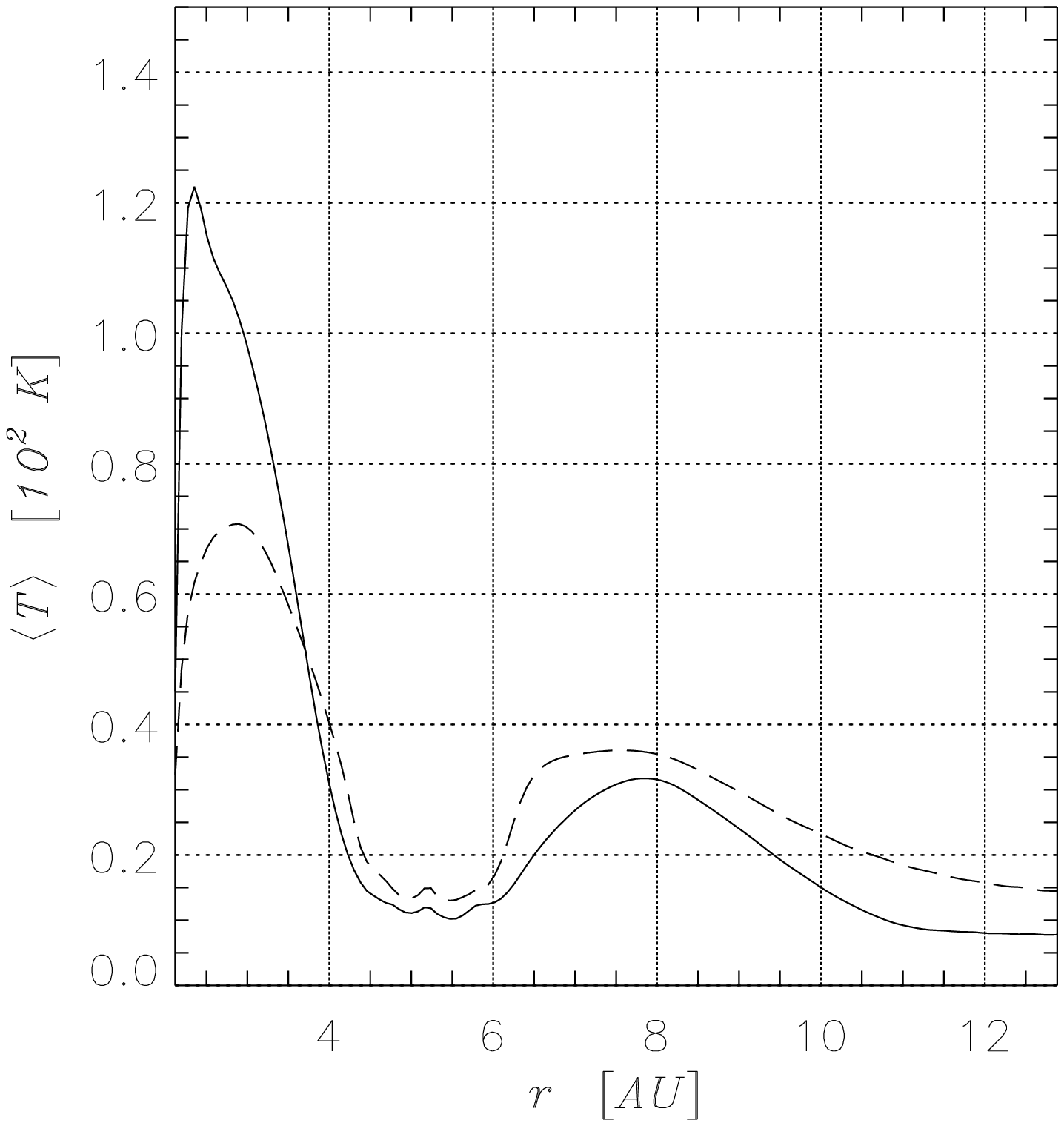}{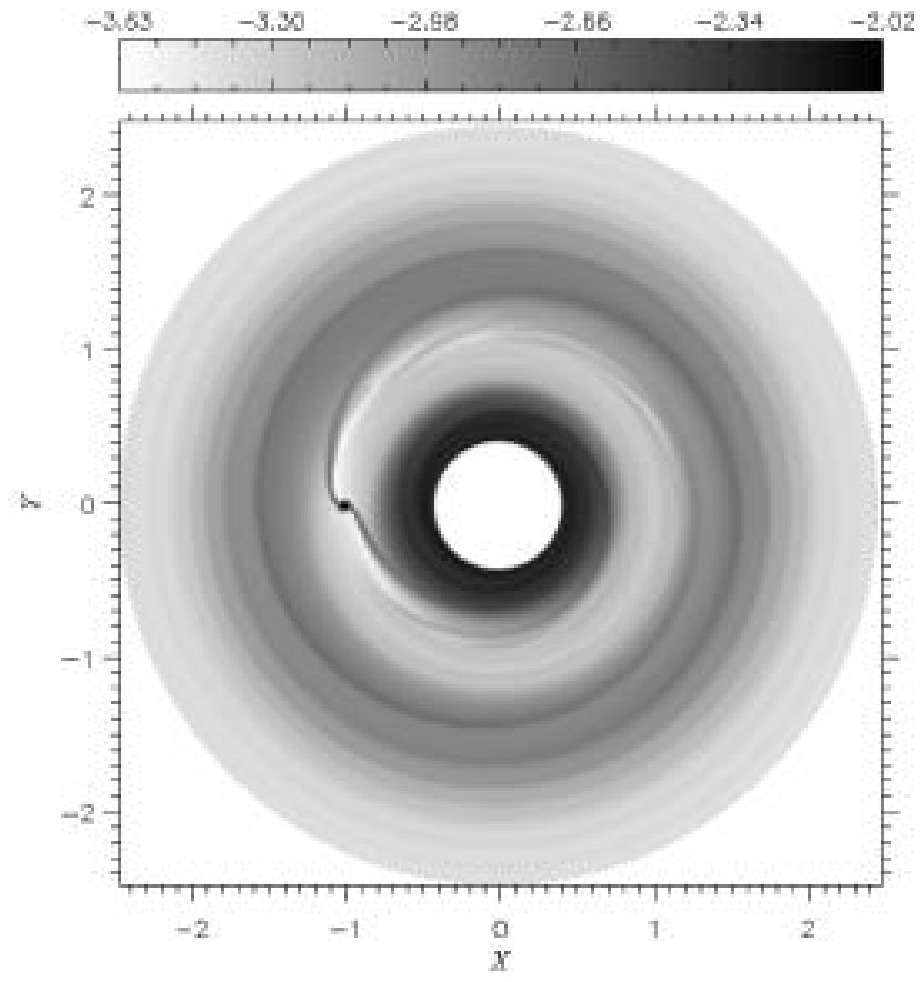}}
\end{center}
\caption{\small{\textit{Left}. Averaged surface density and temperature
       in the Jupiter-mass $\alpha$-viscosity models A3 
       ($\alpha=10^{-3}$, solid
       line) and A2 ($\alpha=10^{-2}$, long-dash line).
       \textit{Right}. 2D-distribution of the surface density (top) and
       temperature (bottom) from models A3. The density scales linearly and
       temperature scales logarithmically. Conversion factors are: 
       $\Sigma=10^{-4}\rightarrow 32.9\;\sdunits$, 
       $T=10^{-2}\rightarrow 198\;\kelvin$.}
\label{f:a3}}
\end{figure*}
So far we have considered disks with a constant $\nu$. However, we
have also investigated the effects of a different viscosity
prescription, namely an $\alpha$-type viscosity \citep{alpha}.
Two Jupiter-mass accreting models were run in this mode, one 
with $\alpha=10^{-2}$ 
(model A2) and the other with $10^{-3}$ (model A3).
Such values are representative of those recently estimated by 
\citet{papa2003} and 
\citet{winters2003} who performed MHD disk simulations, 
with embedded Jupiter-size bodies, and could therefore evaluate the
stress parameter $\alpha$ in a self-consistent fashion. 
Indeed, referring to the constant viscosity models discussed above, 
by making the azimuthal mean
$2\,\pi/\langle\alpha\rangle=(1/\nu)\,\int{c_s\,H\,\ud\varphi}$, 
we get a value of $\langle\alpha\rangle$ ranging from $3\times10^{-2}$
in the inner disk to $4\times10^{-3}$ in the outer disk. Specifically,
C-models yield an overall trend of $\langle\alpha\rangle$ which is the
closest to  $10^{-2}$. And in fact outcomes from $\alpha$-viscosity
model A2 are not much different from those obtained from the
Jupiter-mass C-model.
The quantity $\langle\Sigma\rangle$ is very similar to that illustrated
in left panel of \refFgt{f:gst} (long-dashed line) whereas 
$\langle T\rangle$ is only slightly different (near the inner border)
from the one in the right panel of the same Figure.

Some more important deviations from what has been seen thus far are instead
encountered in the A3-model, as displayed in \refFgt{f:a3}. The
averaged surface density shows a significant hump 
(top-left panel, solid line), 
inside the gap region, not observed in the other viscosity regimes.
This is not caused by the mass accumulation around the planet but
it is produced by \hspace*{0.1em} material \hspace*{0.1em} lingering 
\hspace*{0.1em} around Lagrangian points L4 and L5,
\clearpage
\noindent%
as it appears from the ring-like traces in the middle of the density gap
in the top-right panel \refFgt{f:a3}. 
This also appear to be a persistent feature, 
in fact no change is measured over the last 50 orbital periods.
L4 and L5 points are equilibrium
locations for the restricted three-body problem.
Hence, one can expect that the thermal energy is larger near such points than
anywhere else in the gap.
In turn, viscous torques, which are enhanced by the locally increased 
kinematic viscosity are more efficient in contrasting gravitational torques. 
Thereby, material is more likely forced to evolve on the much longer
viscous time scale
$t_{\mathrm{vis}}\simeq r^2_{\mathrm{p}}/\nu=r_{\mathrm{p}}/(\alpha\,c_s\,h)$.
 
As for the temperature (\refFgp{f:a3}, bottom panels), the largest
discrepancies between models A2 and A3 occur toward the inner border.

\subsection{Gaps as Planet Signatures}
\label{sec:GPS}
\begin{deluxetable}{cccccccc}
\tablecolumns{8}
\tablewidth{0pt}
\tablecaption{Gap Occurrence and Depth.\label{tb:gaps}}
\tablehead{%
\colhead{} & \multicolumn{3}{c}{\textsc{C-Models}} &
\colhead{} & \multicolumn{3}{c}{\textsc{H-Models}} \\
 \cline{2-4}\cline{6-8}
 \multicolumn{1}{c}{\raisebox{1.5ex}[-1.5ex]{$\Mp/\MJup$}}&
 \multicolumn{1}{c}{\textsc{\small Acc.}} &
 \multicolumn{1}{c}{\textsc{\small Non-Acc.}} &
 \multicolumn{1}{c}{\textsc{\small N.F.}} & 
 \colhead{} &
 \multicolumn{1}{c}{\textsc{\small Acc.}} &
 \multicolumn{1}{c}{\textsc{\small Non-Acc.}} &
 \multicolumn{1}{c}{\textsc{\small N.F.}}
}
\startdata
 $1.0\phn$ & $<0.01$          & $<0.01$ &   &   & $0.6$   & $0.8$     & X \\
 $0.5\phn$ & $\phm{<}0.06$    & $\phm{<}0.2\phn$   &   & & $0.9$     & \textsc{NG} & X \\
 $0.2\phn$ & $\phm{<}0.3\phn$ & $\phm{<}0.4\phn$   & X & & \textsc{NG} & \textsc{NG} & X \\
 $0.1\phn$ & $\phm{<}0.6\phn$ & \textsc{NG} & X & & \textsc{NG} & \textsc{NG} & X \\
 $0.06$    & $\phm{<}0.8\phn$ & \textsc{NG} & X & &         &            & X \\
\enddata
 
\tablecomments{Ratio of the minimum density in
       the gap to the density at the gap's taller shoulder (rounded to the
       first significant digit). Values are recovered
       from the azimuthal average of $\Sigma$ around the
       star. The letters ``\textsc{NG}'' stand for ``No Gap''.
        The symbol ``X'' appear whenever at least one of
        criterion~(\ref{eq:gc1}) and (\ref{eq:gc2})
        is not fulfilled (\textsc{N.F.}).
}
\end{deluxetable}

We have seen in the previous two sections that the gap structure 
drastically depends on the viscosity regime.
The standard criterion to open a gap in a disk requires two conditions
\citep{lin1985,lin1993}.
The first is a thermal condition given by:
\begin{equation}
 \left(\frac{\Mp}{\MStar}\right) >  3\,\left(\frac{H}{r}\right)^{3} =
 3\,h^3,
 \label{eq:gc1}
\end{equation}
i.e., the Roche lobe must be larger than the disk scale height.
Because of planet's gravity (see \refeqp{eq:H}) and the low
temperatures established in gap regions, this condition is
always fulfilled when $\Mp\gtrsim 0.1\;\MJup$ (see \refFgp{f:gsh}).  
The second condition concerns the requirement that tidal torques
exceed viscous torques and reads
\begin{equation}
 \left(\frac{\Mp}{\MStar}\right) > \left(\frac{40}{\mathcal{R}}\right),
 \label{eq:gc2}
\end{equation}
where $\mathcal{R}=r^2\,\Omega_\mathrm{K}/\nu$ is the Reynolds number.
The right hand side of \refeqt{eq:gc2} is proportional to $\nu$ and
in our case goes from $\approx4\times10^{-3}$ (H-models) to 
$\approx4\times10^{-4}$ (C-models). Thereby, we should not obtain
gaps in any of the H-models. \refFgt{f:gst} (solid line) decently
agrees with this
prediction, since only a trough is dug in H-model. 
However, condition (\ref{eq:gc2}) is partly violated in the low-viscosity 
regime (\refFgp{f:gvsmass}), because a clear gap is visible for the 
$\Mp=0.2\;\MJup$ case and a trough appears for $\Mp=0.1\;\MJup$.
In \refTab{tb:gaps} we report the gap occurrence and depth for
some of the models.

One reason why the above criterion could not properly apply to
our calculations lies in the fact that it was derived for simple
polytropic disks, while we simulate also the thermal evolution of the
system. 
As implied by the profiles in the top-left panel of \refFgt{f:a3}, it is 
even harder to predict the gap presence when the kinematic viscosity depends 
on the local fluid temperature.
 
\section{Circumplanetary Disks}
\label{sec:pstructure}
The first distinction about the circumplanetary flow  has to be 
made according to whether the planet is accreting or not. Therefore,
we shall carry out separate discussions, based on H- and C-models, 
as well as $\alpha$-viscosity models A2 and A3.

\subsection{Accreting Models}
\label{ssec:acc}
\begin{figure*}[!t]
\epsscale{2.0}
\begin{center}
\mbox{%
\plottwo{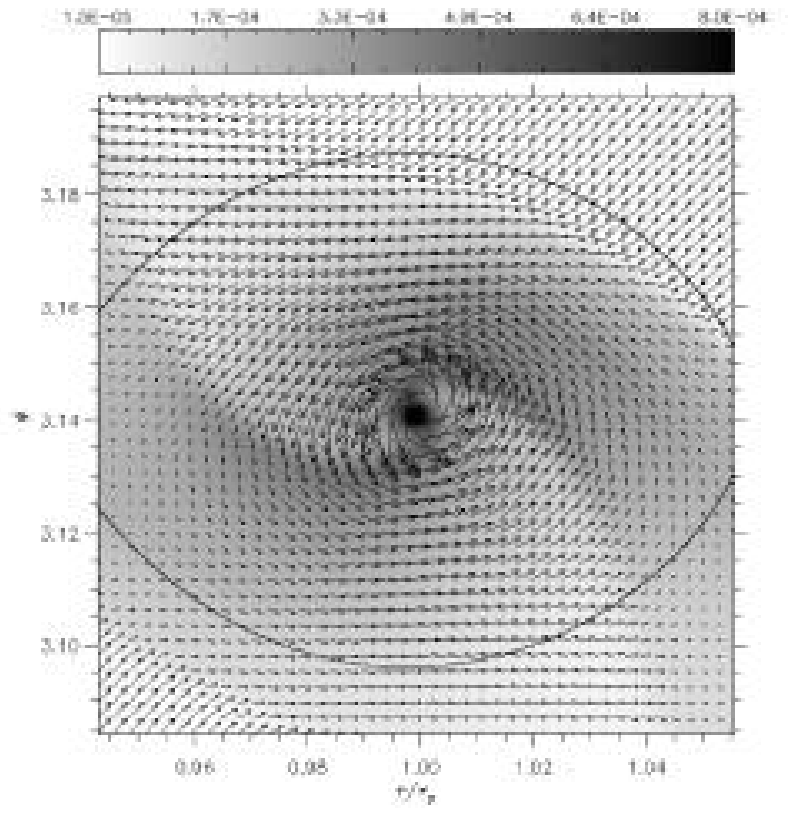}{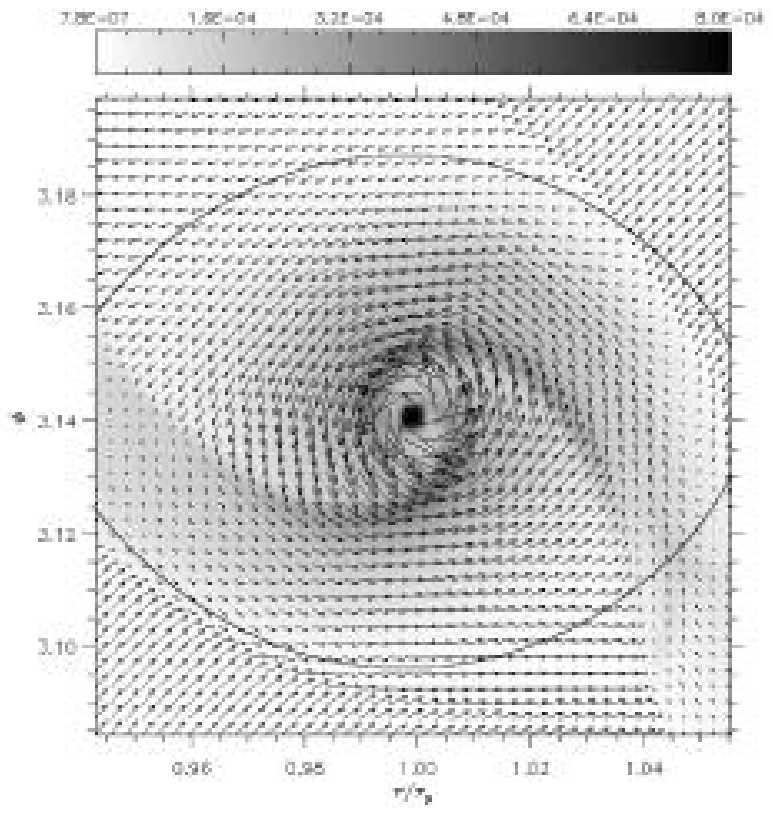}}
\mbox{%
\plottwo{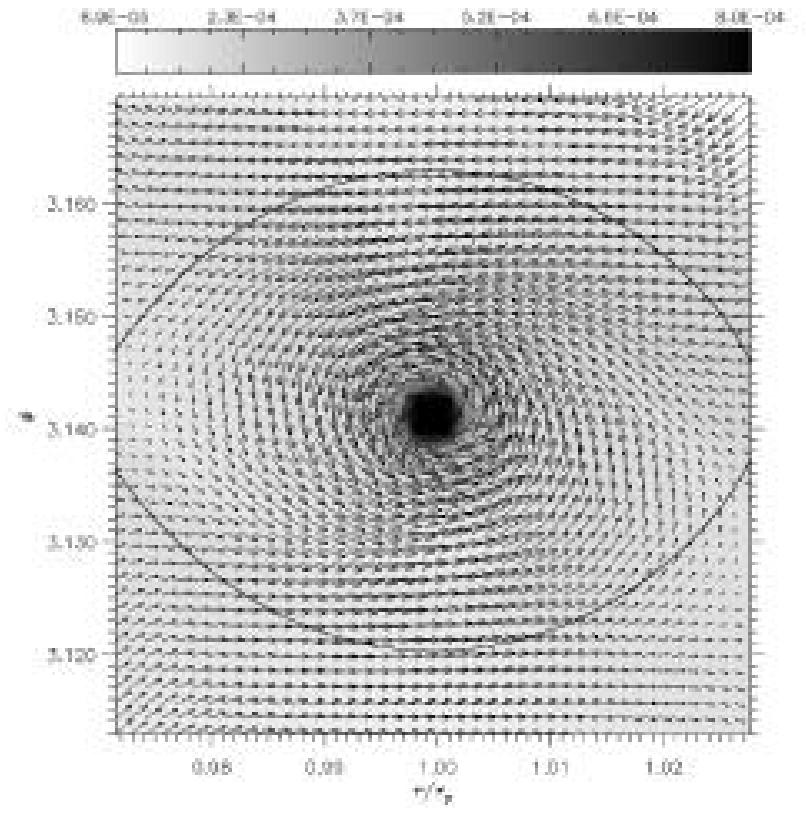}{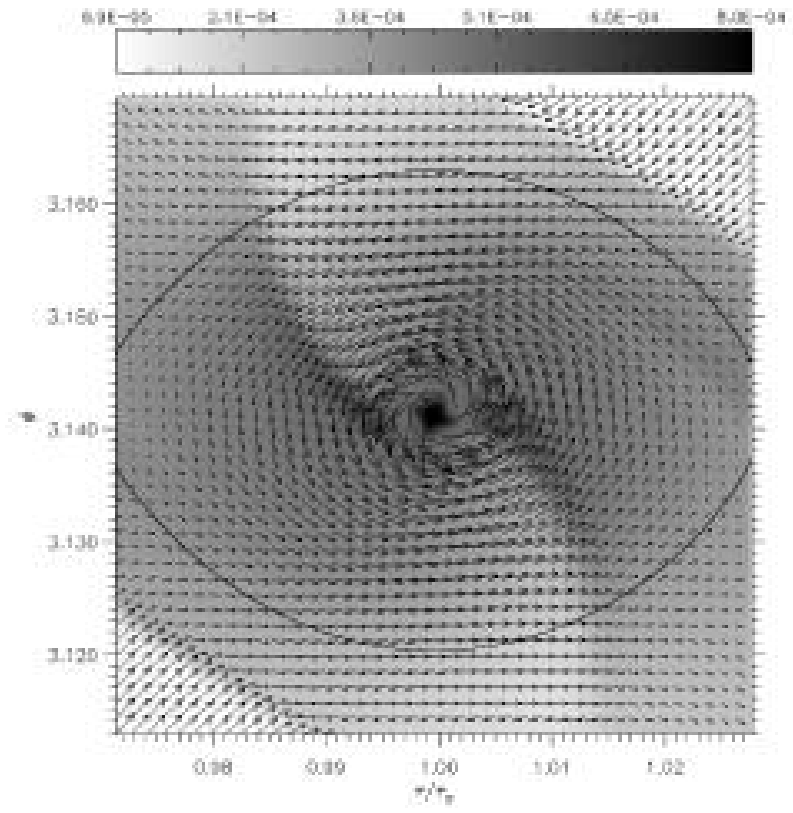}}
\end{center}
\caption{\small{Density distribution and velocity field around $\Mp=1\;\MJup$
       (\textit{top}) and $\Mp=0.1\;\MJup$
       (\textit{bottom}) models. Right panels refer to H-models
       ($\nu=10^{16}\;\nunits$) whereas left ones refer to C-models
       ($\nu=10^{15}\;\nunits$). In each panel, color scales are
       linear. In the units used in the color bar, $10^{-4}$
       corresponds to $\Sigma=32.9\;\sdunits$.}
\label{f:zpic1}}
\end{figure*}
\begin{figure*}[!t]
\epsscale{2.0}
\plottwo{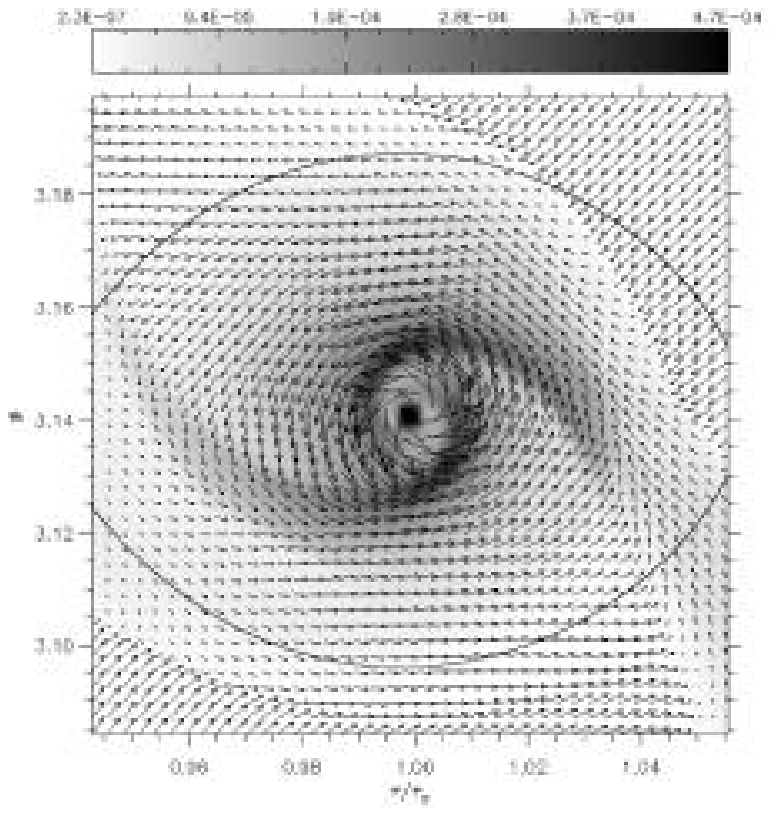}{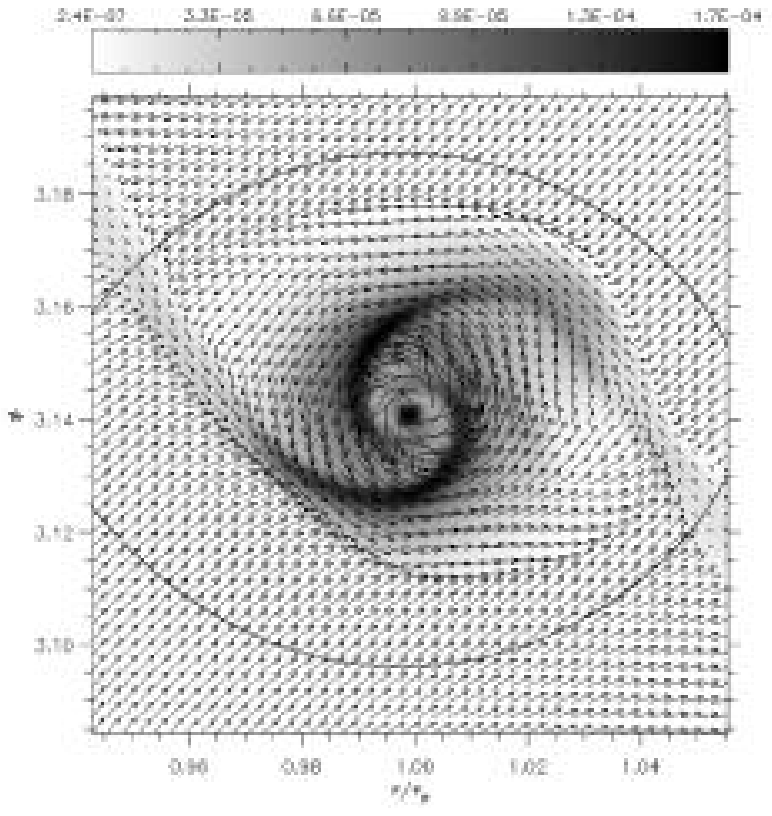}
\caption{\small{Density distribution and velocity field around the Jupiter-mass
       model in which the $\alpha$-viscosity prescription is used and
       set to
       $\alpha=10^{-2}$ (\textit{left}) and $\alpha=10^{-3}$ (\textit{right}).
       The color bar scale is linear and $10^{-4}$ is equal to 
       $\Sigma=32.9\;\sdunits$.}
\label{f:apic}}
\end{figure*}

\refFgt{f:zpic1} indicates that,
inside of the Roche lobe, the general aspect of the flow circulation around 
$1\;\MJup$ and $0.1\;\MJup$ protoplanets resembles that obtained with 
local isothermal models in two dimensions (\gAA). 
However, specific characteristics of the flow do differ.
A circumplanetary disk, extending over a fair fraction of Roche
lobe, can be identified for both masses and viscosity regimes.
The mass of such disks around one Jupiter-mass protoplanets,
within the 80\% of the Hill radius, 
is $1.03\times 10^{-5}\;\MJup$ in the H-model and $5.36\times
10^{-6}\;\MJup$ in the C-model. As regards calculations with
$\Mp=0.1\;\MJup$ planets, the mass measured in the circumplanetary
disk (or sub-disk, for brevity)
of the C-model is $4.03\times 10^{-6}\;\MJup$, while the amount is
nearly two times less in the H-model. 
While the flow dynamics in the Jupiter-mass model with
$\alpha=10^{-2}$ (left panel in \refFgp{f:apic}) is similar that of 
the C-model, the model with $\alpha=10^{-3}$ presents some
peculiarities. As shown in right panel of \refFgt{f:apic}, the size of the
circumplanetary disk is smaller than the Roche lobe. Moreover, 
the streams of matter associated with the spiral waves are narrower.
The sub-disk mass measured in these cases is $3.9\times 10^{-6}\;\MJup$
(A2) and $1.5\times 10^{-6}\;\MJup$ (A3).

\begin{figure}[!b]
\epsscale{1.0}
\plotone{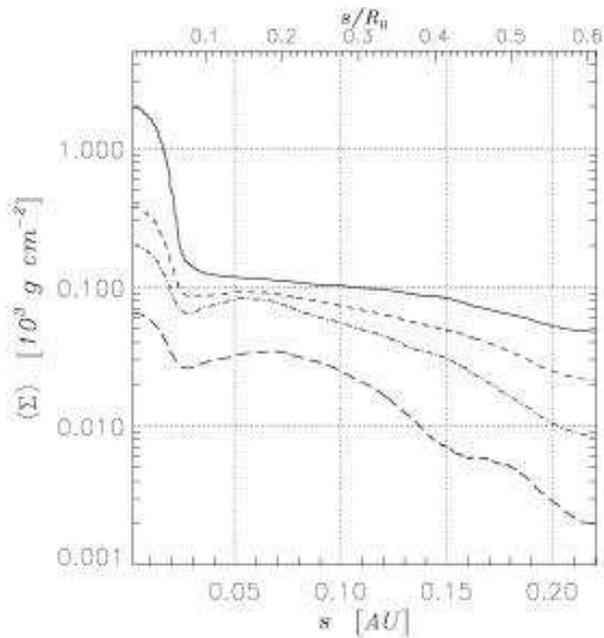}
\caption{\small{Surface density averaged around $\Mp=1\,\MJup$ planets in the
       inner part of the Roche lobe. Considered models are H (solid
       line), C (short-dash line), A2 (Dash-dot line),
       and A3 (long-dash line).}
\label{f:ssd}}
\end{figure}

The azimuthal average of the surface density distribution around the
planet, inside the Roche lobe, can be fitted by a relation of the type
(see \refFgp{f:ssd}):
\begin{equation}
\langle \Sigma \rangle \simeq \langle \Sigma \rangle_0\,
                    \exp{\left[a_1\left(\frac{\dpl}{\RH}\right)\right]},
\label{eq:syntSav}
\end{equation}
where $\dpl/\RH\le 1$ and larger than a threshold length (either $0.1$
or $0.2\;\RH$). 
The values of $\langle \Sigma \rangle_0$ and $a_1$ are reported in
\refTab{tb:syntSav}.
\begin{deluxetable}{cccrcccr}
\tablecolumns{8}
\tablewidth{0pt}
\tablecaption{Fit Parameters for the Averaged Surface Density.
\label{tb:syntSav}}
\tablehead{%
\colhead{} & \multicolumn{3}{c}{\textsc{C-Models}} &
\colhead{} & \multicolumn{3}{c}{\textsc{H-Models}} \\
 \cline{2-4}\cline{6-8}
 \multicolumn{1}{c}{\raisebox{1.5ex}[-1.5ex]{$\Mp/\MJup$}}&
 \multicolumn{1}{c}{$\langle \Sigma \rangle_0\;[\sdunits]$} &
 \multicolumn{1}{c}{$a_1$} &
 \multicolumn{1}{c}{\textsc{Range}} &
 \colhead{} &
 \multicolumn{1}{c}{$\langle \Sigma \rangle_0\;[\sdunits]$} &
 \multicolumn{1}{c}{$a_1$} &
 \multicolumn{1}{c}{\textsc{Range}} 
}
\startdata
  $1.0\phn$ & $2.92\times10^{2}$ &  $-4.4$  &  $[0.2, 1.0]\,\RH$&
                   & $2.14\times10^{2}$ &  $-2.3$  &  $[0.2, 1.0]\,\RH$\\
  $0.5\phn$ & $1.59\times10^{2}$ &  $-2.0$  &  $[0.2, 1.0]\,\RH$&
                   & $1.35\times10^{2}$ &  $-1.3$  &  $[0.2, 1.0]\,\RH$\\
  $0.2\phn$ & $2.32\times10^{2}$ &  $-1.5$  &  $[0.1, 1.0]\,\RH$&
                   & $8.57\times10^{1}$ &  $-0.5$  &  $[0.1, 1.0]\,\RH$\\
  $0.1\phn$ & $1.99\times10^{2}$ &  $-0.9$  &  $[0.1, 1.0]\,\RH$&
                   & $6.75\times10^{1}$ &  $-0.2$  &  $[0.2, 1.0]\,\RH$\\
  $0.06$           & $1.47\times10^{2}$ &  $-0.4$  &  $[0.2, 1.0]\,\RH$&
                   &                    &          &                    \\
 \cline{2-8}
            & \multicolumn{3}{c}{\textsc{A3-Model}} &
            & \multicolumn{3}{c}{\textsc{A2-Model}} \\
 \cline{2-4}\cline{6-8}
  $1.0\phn$ & $1.80\times10^{2}$ &  $-7.5$  &  $[0.2, 0.8]\,\RH$&
                   & $2.36\times10^{2}$ &  $-5.0$  &  $[0.2, 1.0]\,\RH$\\
\enddata
 
\tablecomments{Parameters that enter \refeqt{eq:syntSav}. This
       represents a linear best-fit of the logarithm of the averaged 
       density $\langle \Sigma \rangle$ over the specified interval.}
\end{deluxetable}

The presence of spiral features in the density distribution is an
indication that the circumplanetary flow is Keplerian-like.
Indeed, decomposing the velocity field $\gv{u}$ into the in-fall
velocity (toward the planet) $w\subscr{in}$ and the rotational
velocity (counter-clockwise around the planet)
$w\subscr{rot}$, it turns out that
the former is more than an order of magnitude less than the latter.
In Jupiter-mass models, the rotational velocity within $0.5\,\RH$ of
the planet drops approximately as
$\dpl^{-0.6}$, nearly independently of the viscosity regime
investigated in these computations.
For $0.1\;\MJup$ models, this ratio declines at a slightly higher rate.
Consequently, for $\dpl\in[0.1,0.5]\,\RH$, the ratio of
$w\subscr{rot}$ to the local Keplerian
velocity is always larger than $0.7$.

Comparing the right panels (C-models) of \refFgt{f:zpic1}
to Figure~13 (top and middle right panels) in \gAA, 
one can clearly see that spiral perturbations are less intense,
as we show below, and more open. The latter circumstance is
related to the lower values of the Mach number in the circumplanetary
flow, which governs the inclination of the spiral wave with respect to
the direction of the rotational motion. In fact, in local isothermal 
models $\mathcal{M}_{\mathrm{iso}}\simeq\sqrt{q\,(r_\mathrm{p}/\dpl)}/h$.
If $\Mp=1\;\MJup$, $\mathcal{M}_{\mathrm{iso}}$  drops
from $\approx8$, at $\dpl=0.1\,\RH$, to $2$, at $\dpl=1\,\RH$.  
As comparison, in the circumplanetary disk displayed in the
upper-left panel of \refFgt{f:zpic1}, $\mathcal{M}$ lies between
$2.4$ and $1.4$, whereas the low viscosity Jupiter-model (upper-right panel)
provides values between $3$ and $1.5$. 
Models A2 and A3 yield similar values.
The main reason for this difference resides in the larger value of the
sound speed. 
Hence, perturbations can travel faster and are less
distorted by the background motion of the flow.
In the Jupiter-mass cases, illustrated in \refFgt{f:zpic1}, the azimuthally
averaged $\mathcal{M}$ is approximately proportional to either
$\sim s^{-0.1}$ (H-model) or to $\sim s^{-0.2}$ (C-model).
In the limit of a nearly constant Mach number, wave perturbations
assume the form of Archimedes' spirals.

\begin{figure}[!b]
\epsscale{1.0}
\plotone{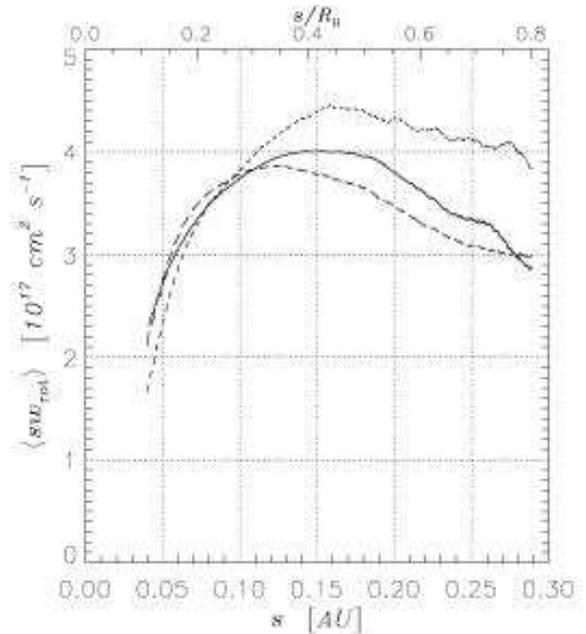}
\caption{\small{Specific angular momentum ($\dpl\,w_{rot}$), averaged around 
       $\Mp=1\,\MJup$ planets modeled with different geometries and
       thermal structures. The solid line belongs to the 2D
       C-model presented in this paper. The short-dash line derives
       from a locally isothermal 2D model ($T\simeq 90\;\kelvin$) as 
       described in \gAA.
       The long-dash line comes from a locally isothermal 3D model
       ($T\simeq 90\;\kelvin$) presented in \gApJ. In this last case, 
       the specific angular momentum is evaluated in the disk midplane
       ($z=0$).}
\label{f:sam}}
\end{figure}
As pointed out in \gApJ, in a three-dimensional space circumplanetary
spirals are weakened \citep[see also][]{bate2003}. In fact, because of the
flow circulation above the disk midplane (see \gApJ, Figure~3, middle
and bottom panels), 
less angular momentum is carried inside the Roche lobe by the midplane flow.
As anticipated above, the spiral density features look smoother in the
models presented here. This is quantitatively demonstrated in 
\refFgt{f:sam}. In the Figure the average specific angular momentum 
$\dpl\,w\subscr{rot}$ obtained from the $\Mp=1\;\MJup$ C-model (solid
line) is compared  to that of a 2D Jupiter-mass \textit{local-isothermal} model 
(short-dash line) and to that of a 3D Jupiter-mass model from \gApJ\ 
(long-dash line), which also assumes a \textit{quasi-isothermal gas} 
around the planet. In the 3D case, the quantity $\dpl\,w\subscr{rot}$
is computed at $z=0$, i.e., in the disk midplane.
The latter two models are quasi-isothermal in the sense that the
temperature distribution is fixed and depends only on the
distance distance from the star $r$ according to the relation
$T\propto (\mu/\gamma)\,h^2\,\MStar/r$.
Thus, around the planet's location 
$r=r_{\mathrm{p}}$, $T\approx90\;\kelvin$ 
if the disk aspect ratio $h$ is $0.05$, $\mu=2.39$, and $\gamma=1.4$.
From \refFgt{f:sam} one can see that
in the outer part of the Roche lobe ($\dpl\gtrsim 0.3\;\RH$), where 
spiral perturbations are strongest, the discrepancy between
the two \textit{quasi-isothermal} models lies between 20 and 30\%, whereas
the two-dimensional \textit{thermal} model provides a specific angular
momentum only 10\% larger than that of the three-dimensional
\textit{quasi-isothermal} model.
This occurrence can be attributed to the better pressure structure achieved 
by the calculations that account for the thermal energy budget.

\begin{figure}[!t]
\epsscale{1.0}
\plotone{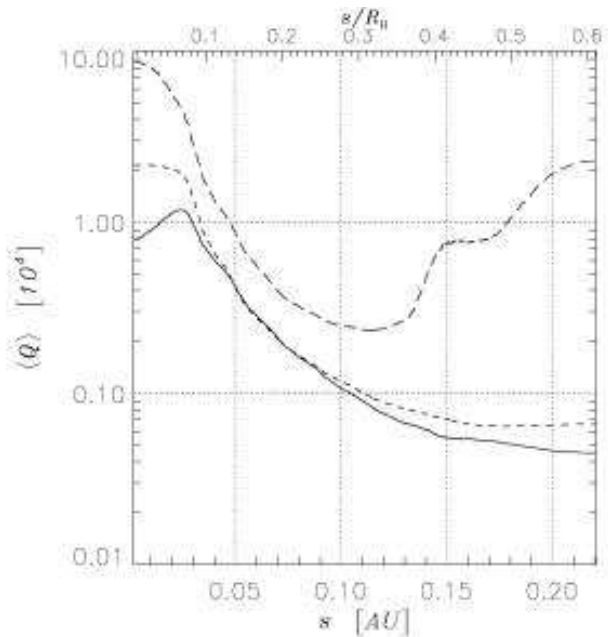}
\caption{\small{Toomre parameter $Q=c_s\,w_{rot}/(\pi\,G\,\dpl\,\Sigma)$ 
       averaged around $\Mp=1\,\MJup$ bodies in models H (solid line), 
       C (short-dash line), and A3 (long-dash line). The large value
       of $\langle Q\rangle$ implies that circumplanetary disks around
       Jupiter-mass planets are gravitationally stable.}
\label{f:toom}}
\end{figure}
Circumplanetary disks should not suffer from gravitational
instabilities since the Toomre parameter $Q$ is orders of magnitude
larger than unity, as proved by \refFgt{f:toom}. This arises from
the small value of the ratio between the mass of the circumplanetary
disk and that of the planet.

\begin{figure*}[!t]
\epsscale{2.0}
\begin{center}
\mbox{%
\plottwo{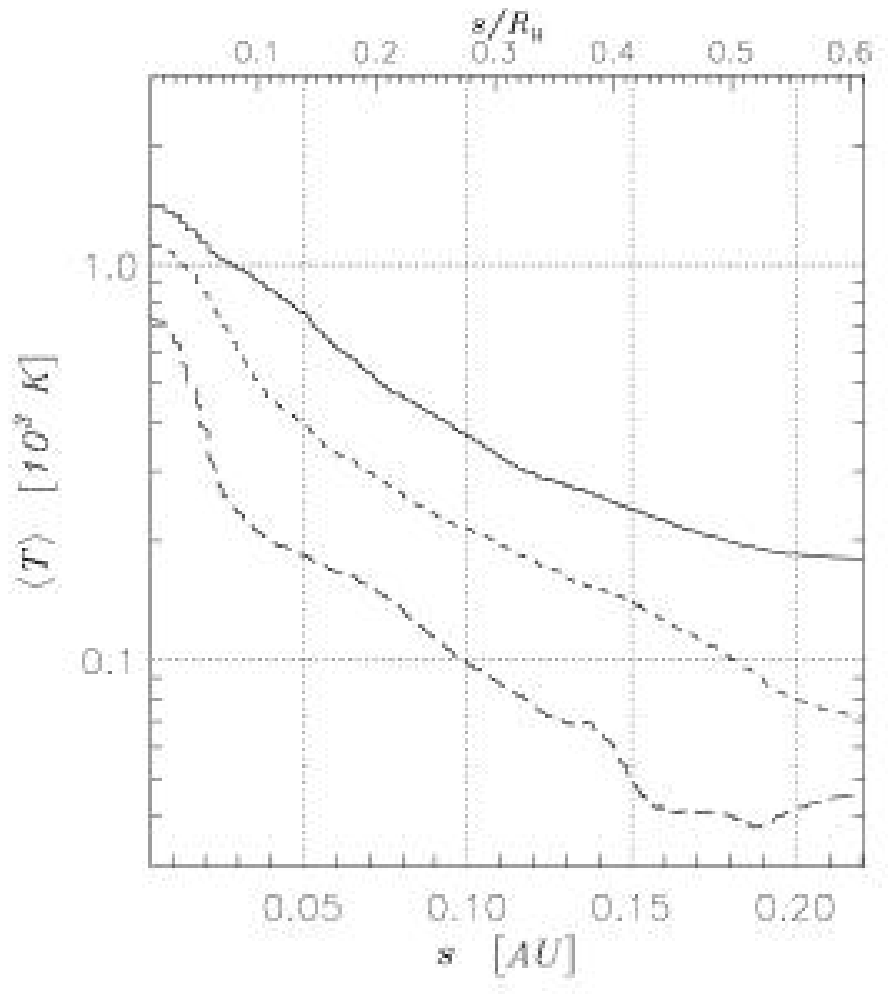}{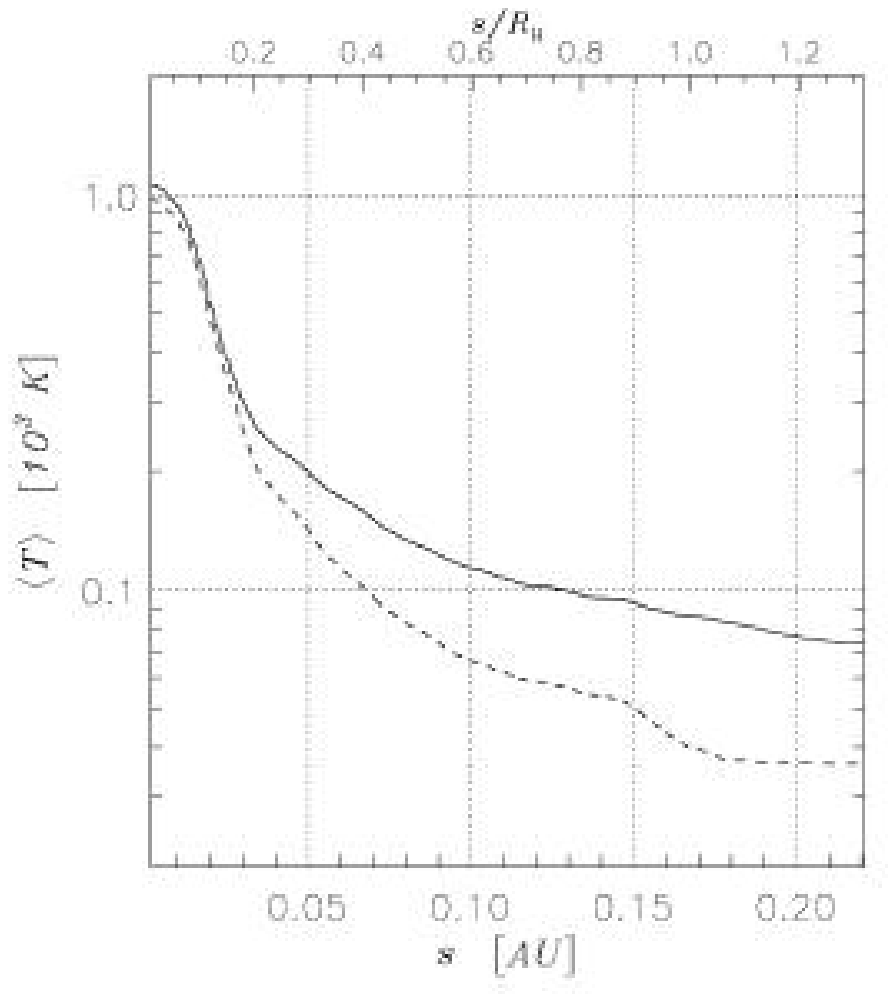}}
\mbox{%
\plottwo{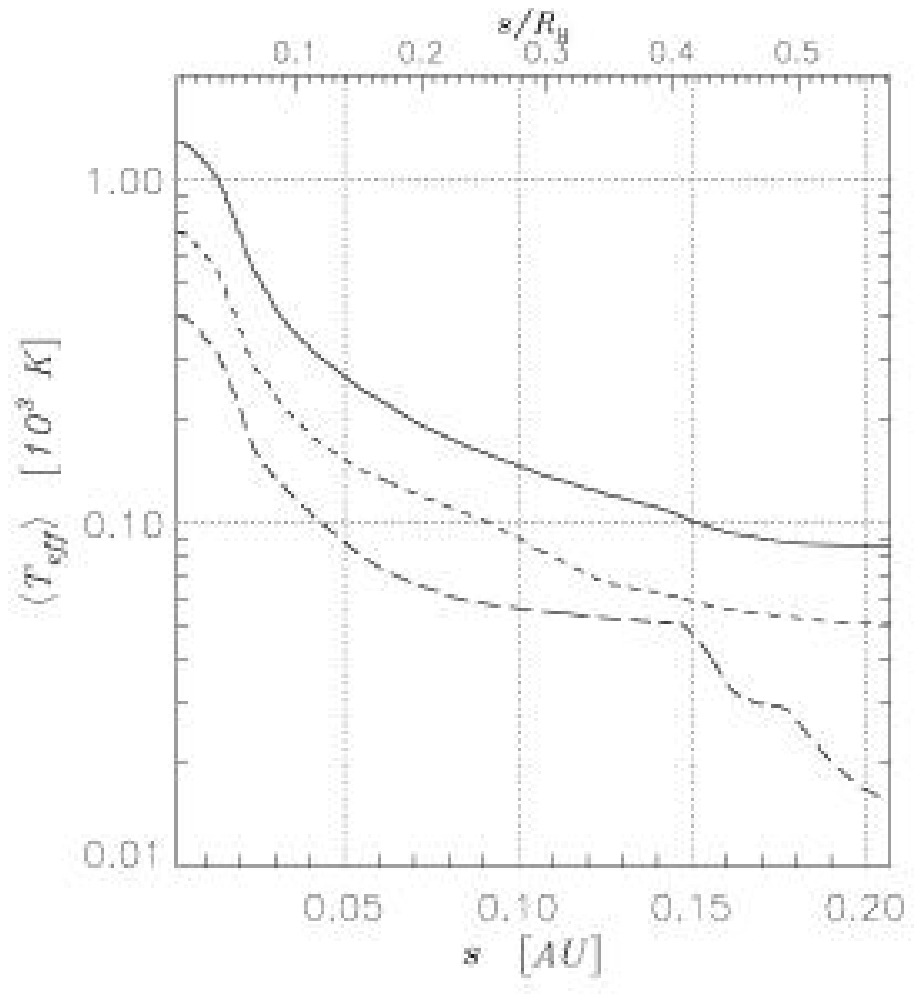}{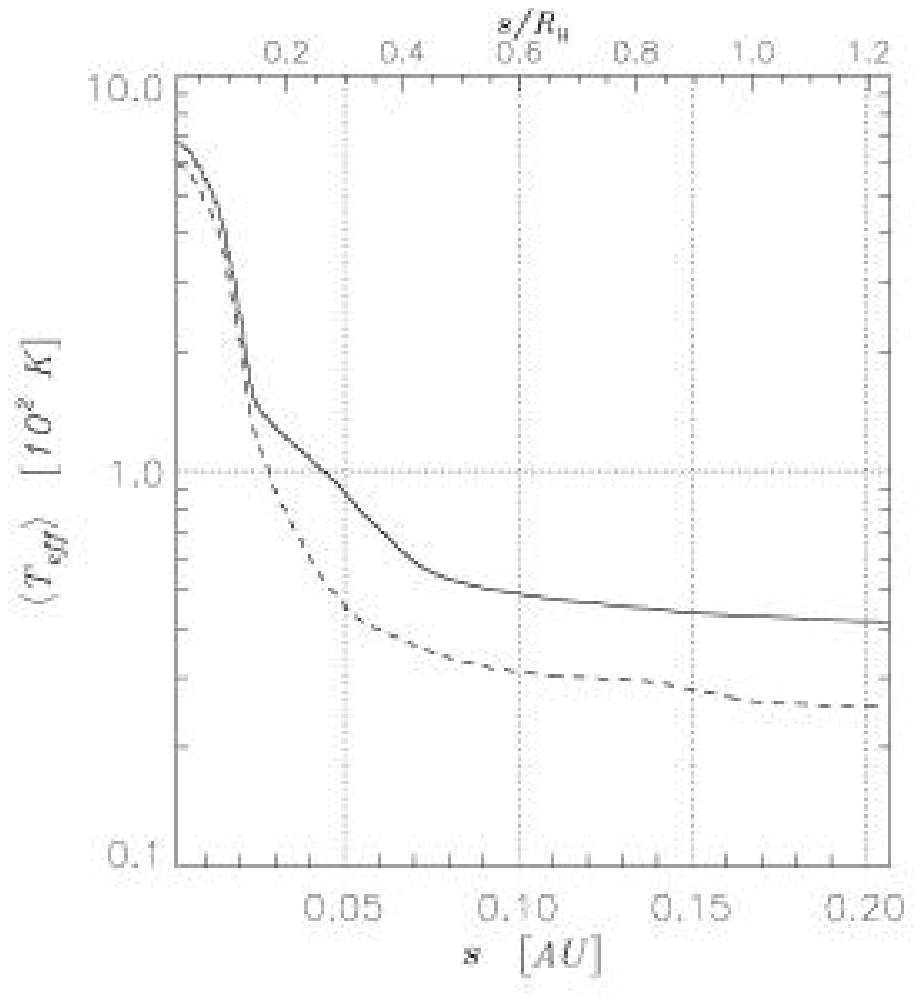}}
\end{center}
\caption{\small{Azimuthally averaged temperature (\textit{top}) and effective
       temperature (\textit{bottom}) distribution,
       around Jupiter-size (\textit{left}) and $0.1\;\MJup$ (\textit{right})
       accreting models.  
       Solid, short-dash, and long-dash lines refer to models H, C,
       and A3, respectively.}
\label{f:zpic2}}
\end{figure*}

Two-dimensional temperature distributions are rather symmetric with
respect to the planet's position, though they are marked by weak spiral
perturbations.
\refFgt{f:zpic2} (top panels) shows the azimuthal averaged temperature around
protoplanet with $\Mp=1\;\MJup$ (left) and $\Mp=0.1\;\MJup$ (right).  
In the Jupiter-mass case, maximum temperatures range
from roughly $1500$ to $1000\;\kelvin$, whereas $T$ reaches the value
of the ambient medium ($\lesssim50\;\kelvin$) toward the border of
the Hill sphere.
Comparing H- and C-models, one can realize that the average
temperature in the inner part of the Roche lobe differ by less
than a factor two, whereas the $\alpha$-viscosity model A3 shows 
significant lower temperatures. 
In the low-mass case ($\Mp=0.1\;\MJup$), the peak temperature is
around $1000\kelvin$ and between $\approx80$ and 
$\approx40\;\kelvin$ at $\dpl=\RH$, respectively for both H- and C-model. 
The average temperature differs by less than $\approx 30\;\kelvin$ 
(\refFgp{f:zpic2}, upper-right panel).

Over the entire sub-disk domain, $\langle T \rangle$ can be
well fitted by the curve: 
\begin{equation}
\langle T \rangle \simeq \langle T \rangle_{\bar{\dpl}}\,
                     \left(\frac{\bar{\dpl}}{\dpl}\right)^{\xi},
\label{eq:syntT}
\end{equation}
in which the length $\bar{\dpl}$ is set to $0.1\,\RH$, for convenience.
Fitting parameters are reported in \refTab{tb:syntTav}, together with
the validity range of the fitting function. From entries in the Table,
it turns out that the temperature generally falls off as $1/\dpl$. 
\begin{deluxetable}{crrrcrrr}
\tablecolumns{8}
\tablewidth{0pt}
\tablecaption{Fit Parameters for the Averaged Temperature Distribution:
       Accreting Models.\label{tb:syntTav}}
\tablehead{%
\colhead{} & \multicolumn{3}{c}{\textsc{C-Models}} &
\colhead{} & \multicolumn{3}{c}{\textsc{H-Models}} \\
 \cline{2-4}\cline{6-8}
 \multicolumn{1}{c}{\raisebox{1.5ex}[-1.5ex]{$\Mp/\MJup$}}&
 \multicolumn{1}{c}{$\langle T \rangle_{\bar{\dpl}}\;[\kelvin]$} &
 \multicolumn{1}{c}{$\xi$} &
 \multicolumn{1}{c}{\textsc{Range}} &
 \colhead{} &
 \multicolumn{1}{c}{ $\langle T \rangle_{\bar{\dpl}}\;[\kelvin]$}&
 \multicolumn{1}{c}{$\xi$} &
 \multicolumn{1}{c}{\textsc{Range}} 
}
\startdata
  $1.0\phn$ & $5.405\times10^{2}$ &  $0.97$  &  $[0.05, 1.0]\,\RH$&
                   & $9.534\times10^{2}$ &  $0.96$  &   $[0.1, 1.0]\,\RH$\\
  $0.5\phn$ & $5.304\times10^{2}$ &  $0.96$  &  $[0.05, 1.0]\,\RH$&
                   & $8.425\times10^{2}$ &  $0.95$  &   $[0.1, 1.0]\,\RH$\\
  $0.2\phn$ & $5.713\times10^{2}$ &  $1.12$  &  $[0.05, 1.0]\,\RH$&
                   & $6.713\times10^{2}$ &  $0.90$  &  $[0.05, 1.0]\,\RH$\\
  $0.1\phn$ & $4.395\times10^{2}$ &  $1.03$  &  $[0.05, 1.0]\,\RH$&
                   & $4.868\times10^{2}$ &  $0.79$  &  $[0.05, 1.0]\,\RH$\\
  $0.06$           & $4.634\times10^{2}$ &  $1.07$  &  $[0.05, 1.0]\,\RH$&
                   &                     &          &                    \\
 \cline{2-8}
            & \multicolumn{3}{c}{\textsc{A3-Model}} &
            & \multicolumn{3}{c}{\textsc{A2-Model}} \\
 \cline{2-4}\cline{6-8}
  $1.0\phn$ & $2.484\times10^{2}$ &  $1.04$  &  $[0.05, 0.8]\,\RH$&
                   & $4.234\times10^{2}$ &  $0.95$  &  $[0.1, 1.0]\,\RH$\\
\enddata
 
\tablecomments{Parameters that enter \refeqt{eq:syntT}. This is a
       linear best-fit of $\log{\langle T \rangle}$,
       computed inside of the Hill sphere. The column
       ``\textsc{Range}'' refers to the validity range of the fit.
}
\end{deluxetable}

\begin{figure*}[!t]
\epsscale{2.0}
\plottwo{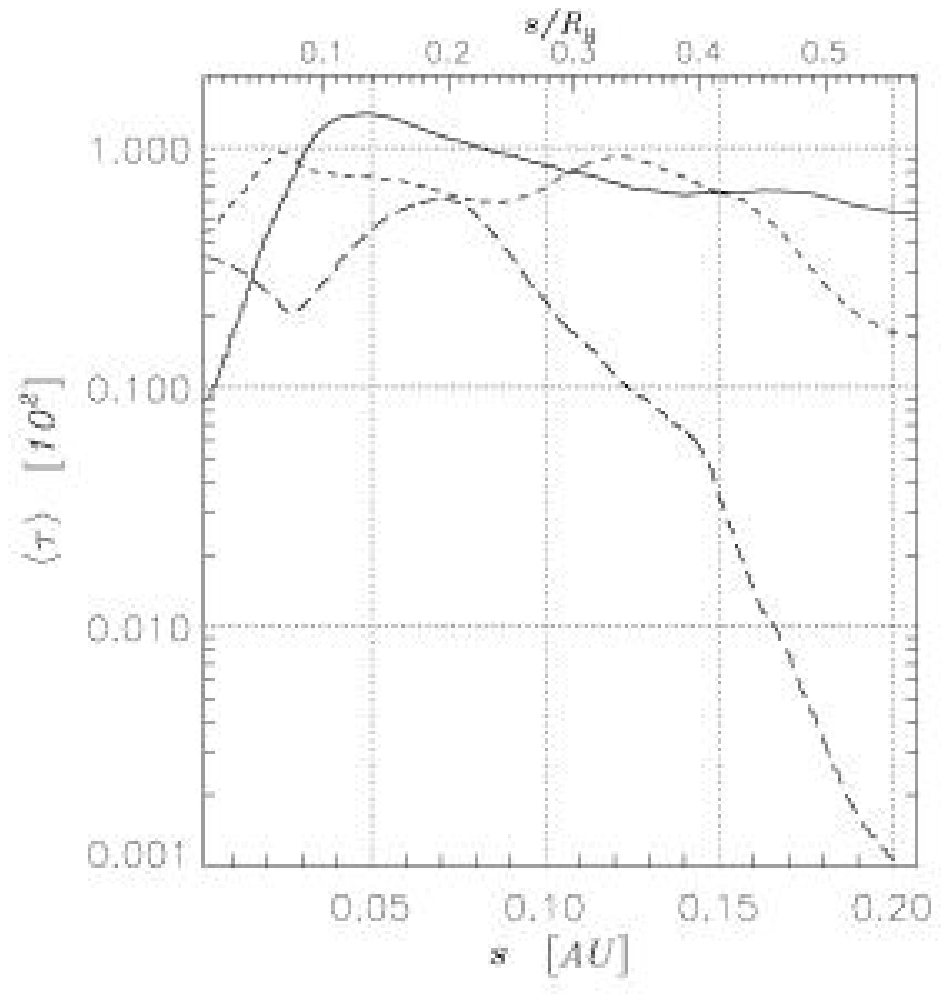}{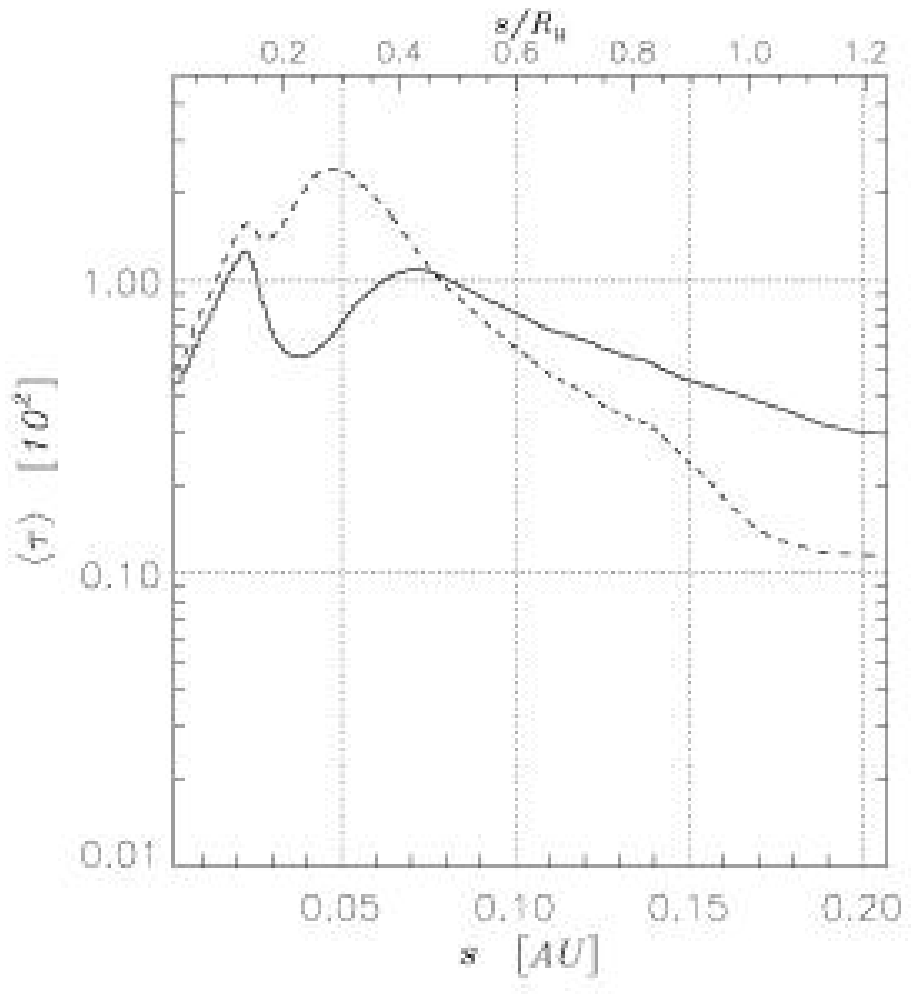}
\caption{\small{Average optical thickness around $\Mp=1\;\MJup$
       (\textit{left}) and $\Mp=0.1\;\MJup$ (\textit{right}) protoplanets.
       The solid, short-dash, and long-dash lines reproduce the H-,
       C-, and A3-model, respectively.}
\label{f:avprofac}}
\end{figure*}

Bottom panels of \refFgt{f:zpic2} display the average effective 
temperature $\langle T_{\mathrm{eff}}\rangle$ (see
\refeqp{eq:lambda3}) in the vicinity of protoplanets. 
\refFgt{f:avprofac} shows that the magnitude of the optical thickness 
$\langle \tau\rangle$ is between $10$ and $100$, hence
\refeqt{eq:Lambda} yields 
$T_{\mathrm{eff}}\approx T/\tau^{0.25}$
and thus there is, at most, a factor three between the two
temperatures.
It should be noted that only the model A3 presents optically
thin regions toward the edges of the sub-disk.

Accreted matter also contributes to the energy budget of the
circumplanetary disk via dissipation of gravitational energy into
heat. This source of energy is not treated explicitly
in these simulations. However, it is considered implicitly as a
compression work term arising from the removal of matter in the
accreting region. 
\citet{tanigawa2002} assumed the dissipation of gravitational energy
as the only heating source and derived temperature profiles of
sub-disks with different sound-speed regimes, performing local
simulations with an isothermal equation of state. 
They found that the temperature would scale as $1/\sqrt{\dpl}$, 
independently of the planetary mass.

We have to point out that, since radial transport of radiation is not
treated here, one should expect large temperature gradients
to be smoothed out. But we have seen that temperature profiles are not
extremely steep, in the range of distances covered by these models, hence
correction by radial transfer should not be dramatically relevant, as
we are going to demonstrate.
\begin{figure}[!b]
\epsscale{1.0}
\plotone{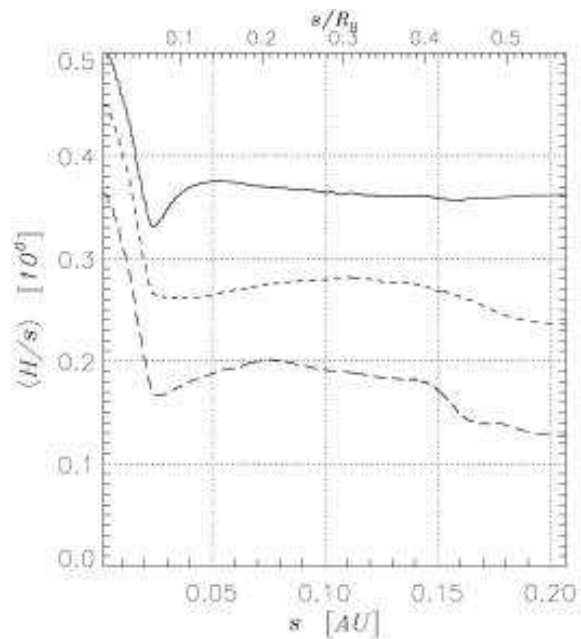}
\caption{Average circumplanetary disk aspect ratio around 
       $\Mp=1\;\MJup$ protoplanets.
       The solid, short-dash, and long-dash lines reproduce the H-,
       C-, and A3-model, respectively.
\label{f:avhs}}
\end{figure}
\subsubsection{Radial Radiation Transfer in Circumplanetary Disks}
\label{sssec:localtransf}

Since the temperature gradient in the vicinity of a protoplanet may be
quite steep, one of the hypothesis used in \refsec{ssec:energyeq}
might locally break down. We refer to the assumption that the 
radiative flux in the vertical direction overwhelms those in the 
horizontal direction. With a semi-analytical  analysis it is 
possible to make an \textit{a posteriori} check of that hypothesis and
shed some light over this matter.
For simplicity we will suppose that the temperature distribution has a
cylindrical symmetry around the planet, which is indeed what is 
observed in all of the models.
Hence the total flux of energy
transferred via radiation is
\begin{equation}
\Lambda \simeq \int_{-\infty}^{+\infty} \frac{\partial F_z}{\partial z}\,
          \ud z
       + \int_{-\infty}^{+\infty}\frac{1}{\dpl}\frac{\partial}{\partial\dpl}[\dpl\,F_\dpl]\,
          \ud z
      = \Lambda_z + \Lambda_{\dpl},
\label{eq:Lambda2d}
\end{equation}
where $\dpl$ is the distance from the planet.
The two terms on the right-hand side can be written as
\begin{equation}
\Lambda_z \simeq -\frac{128\,\sigma_{\mathrm{R}}}{3\,\kappa\,\rho}\, \frac{T^4}{H},
\label{eq:Lambda_z}
\end{equation}
and
\begin{equation}
\Lambda_{\dpl}\simeq
 -\xi^2\,\frac{128\,\sigma_{\mathrm{R}}}{3\,\kappa\,\rho}\,H\,%
  \frac{T^4}{\dpl^2}.
\label{eq:Lambda_s}
\end{equation}
\refEqt{eq:Lambda_z} was obtained by setting 
$\partial T/\partial z\approx T/H$ in \refeqt{eq:Lambda2d}, 
whereas \refeqt{eq:Lambda_s} was
derived by adopting \refeqt{eq:syntT}, which we have checked to hold
as long as $\bar{\dpl}\lesssim\dpl\le\RH$.
The validity of the relation~(\ref{eq:lambda1}),
namely that $\Lambda\simeq\Lambda_z$, depends
on the ratio of the left-hand sides of
\refeqs{eq:Lambda_z}{eq:Lambda_s}:
\begin{equation}
\frac{\left|\Lambda_{\dpl}\right|}{\left|\Lambda_{z}\right|}
= \xi^2\,\left(\frac{H}{\dpl}\right)^2.
\label{eq:Lratio0}
\end{equation}

According to \refeqt{eq:Lratio0}, only when $H<\dpl$ the
radiative cooling (\refeqp{eq:Lambda}) included in the
energy equation is also locally a good approximation.
As shown in \refFgt{f:avhs}, this seems indeed to be 
the case in models involving Jupiter-size planets. 
To cast the above ratio in a more explicit form, one can use the expression
for $H$ (\refeqp{eq:H}) in the limit $\dpl\ll r$, which yields
\begin{equation}
\left(\frac{H}{\dpl}\right)^2=\frac{c_s^2}{\gamma}\,\left(\frac{\dpl}{G\,\Mp}\right).
\label{eq:Hs0}
\end{equation}
By using the form of the sound speed in \refeqt{eq:cs}, one gets
\begin{equation}
\left(\frac{H}{\dpl}\right)^2=\left(\frac{k\,T}{\mu\,m_\mathrm{H}}\right)\,%
                         \left(\frac{\dpl}{G\,\Mp}\right).
\label{eq:Hs1}
\end{equation}
Therefore, because of the assumption made above on the temperature
profile, \refeqt{eq:Lratio0} becomes
\begin{equation}
\frac{\left|\Lambda_{\dpl}\right|}{\left|\Lambda_{z}\right|}
= \xi^2\,\left(\frac{k\,\langle T \rangle_{\bar{\dpl}}}{\mu\,m_\mathrm{H}}\right)\,%
        \left(\frac{\bar{\dpl}}{G\,\Mp}\right)\,%
        \left(\frac{\bar{\dpl}}{s}\right)^{\xi-1}.
\label{eq:Lratio}
\end{equation}
Since the distance $\dpl$ is constrained by the applicability
range of \refeqt{eq:syntT}, the ratio given by \refeqt{eq:Lratio} is
supposedly meaningful roughly for $\dpl\in[0.1\,\RH, \RH]$, for the
models considered here. 
With the appropriate values, the ratio~(\ref{eq:Lratio}) becomes 
\begin{equation}
\frac{\left|\Lambda_{\dpl}\right|}{\left|\Lambda_{z}\right|}
= 0.145\,\xi^2\left(\frac{\langle T \rangle_{\bar{\dpl}}}{1000\,\kelvin}\right)%
        \left(\frac{\MJup}{\Mp}\right)^{2/3}%
        \left(\frac{\bar{\dpl}}{s}\right)^{\xi-1}.
\label{eq:Lratiosim}
\end{equation}
\refEqt{eq:Lratiosim} is an evaluation of \refeqt{eq:Lratio} at a
distance $\dpl$ from the planet, whose length shortens as $\Mp^{1/3}$. Since
the squared sub-disk aspect ratio grows as $1/\Mp$, the flux ratio goes as
$\Mp^{2/3}$. 
From \refTab{tb:syntTav}, we see that
$|\xi-1|$ is between $0.03$ and $0.12$ in C-models and between $0.04$
and $0.21$ in H-models. Therefore, the right-most term in 
\refeqt{eq:Lratiosim}
can be considered as a unity term. Thus, around
$\dpl\approx\bar{\dpl}=0.1\,\RH$,  
$\left|\Lambda_{\dpl}\right|/\left|\Lambda_{z}\right|=0.13$ for the
Jupiter-mass H-model and a factor two as small for the C-model. 
Similar small numbers are obtained from both the A2 and A3 model.
For $\Mp/\MJup=0.1$, the ratio
increases, respectively, to $0.2$ and $0.3$, which is consistent with the
applicability limit of the two-dimensional approximation. 
Thereby, apart from regions closer than $\sim 0.1\;\RH$ to the
accreting planet which, probably, cannot be properly investigated 
by these computations, radiation 
transport in the radial direction does not play a major role in the
energy budget of circumplanetary disk material. This is in agreement
with the standard scenario of the late stages of protojovian disks
\citep[e.g.,][]{coradini1989,canup2002}. 

\subsection{Non-accreting Models}
\label{ssec:noacc}

\begin{figure*}[!t]
\epsscale{2.0}
\begin{center}
\mbox{%
\plottwo{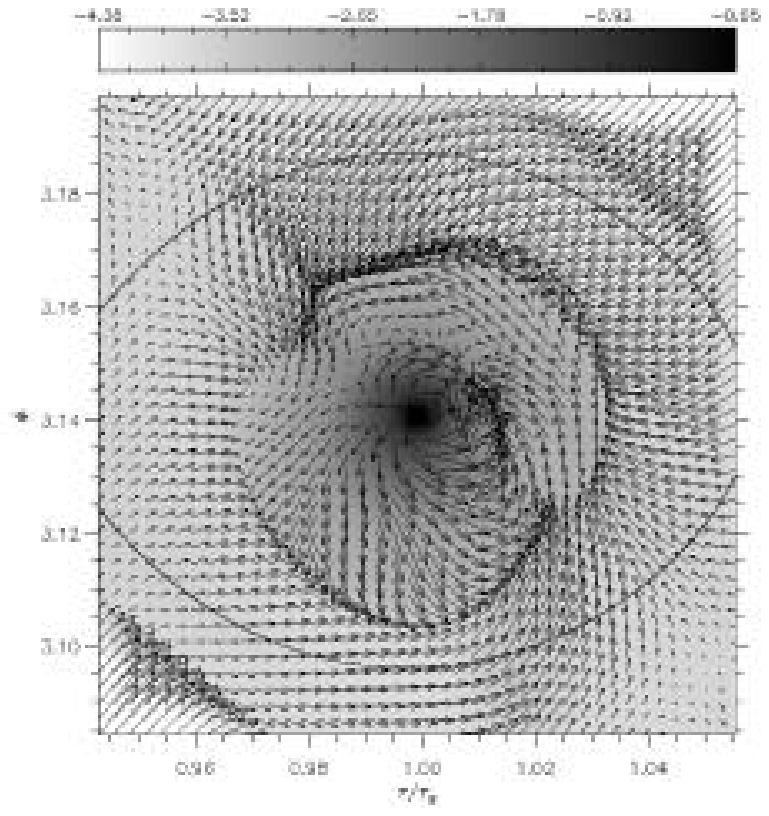}{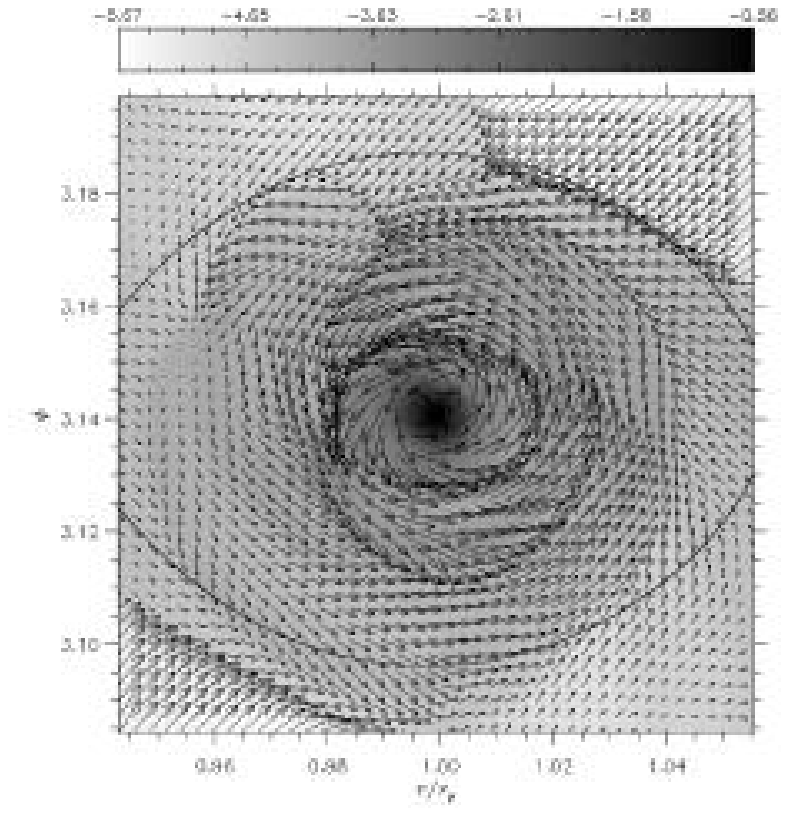}}
\mbox{%
\plottwo{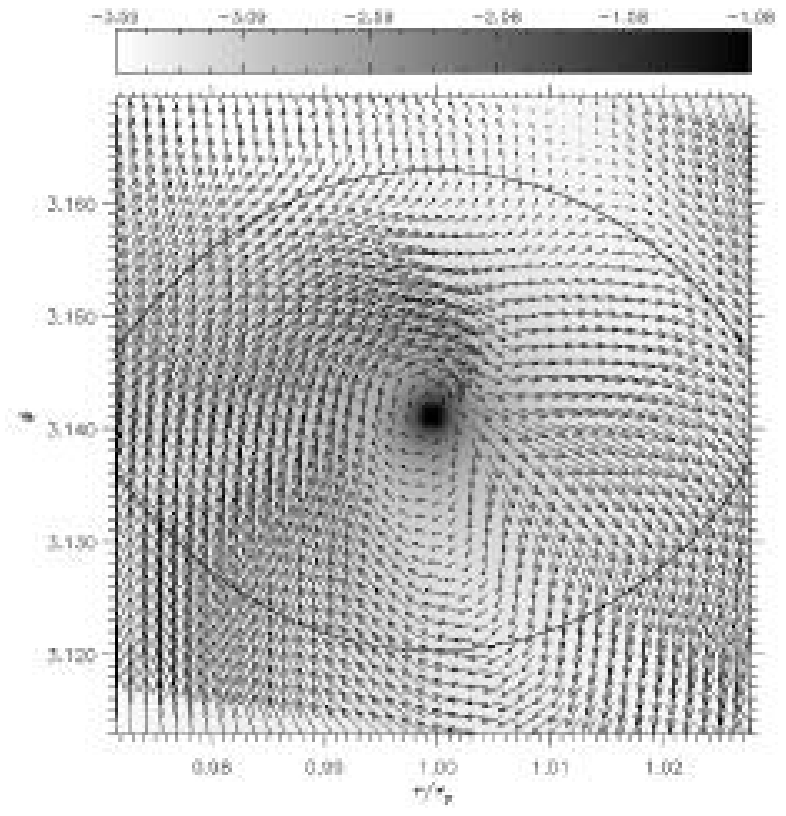}{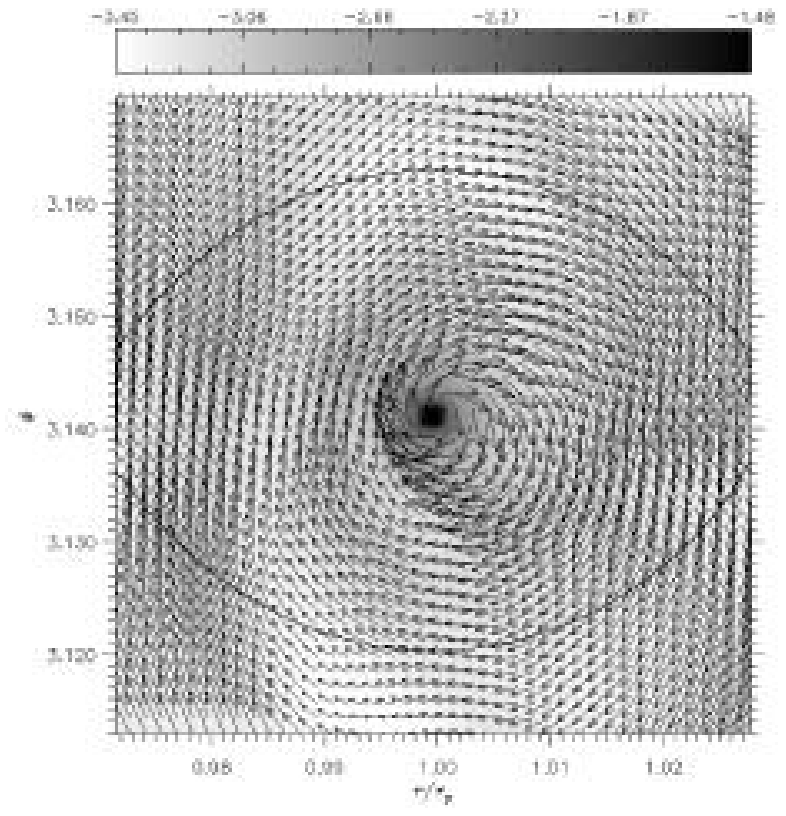}}
\end{center}
\caption{\small{Density distribution and velocity field around $\Mp=1\;\MJup$
       (\textit{top}) and $\Mp=0.1\;\MJup$
       (\textit{bottom}) non-accreting models.
       Left panels show H-models while right ones show to C-models.
       Colors scale logarithmically in all of the panels.
       In the units used in the color bar, $-3$ corresponds to
       $\Sigma=328.7\;\sdunits$.}
\label{f:zpic3}}
\end{figure*}
Accreting models assume that the planetary accretion rate can adjust to
the rate at which matter can be supplied by the surrounding
environment. We now investigate the opposite situation, namely when
no further accretion onto the protoplanet is allowed.

The overall flow structure and density distribution of non-accreting
protoplanets are drastically different from those of accreting ones.
This can be clearly seen in \refFgt{f:zpic3}, where we show the
surface density and the flow field for $\Mp=1\;\MJup$ (upper panels) and
$0.1\;\MJup$ (lower panels) objects.
The standard scenario of a Keplerian-like sub-disk is completely
altered. This is a consequence of the large pressure gradient, built
up by the large density gradient, which
allows matter to move on orbits not constrained by the centrifugal
balance. Therefore, the two-arm spiral feature is replaced by a more
complex system of multiple shock fronts across which material is first
deflected toward the planet and then ejected outward, as also indicated by
the sign of the in-fall velocity $w\subscr{in}$ within the region.

\begin{figure*}[!t]
\epsscale{2.0}
\begin{center}
\mbox{%
\plottwo{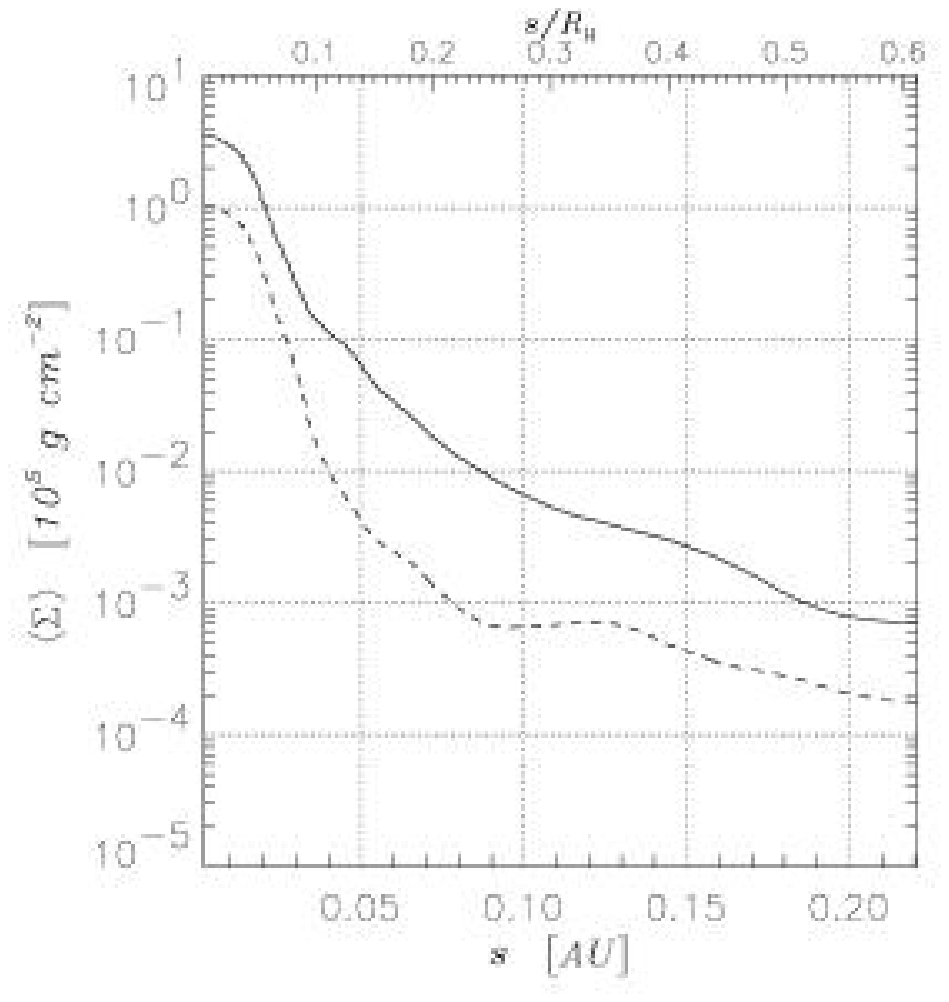}{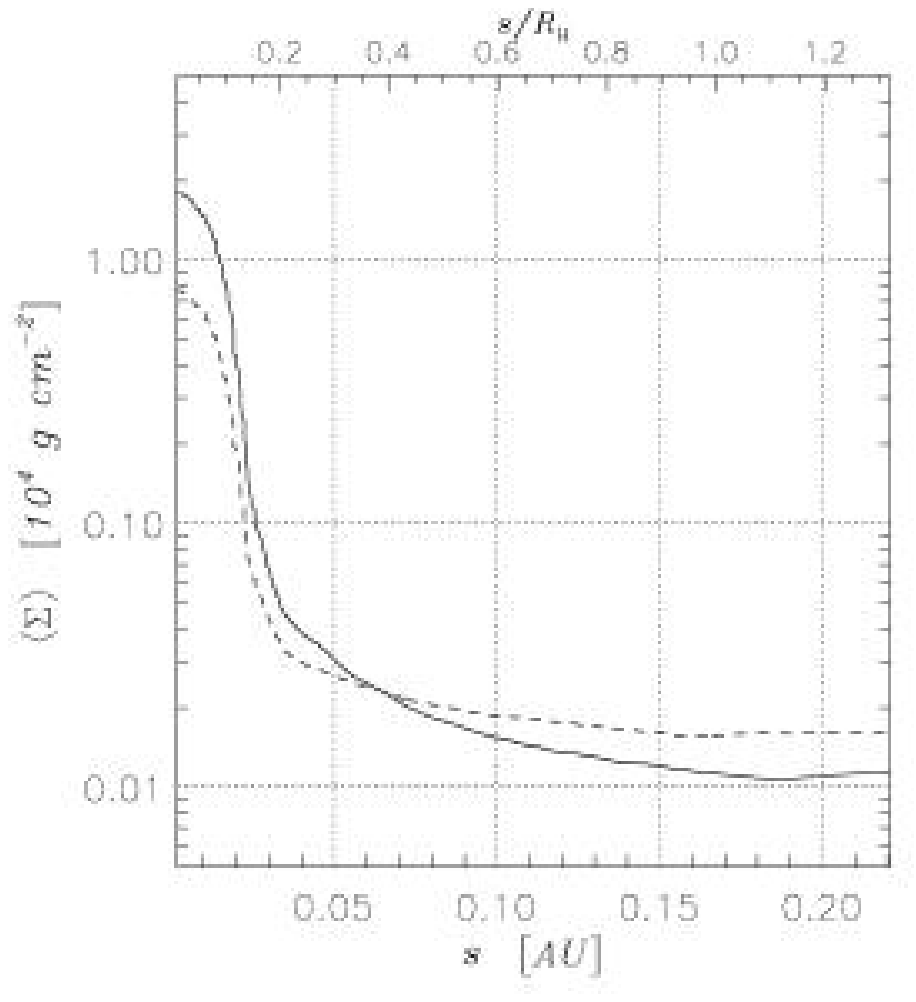}}
\mbox{%
\plottwo{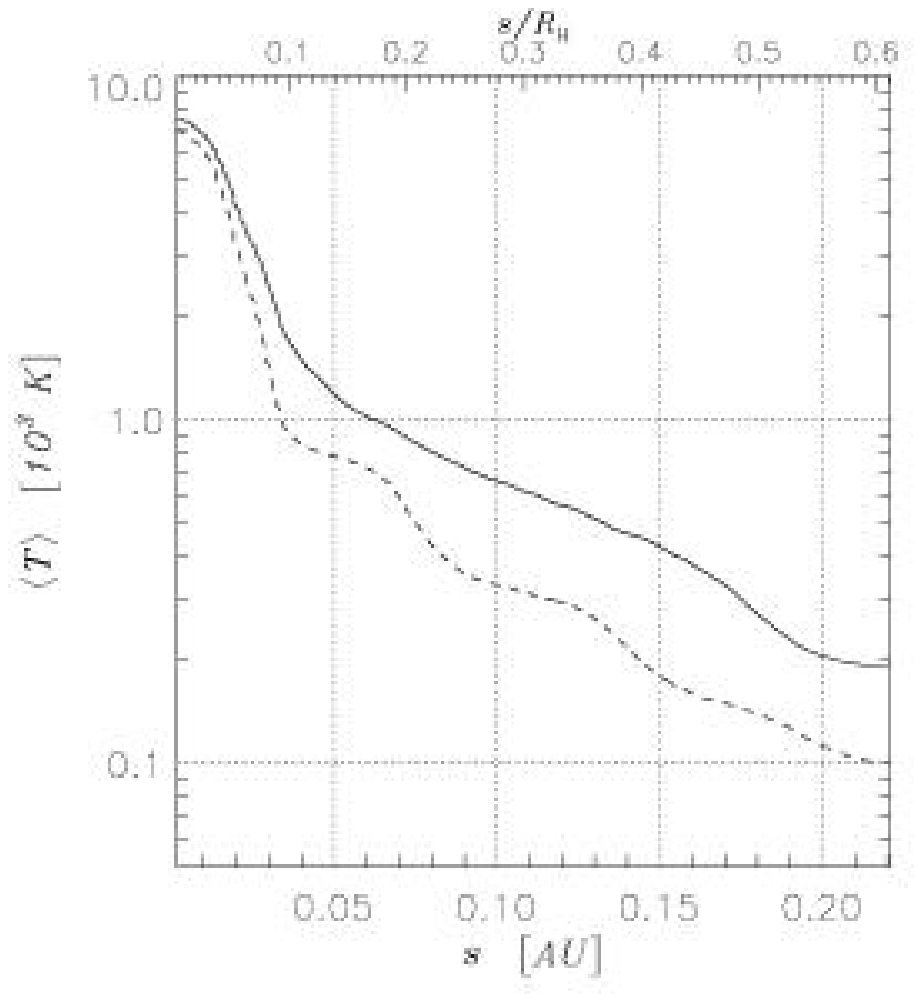}{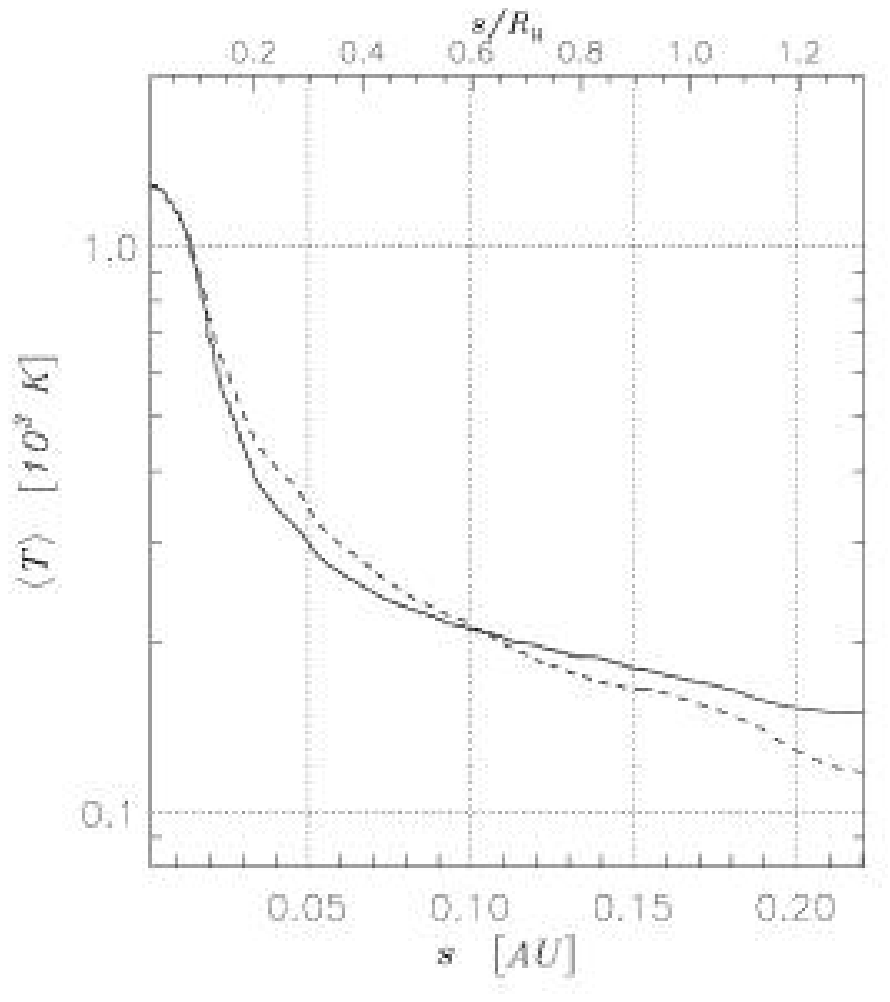}}
\end{center}
\caption{\small{Average surface density (\textit{top}) and temperature
       (\textit{bottom}) around $1\;\MJup$
       (\textit{left}) and $0.1\;\MJup$ (\textit{right}) non-accreting
       protoplanets.
       Solid lines refer to the H-models whereas short-dash
       and long-dash lines belong to C-models.}
\label{f:avprofna}}
\end{figure*}

The flow
strongly diverges in a zone that has a radius equal to $\approx
0.4\,\RH$ in the upper-left 
panel (H-Jupiter model) and to $\approx 0.2\,\RH$ in the upper-right
panel (C-Jupiter model). The average surface density inside of $\RH/3$
is between one and two and a half orders of magnitude larger than 
that measured in accreting models. The mass collected in $\dpl\le 0.8\;\RH$ is
$1.97\times 10^{-4}\;\MJup$  and $4.30\times 10^{-5}\;\MJup$ for the
Jupiter-mass H-model and the C-model, respectively.
The azimuthal average of the circumplanetary density, around
Jupiter-size objects, can be approximated to 
$\langle\Sigma\rangle\simeq
1.2\times10^{4}\,\left(0.1\,\RH/\dpl\right)^{2.8}\;\sdunits$ (H-model) and to 
$\langle\Sigma\rangle\simeq
1.5\times10^{3}\,\left(0.1\,\RH/\dpl\right)^{2.7}\;\sdunits$ (C-model).
The top panel of \refFgt{f:avprofna} shows the averaged surface
density profiles around $\Mp=1\;\MJup$ (left).

A major difference can be also observed in the sub-disk circulation of
low-mass models (\refFgp{f:zpic3}, lower panels). In fact, the fluid
rotates in a clockwise direction. In general, the direction of
rotation in sub-disks 
is determined by the balance among the Coriolis force, the pressure gradient,
and the gravitational attraction by the planet.
Basically, referring to the lower-right panel of \refFgt{f:zpic3},
the Coriolis force deflects fluid elements, orbiting at 
$r\lesssim r_{\mathrm{p}}$, rightward. 
Therefore, it forces matter to cross the
gap and to reconnect to the other side, while still moving in a position
upstream of the perturber body.
In contrast, the term
$-\ud P/\ud\dpl$, supposedly positive, opposes to reconnection when
fluid is
upstream of the planet's location but favors it when matter is 
downstream of the planet.
The prograde rotation that has been encountered so far indicates that the
Coriolis deflection overwhelms the pressure gradient. For this reason
reconnection across the gap always occurs upstream of the planet. 
Evidently, in the
low-mass models ($\Mp=33\;\MEarth$) displayed in the bottom panels of
\refFgt{f:zpic3}, 
the pressure gradient
drives material, flowing from upstream, past the planet's position and
forces it to circulate clockwise around the perturber. Consequently,
matter entering the Hill sphere has an angular momentum anti-parallel
(in the planet's frame) to that of the circumstellar disk.
In the range from $0.1$ to $1\,\RH$, the magnitude of the
rotational velocity
$|\langle w\subscr{rot}\rangle|$ increases roughly as $\sqrt{\dpl}$.
The behavior of $\langle \Sigma \rangle$ resembles that of a
power-law, as in Jupiter-mass cases, with powers $0.9$ and $0.5$,
respectively for the H- and C-model (\refFgp{f:avprofna}, right
panel). Interestingly, no large differences are seen in the
two density profiles.

The high pressure gradient is also caused by the large temperature
gradient. Parameters for analytic approximations of the average temperature 
$\langle T \rangle$, according to \refeqt{eq:syntT}, are reported in
\refTab{tb:syntTavacc}.
\begin{deluxetable}{crrrcrrr}
\tablecolumns{8}
\tablewidth{0pt}
\tablecaption{Fit Parameters for the Averaged Temperature Distribution:
       Non-Accreting Models.\label{tb:syntTavacc}}
\tablehead{%
\colhead{} & \multicolumn{3}{c}{\textsc{C-Models}} &
\colhead{} & \multicolumn{3}{c}{\textsc{H-Models}} \\
 \cline{2-4}\cline{6-8}
 \multicolumn{1}{c}{\raisebox{1.5ex}[-1.5ex]{$\Mp/\MJup$}}&
 \multicolumn{1}{c}{$\langle T \rangle_{\bar{\dpl}}\;[\kelvin]$} &
 \multicolumn{1}{c}{$\xi$} &
 \multicolumn{1}{c}{\textsc{Range}} &
 \colhead{} &
 \multicolumn{1}{c}{ $\langle T \rangle_{\bar{\dpl}}\;[\kelvin]$}&
 \multicolumn{1}{c}{$\xi$} &
 \multicolumn{1}{c}{\textsc{Range}} 
}
\startdata
  $1.0\phn$ & $1.187\times10^{3}$ &  $1.29$  &  $[0.05, 1.0]\,\RH$&
                   & $1.909\times10^{3}$ &  $1.11$  &  $[0.05, 1.0]\,\RH$\\
  $0.5\phn$ & $1.132\times10^{3}$ &  $0.81$  &  $[0.05, 1.0]\,\RH$&
                   & $1.458\times10^{3}$ &  $1.00$  &  $[0.05, 1.0]\,\RH$\\
  $0.2\phn$ & $9.256\times10^{2}$ &  $1.18$  &  $[0.05, 1.0]\,\RH$&
                   & $9.335\times10^{2}$ &  $0.76$  &  $[0.05, 1.0]\,\RH$\\
  $0.1\phn$ & $7.570\times10^{2}$ &  $0.70$  &  $[0.05, 1.0]\,\RH$&
                   & $6.242\times10^{2}$ &  $0.59$  &  $[0.05, 1.0]\,\RH$\\
  $0.06$           & $6.103\times10^{2}$ &  $0.60$  &  $[0.05, 1.0]\,\RH$&
                   &                     &          &                    \\
\enddata
 
\tablecomments{Parameters that enter \refeqt{eq:syntT} when applied to
        non-accreting models.
}
\end{deluxetable}
Maximum temperatures, at $\dpl=0.1\,\RH$, reach $1900\;\kelvin$ and
$1200\;\kelvin$, in $\Mp=1\;\MJup$ H- and C-models respectively 
(bottom-left panel in \refFgp{f:avprofna}). 
At the limit of the Roche lobe the temperature is somewhat higher than that
measured in accreting models and depends on the gap structure. They
vary from $\sim100$ to $\sim200\;\kelvin$. 
Low mass models furnish temperatures that are very similar in both
viscosity regimes.

It is worthy to notice that in either planetary-mass and viscosity
case, the resulting temperature profiles are steeper than the ones
shown in \refFgt{f:zpic2} only when $\dpl < 0.1\,\RH$, 
but they are nearly as steep past this distance.
Actually, from \refTab{tb:syntTavacc} it appears the toward low
planetary masses temperature profiles are even flatter.
Performing the same check as done in the previous section, it turns
out that $\left|\Lambda_{\dpl}\right|/\left|\Lambda_{z}\right|$ is at
most $0.3$. In fact, around Jupiter-mass planets, the ratio $H/\dpl$
varies  between $0.3$ and $0.5$, for distances larger than $0.1\,\RH$. 
Hence, even in these circumstances the type of energy equation
implemented here yields a reasonable description of the thermal structure of
sub-disks, down to $\sim 0.1\,\RH$.

\section{Accretion and Migration}
\label{sec:accmig}
\begin{figure}[!t]
\epsscale{1.0}
\plotone{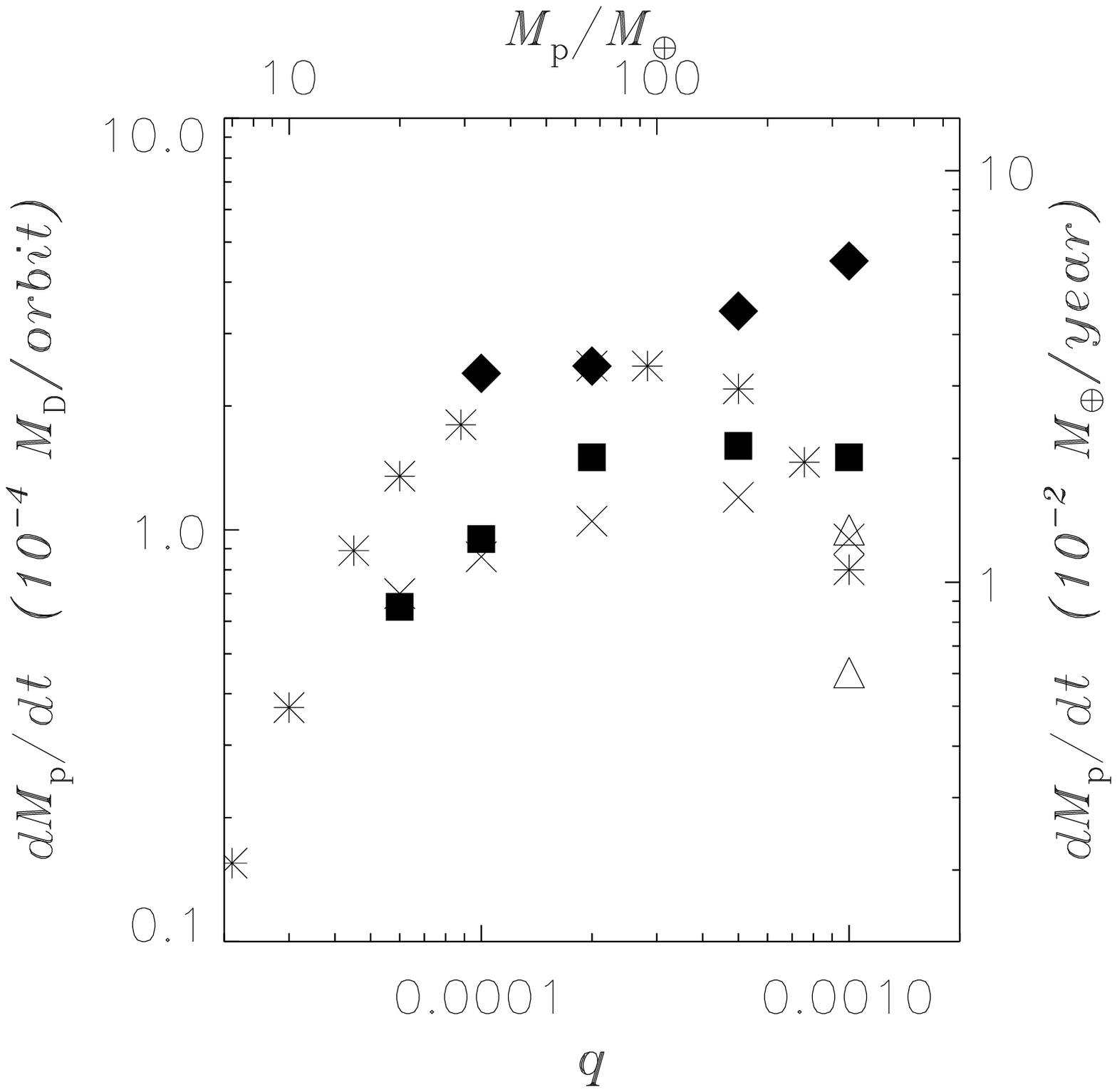}
\caption{\small{Comparison of the accretion rates obtained from different types
       of simulations. Filled diamonds and squares indicate estimates
       provided by H- and C-models. Open triangles show data points
       from $\alpha$-viscosity models A2 (larger value) and A3. 
       Asterisks and crosses represent the rates furnished by local
       isothermal computations in two (\gAA) and three
       (\gApJ) dimensions, respectively. These last two
       sets of values are rescaled to match the disk mass
       $\Md=4.8\times10^{-3}\;\MStar$, used in these
       simulations.}
\label{f:mpdot}}
\end{figure}

The evaluation of the mass accretion rate of protoplanets is carried out
by reducing the surface density around the planet, according to a relation of 
the type $\Delta \Sigma/\Sigma=\Delta t/\tau_{\mathrm{ev}}$ 
(see, \gAA\ for details).  
This reduction process has a time scale $\tau_{\mathrm{ev}}$ that is much
smaller than the integration time step $\Delta t$. 
In \refFgt{f:mpdot} the planetary accretion rate $\dMp$ evaluated in
these computations is compared to that achieved with local-isothermal
2D (\gAA) and 3D (\gApJ) models, where the disk mass was
rescaled to match the values adopted here (see \refsec{sec:parameters}). 

In the context of the accretion history of protoplanets,
it is worthwhile to stress that 2D calculations are only able to 
simulate the late stages of the accretion process, i.e.,
when it mainly occurs via mass transfer in the circumplanetary disk.
This is consistent with the parameterization adopted here to describe
the circumstellar environment.

From \refFgt{f:mpdot}, one can see that C-models (filled squares) 
provide accretion rates quite similar to those obtained from models
with a fixed temperature distribution (crosses). 
This occurrence is related to the quasi-Keplerian, circumplanetary
flow observed in both kinds of computations.
Jupiter-mass models with an $\alpha$-type viscosity (open triangles) 
give rates which are similar to that of the C-model when
$\alpha=10^{-2}$ but somewhat lower 
($\dMp=6\times10^{-3}\;\MEarth\,\mathrm{yr}^{-1}$) when
$\alpha=10^{-3}$, consistently with the low sub-disk mass measured in
this case.
H-models provide values of $\dMp$ (filled diamonds) that, in case of
Jupiter-size planets, fall above the estimates from all of the other
models. 
According to the accretion procedure, the larger the amount of matter 
contained in the accretion region $\dpl=\kappa_{\mathrm{ac}}$, the
larger the accreted mass is. 
Within $\dpl = 0.1\,\RH$ of a Jupiter-mass planet, the
mean density computed in the H-model is roughly three times
larger than that in the C-model (\refFgp{f:zpic1}). 
This is related to the background gap density that is much higher in
H-models, as reported in \refTab{tb:gaps} (see also \refFgp{f:gst}),
due to the larger fluid viscosity.

\begin{figure}[!b]
\epsscale{1.0}
\plotone{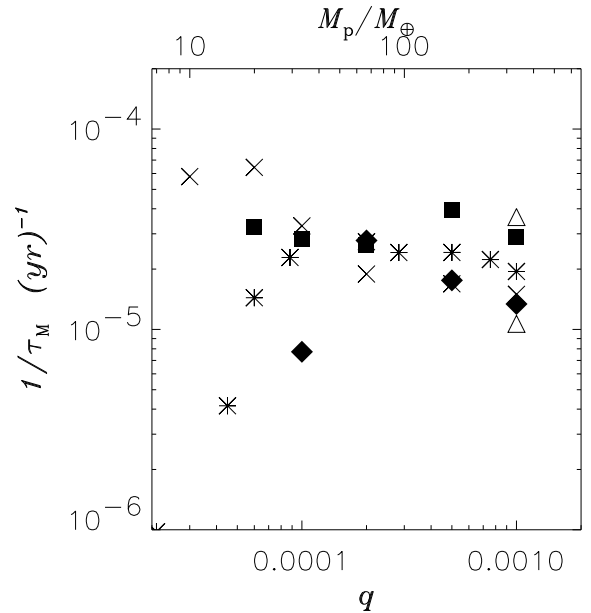}
\caption{\small{Migration time scales plotted against the planetary
       mass. Notations are the same as in \refFgt{f:mpdot}.
       Data points indicate the magnitude of the total
       gravitational torque when this is computed excluding the
       contributions of matter residing inside of the sphere
       $\dpl=\RH$. Only accreting models are considered.
       As for $\alpha$-viscosity models (open triangles), A2
       gives the faster migration rate of the two.}
\label{f:taumig}}
\end{figure}
\refFgt{f:taumig} shows the total gravitational torque
$\mathcal{T}_{\mathrm{D}}$ exerted by 
disk material onto accreting planets, as the inverse of the migration
time scale 
($\mathcal{T}_{\mathrm{D}}\propto |\dot{r}_{\mathrm{p}}|/r_{\mathrm{p}}\equiv%
1/\tau_{\mathrm{M}}$).
This quantity is computed from the gravitational force acting on
the planet, as explained in \gAA. 
The length $\beta$ represents the radius of the
circular region around the planet's position whose contribution is not
taken into account. In \refFgt{f:taumig}, this length is
set equal to $\RH$. 
Outcomes from C-models, and generally those from H-models, compare
well to the local-isothermal computations.
Models with $\alpha$-viscosity provide migration rates on the same
magnitude as well.
C-models show a rather constant value of $\tau_{\mathrm{M}}$
as a function of the planet's mass, whereas H-models present a minimum
at $\Mp=0.2\;\MJup$.  
Migration time scales of H-models differ by less than a factor two
from those of C-models. Only for $\Mp=33\;\MEarth$, $\tau_{\mathrm{M}}$
is appreciably longer. One may attribute this effect to the absence of a
real gap in this model (see \refTab{tb:gaps}), and thus to an early onset
of the Type~I drifting regime.   

\begin{deluxetable}{lrrrr}[!t]
\tablecolumns{5}
\tablewidth{0pt}
\tablecaption{Torques From Different Disk Regions: Accreting Models.
\label{tb:zonaltor}}
\tablehead{%
\colhead{} & \multicolumn{4}{c}{\textsc{Normalized Torques}} \\
 \cline{2-5}
 \multicolumn{1}{c}{\raisebox{1.5ex}[-1.5ex]{\textsc{Model}}}&
 \multicolumn{1}{c}{$\mathcal{A}$} &
 \multicolumn{1}{c}{$\mathcal{B}$} &
 \multicolumn{1}{c}{$\mathcal{C}$} &
 \multicolumn{1}{c}{$\mathcal{D}$} 
}
\startdata
  $1.0\;\MJup$ C & $0.0987$ & $-1.9416$ & $1.3707$ & $-0.5278$ \\
  $1.0\;\MJup$ H & $0.3247$ & $-13.538$ & $12.828$ & $-0.6147$ \\
 $1.0\;\MJup$ A3 & $0.9857$ & $-4.7780$ & $4.9240$ & $-2.1317$ \\
 $1.0\;\MJup$ A2 & $0.0913$ & $-2.1357$ & $1.3905$ & $-0.3461$ \\
  $0.5\;\MJup$ C & $0.1962$ & $-3.6112$ & $2.9099$ & $-0.4949$ \\
  $0.5\;\MJup$ H & $0.3301$ & $-14.940$ & $14.084$ & $-0.4732$ \\
  $0.2\;\MJup$ C & $0.4280$ & $-13.852$ & $13.176$ & $-0.7520$ \\
  $0.2\;\MJup$ H & $0.2406$ & $-14.827$ & $13.840$ & $-0.2536$ \\
  $0.1\;\MJup$ C & $0.4070$ & $-17.989$ & $17.211$ & $-0.6290$ \\
  $0.1\;\MJup$ H & $0.6315$ & $-36.823$ & $35.768$ & $-0.5765$ \\
 $20\;\MEarth$ C & $0.3471$ & $-17.656$ & $16.800$ & $-0.4911$ \\
\enddata
 
\tablecomments{\small{%
            Gravitational torques exerted by the inner disk
            ($\mathcal{A}$) $r\le r_{\mathrm{p}}-2.5\,\RH$,
            and the outer disk 
            ($\mathcal{D}$) $r\ge r_{\mathrm{p}}+2.5\,\RH$.
            The remaining domain 
            ($r_{\mathrm{p}}-2.5\,\RH < r < r_{\mathrm{p}}+2.5\,\RH$) 
            is further divided in the trailing gap region
            ($\mathcal{B}$) $\varphi\le \varphi_{\mathrm{p}}$,
            and the leading gap region
            ($\mathcal{C}$) $\varphi > \varphi_{\mathrm{p}}$.
            The width of the gap region is chosen according to
            the average width observed in \refFgt{f:gst}.
            All torques are normalized to the absolute value of
            the total torque $\mathcal{T}_{\mathrm{D}}$. 
            Furthermore, they exclude the contribution from a circle 
            of radius $\RH$ centered on the planet. 
}}
\end{deluxetable}
In \refTab{tb:zonaltor}, we separate the contributions of different
disk regions to the total torque $\mathcal{T}_{\mathrm{D}}$,
introducing the inner disk ($\mathcal{A}$), the outer disk
($\mathcal{D}$) and the gap region ($\mathcal{B}$ and $\mathcal{C}$). 
Further details on the regions are reported in the Table's note. 
All contributions are normalized to $|\mathcal{T}_{\mathrm{D}}|$.
The width of the gap region is fixed to $5\,\RH$ (see \refFgp{f:gst}) 
and the zone behind the planet ($\mathcal{B}$) is distinguished from 
the one ahead of it ($\mathcal{C}$). 
Torques exerted from inside the Hill sphere are not accounted for.
A large fraction of the contributions arising from zones $\mathcal{B}$
and $\mathcal{C}$ is actually generated in the coorbital regions, 
within a few Hill radii from the planet.
From the entries in \refTab{tb:zonaltor} one can argue that coorbital
torques in H-models are larger in magnitude than they are in C-models.
This is a consequence of the shallower gap of H-models. 
The situation is reversed in the $\alpha$-viscosity models because of 
the density humps at L4 and L5 Lagrangian points. 

\begin{figure*}[!t]
\epsscale{2.0}
\begin{center}
\mbox{%
\plottwo{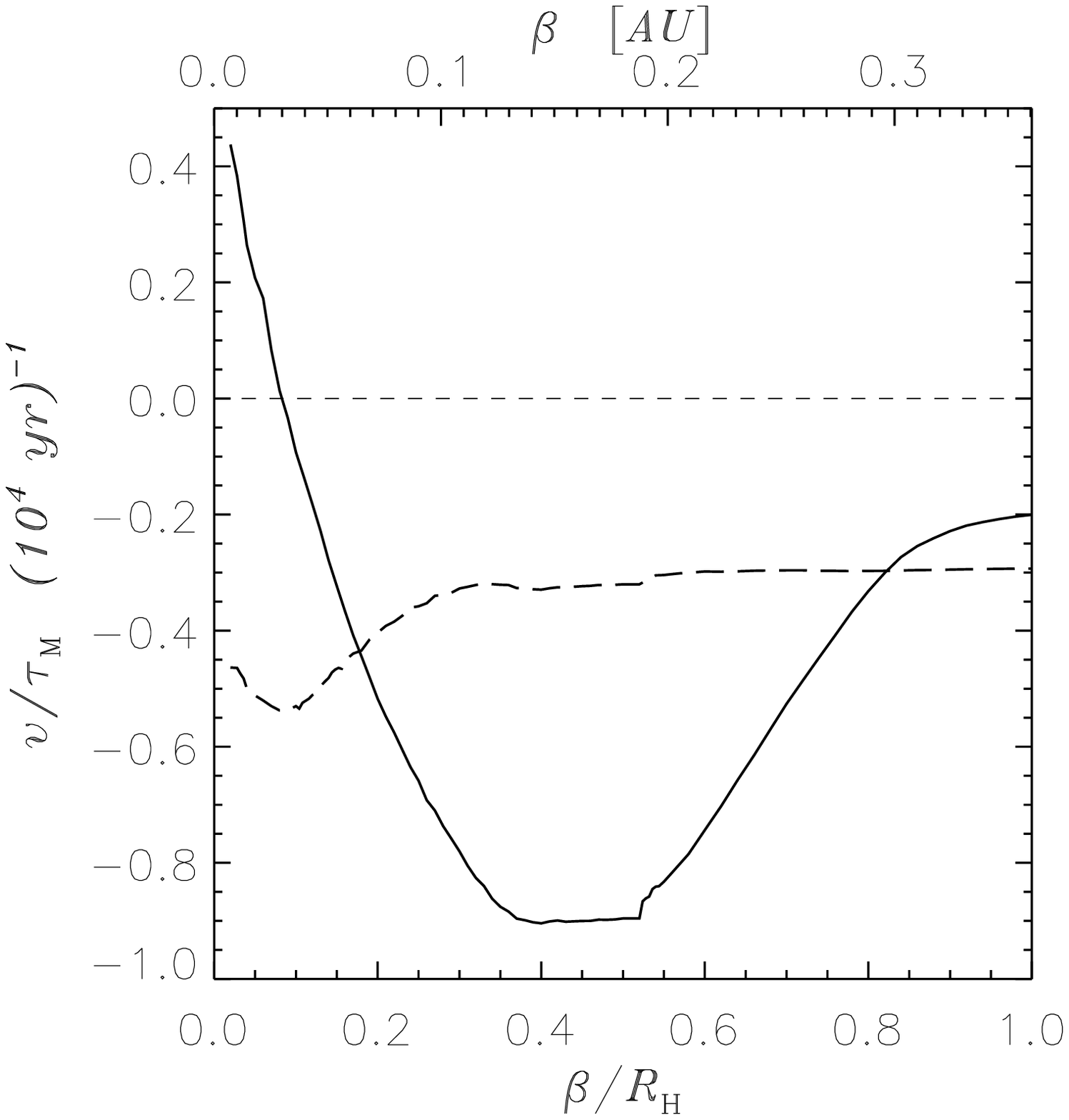}{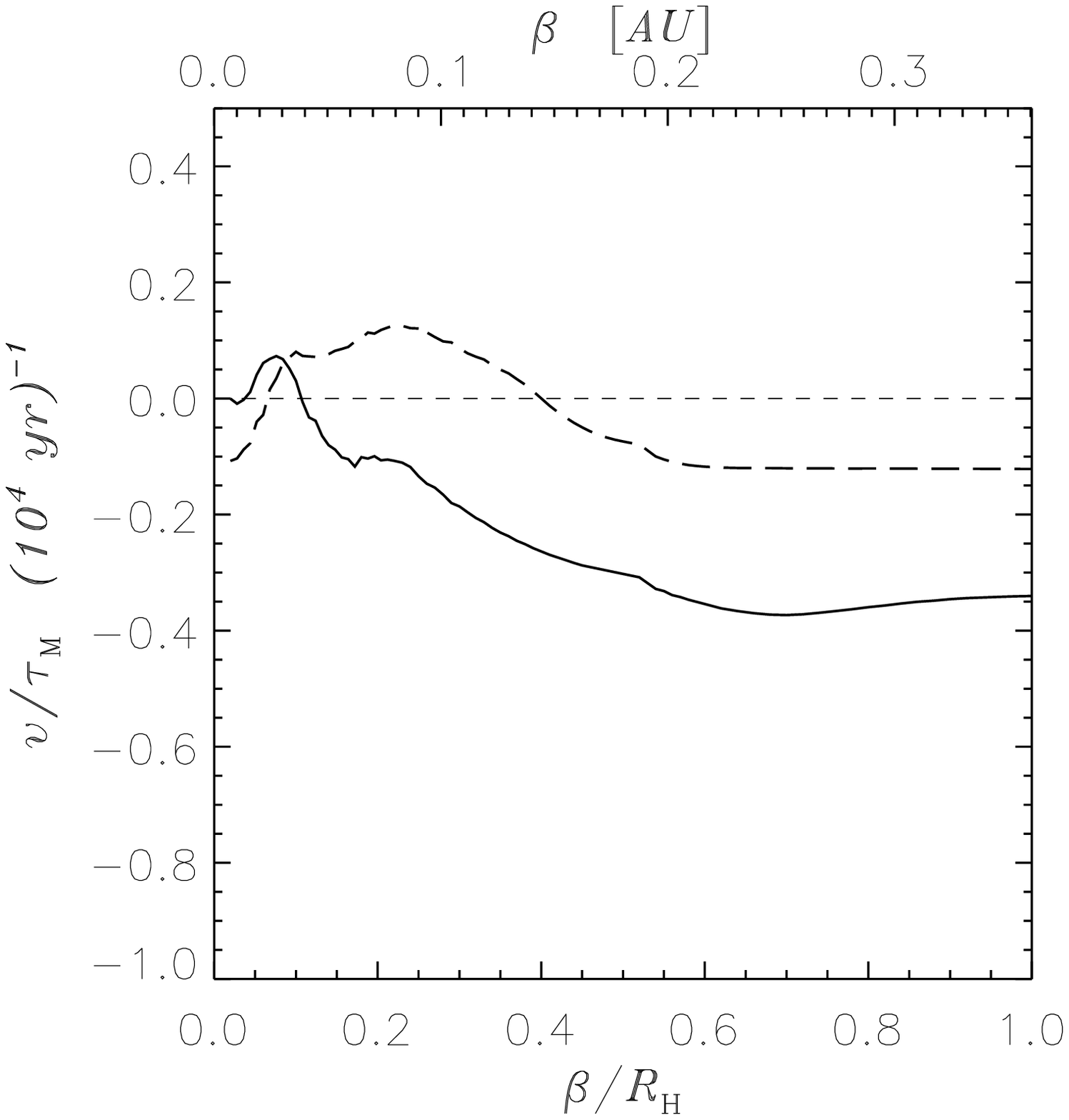}}
\mbox{%
\plottwo{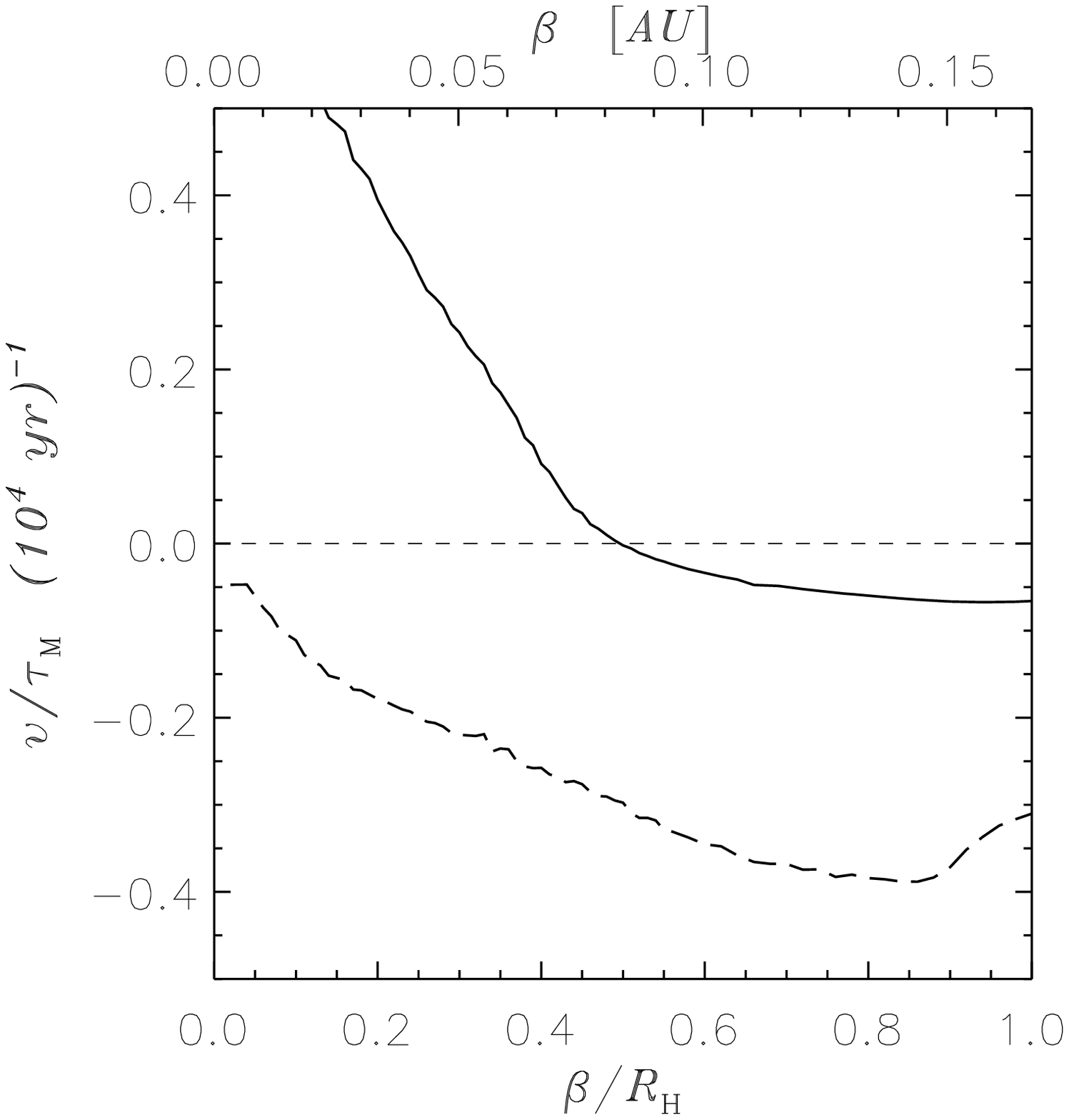}{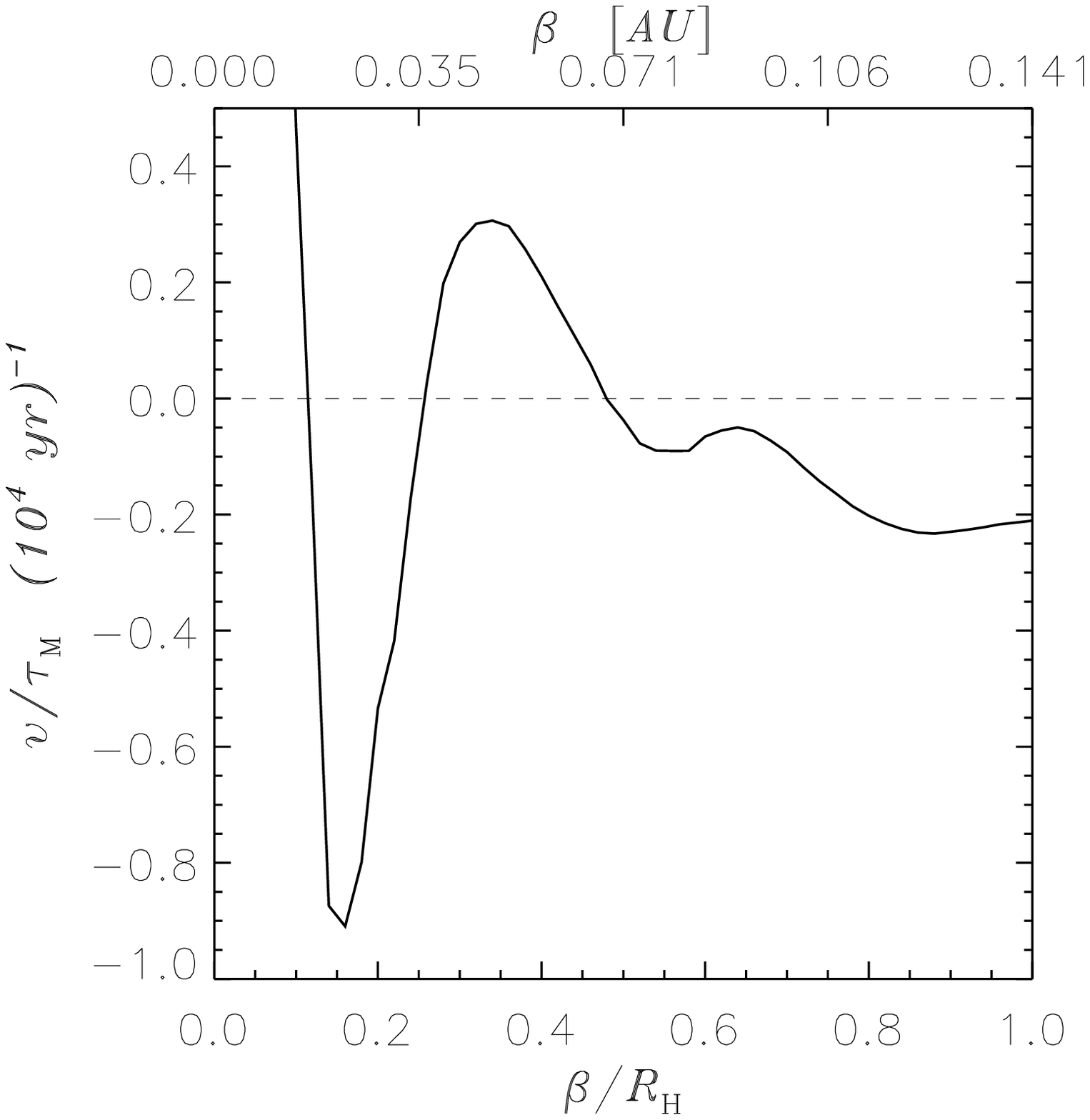}}
\end{center}
\caption{\small{Inverse of the migration time scale versus the radius $\beta$
        of the region excluded from the torque calculation. The
        quantity $v=\dot{a}/|\dot{a}|$ indicates the direction of the planet
        migration. Solid lines refer to models H or A2 while dashed
        lines refer models C or A3. \textit{Top-left panel}. 
        Jupiter-mass models H and C.
        \textit{Top-right}. Jupiter-mass models with $\alpha$-viscosity 
        A2 and A3. 
        \textit{Bottom-left}. $\Mp=0.1\;\MJup$ model H and C.
        \textit{Bottom-right}. $\Mp=20\;\MEarth$ C-model.}
\label{f:tvsd}}
\end{figure*}

As for the influence of the parameter $\beta$ on the total torque,
it turns out that H-models are generally more sensitive than C-models. 
By examining the function
$\mathcal{T}_{\mathrm{D}}=\mathcal{T}_{\mathrm{D}}(\beta)$, 
some more insight can be gained.
For $\Mp=1\;\MJup$ models (\refFgp{f:tvsd}, top-left panel), 
no significant net torque arises from within the Hill sphere in 
the C-model (dashed line). 
The situation appears to be different for the H-model (solid line) because
$\mathcal{T}_{\mathrm{D}}$ decreases to a minimum, around
$\beta=\RH/2$, and then increases and changes sign around
$\beta=0.1\,\RH$. 
Jupiter-mass $\alpha$-viscosity models (\refFgp{f:tvsd}, top-right
panel) display a smooth behavior since the total torque is 
rather constant down to $\beta=\RH/2$.
When $\beta$ diminishes, $\mathcal{T}_{\mathrm{D}}$ increases
(i.e., it has smaller absolute values) and the model with
$\alpha=10^{-3}$ (dashed line) shows a change of sign at $\beta=0.4\,\RH$.
When $\Mp=0.1\;\MJup$ (\refFgp{f:tvsd}, bottom-left panel), 
in either case, $\ud\mathcal{T}_{\mathrm{D}}/\ud\beta<0$, 
because of the positive torques arising within the sub-disk. 
The rate of change is more drastic in the H-model, where the total
torque becomes positive and keeps increasing when $\beta\lesssim \RH/2$.
Torques acting on the $20\;\MEarth$ planet (\refFgp{f:tvsd},
bottom-right panel) are strongly dependent on $\beta$, as  
positive torques are exerted down to
$\approx 0.4\,\RH$ and large negative torques arise between 
$\approx 0.2\,\RH$ and $\approx 0.4\,\RH$.

\section{Conclusions}
\label{sec:conclusions}
In this paper we presented two-dimensional simulations of
disk-planet interactions in which we considered the joint
thermo-hydrodynamics evolution of the system. In order to do that,
we solved an energy equation that accounts for the major
processes responsible for the energetic balance of fluid parcels.
We restricted the treatment of the problem to a flat geometry in order
to make some simplifying 
assumptions on the radiative part of the energy budget, i.e., 
to treat the radiation transfer as a cooling term.
A nested-grid technique was used to resolve different length scales,
therefore both the global and local features could be investigated.

The two-dimensional approximation has been proved to work reasonably
well as long as the planetary mass is larger than a tenth of the mass
of Jupiter (\gApJ). 
Because of this, only planets more massive than $\approx0.1\;\Mp$
have been simulated. 
In order to compromise with the lingering uncertainty on the viscosity
magnitude and to inquire into different temperature regimes,
we considered different constant viscosity and $\alpha$-viscosity 
models. Both accreting and non-accreting planets were simulated.

From the global point of view, and roughly speaking, we can conclude
that circumstellar disk models with a fixed temperature distribution provide 
a reasonably good description of the system, though details of the gap
structure and shape of the disk spirals differ. 
Around Jupiter-mass protoplanets, only with kinematic viscosities as 
low as $10^{15}\;\nunits$, a wide and deep gap is carved in whereas 
it reduces to a shallow density trough when the viscosity is ten times
as large. 
Mean temperatures in the gap can be as low as $20$--$50\;\kelvin$, 
depending on the viscosity regime. The most diluted and cold gap material 
is located where the circumplanetary disk merges into the gap.
The aspect ratio appears smaller
than $0.04$ and it is also marked by a gap-like feature, which shades 
the protoplanet's environment against stellar radiation.
The models with $\alpha=10^{-3}$ show a ring of matter within the gap
that evolves on a viscous time scale.
Regarding smaller planetary masses ($\Mp\approx0.1\;\MJup$), both 
gap and spiral perturbations become quite faint features. Likely, 
only Jupiter-size bodies are capable of digging gaps deep enough
to be observed by present-day instrumentation.

From the local point of view, we obtained the distribution of 
density, temperature, and other thermodynamical quantities 
in circumplanetary disks. Except for the $\alpha$-viscosity model with
$\alpha=10^{-3}$, rather comparable outcomes are provided by models
with different viscosity regimes.
The density, averaged around the planet, shows an exponential radial 
decline.
Sub-disks have masses on the order of $10^{-5}$--$10^{-6}\;\MJup$, which
makes them gravitationally stable.
Temperatures range from several hundred Kelvin degrees, at distances 
$\dpl\lesssim 0.1\RH$ from the planet,
to values between $50$ and $100\;\kelvin$, at the border of the Hill
sphere. Azimuthal averages yield profiles dropping nearly as the inverse of
the distance from the planet. 
Circumplanetary disks have aspect ratios $H/\dpl$ around a few tenths and
they tend to be optically thick. 
All of these properties appear to be consistent with the ``fast-inflow'' 
analytical models of protojovian disks recently developed by
\citet{canup2002}.
By means of a semi-analytical treatment, we showed that, 
also in the sub-disk,
the neglect of radial radiation transfer should not crucially affect our
results. 
Non-accreting models furnish a rather different scenario. 
Around Jupiter-mass objects, circulation in the Roche lobe is not 
Keplerian-like because of the large pressure built up mainly by 
the density gradient and partly by the temperature gradient. 
Around $\Mp=0.1\;\MJup$,
clockwise rotation establishes as a result of the weakening of the
Coriolis force with respect to the pressure gradient term.

Accretion and migration rates inferred by these simulations are
comparable to those evaluated with previous 
purely hydrodynamical local-isothermal computations. 
An analysis of the zonal torques indicates that coorbital torques
increase as viscosity increases. Moreover, negative coorbital
torques, arising from material outside of the Roche lobe and lagging
behind the planet, are generally larger than those from material
ahead of the planet.   
We also found that, in the two-dimensional
approximation, the material lying inside of the Roche lobe is able to 
exert strong gravitational torques on the planet.
Since the density gap is usually filled in H-models (see \refTab{tb:gaps}),
Type~I migration regime might extend to larger planetary
masses (see \refFgp{f:taumig}).

In order to address the issue of how the third dimension affects the
thermo-hydrodynamics of the system, both globally and locally around 
protoplanets,
we are extending this study to 3D simulations in which radiation
transfer is treated in the flux-limited diffusion approximation
\citep{fld}.

\acknowledgments

We thank Udo Ziegler for having made available to us an early FORTRAN
Version of his code \textsc{Nirvana}.
G.~D.\ is deeply grateful to all the friends at the INAF-OAC (Naples) and
RSMAS/MPO (University of Miami) for their warm hospitality.
The paper benefited from the useful comments of an anonymous referee. 
This work was supported by the German Science Foundation and by the  
UK Astrophysical Fluids Facility (UKAFF).
The numerical computations reported here were carried out at the 
Computer Center of the University of Jena and at the UKAFF.

\end{document}